\documentclass[fleqn,10pt]{wlscirep}

\usepackage[utf8]{inputenc}
\usepackage[T1]{fontenc}
\usepackage[labelfont={sc,bf}]{caption} 
\usepackage[labelfont=bf]{subcaption}
\usepackage[flushleft]{threeparttable}

\usepackage{lineno}

\usepackage{xr}
\usepackage{pifont}
\newcommand{\xmark}{\ding{55}}

\newcommand{\beginsupplement}{%
        \setcounter{table}{0}
        \renewcommand{\thetable}{S\arabic{table}}%
        \setcounter{figure}{0}
        \renewcommand{\thefigure}{S\arabic{figure}}%
        \renewcommand{\thesubsection}{S\arabic{subsection}}
                        \setcounter{subsection}{0}
                \renewcommand{\thesection}{S\arabic{section}}
                \setcounter{section}{0}
     }

\usepackage{amsmath, amsthm, amssymb, amsfonts}
\usepackage{bm}

\title{Knowledge and Social Relatedness Shape Research Portfolio Diversification}

\author[1,*]{Giorgio Tripodi}
\author[2,3,*]{Francesca Chiaromonte}
\author[1,4,*]{Fabrizio Lillo}

\affil[1]{\footnotesize Scuola Normale Superiore, 56126 Pisa, Italy}
\affil[2]{\footnotesize 
Institute of Economics and EMbeDS, Scuola Superiore Sant’Anna, 56127 Pisa, Italy}
\affil[3]{\footnotesize Department of Statistics, The Pennsylvania State University, University Park, PA 16802}
\affil[4]{\footnotesize Department of Mathematics, University of Bologna, 40126 Bologna, Italy}
\affil[*]{\footnotesize Correspondence to: Giorgio Tripodi, \url{giorgio.tripodi@sns.it};
Francesca Chiaromonte, \url{fxc11@psu.edu};
Fabrizio Lillo, \url{fabrizio.lillo@unibo.it}}

\begin{abstract}
Scientific discovery is shaped by scientists' choices and thus by their career patterns. The increasing knowledge required to work at the frontier of science makes it harder for an individual to embark on unexplored paths. Yet collaborations can reduce learning costs -- albeit at the expense of increased coordination costs. In this article, we use data on the publication histories of a very large sample of physicists to measure the effects of knowledge and social relatedness on their diversification strategies. Using bipartite networks, we compute a measure of topic similarity and a measure of social proximity. We find that scientists' strategies are not random, and that they are significantly affected by both. Knowledge relatedness across topics explains $\approx 10\%$ of logistic regression deviances and social relatedness as much as $\approx 30\%$, suggesting that science is an eminently social enterprise: when scientists move out of their core specialization, they do so through collaborations. Interestingly, we also find a significant negative interaction between knowledge and social relatedness, suggesting that the farther scientists move from their specialization, the more they rely on collaborations. Our results provide a starting point for broader quantitative analyses of scientific diversification strategies, which could also be extended to the domain of technological innovation -- offering insights from a comparative and policy perspective.
\end{abstract}

\begin{document}

\flushbottom
\maketitle

\thispagestyle{empty}

\section{Introduction}

\noindent
The activities of scientists and innovators often span several areas, with choices of research endeavours driven by a variety of factors.
The "essential tension" between exploration and exploitation described by Kuhn certainly characterizes research careers \cite{kuhn1979essential}, but scientists can evolve ways to handle this trade-off.
On the one hand, advances in science and technology create a "burden of knowledge" \cite{Jones2009}; the sheer amount of information required to move forward has grown, and larger educational costs may force scientists and innovators towards a narrower specialization.
On the other hand,  contemporary science is dominated by teams that bring together different expertise -- albeit at a cost in terms of coordination and credit sharing \cite{wuchty2007}.  
This article focuses on the analysis of scientists' research portfolio, investigating the roles of knowledge relatedness (among research topics) and social relatedness (among authors), as well as their interaction, as drivers of diversification.

Recent efforts to better characterize patterns in research and innovation activities produced valuable insights. 
For instance, based on a knowledge network created using MEDLINE articles annotated with chemical entities, Foster \textit{et al.} \cite{foster2015tradition} quantitatively analyzed the dichotomy between exploration and exploitation. According to their taxonomy, each new article can expand or consolidate the knowledge space by generating a new chemical relationship (i.e., a new combination) or contribute to an existing one. Results show that research strategies (i.e., the types of articles produced) are stable over time and exploitation is preferred over exploration, despite a growing number of opportunities. Exploration is riskier, with rewards (i.e., citations) that are higher but insufficient to compensate the risk.  
In the domain of physics, Pan \textit{et al.}\cite{pan2012evolution} focused on the temporal evolution of interdisciplinary research. 
The authors constructed and analyzed yearly snapshots of the connections among physics sub-fields uniquely identified through PACS codes.
Results show that connectivity, and thus interdisciplinarity within physics, increased -- but in a non-random way that reflects the hierarchical structure of sub-fields. In particular, \textit{condensed matter} and \textit{general physics} acted as hubs for the increasing number of connections. 
Recently, Sun \textit{et al.}\cite{Sun2020TheEO} proposed a novel framework, based on time-varying networks, to track knowledge flows within and across physics sub-fields. Such a method is able to highlight the increasing general trend towards interdisciplinary research as well as identify interesting patterns of influence among sub-fields over time.
More directly related to our purposes, recent works \cite{battiston2019taking, aleta2019explore, jia2017} 
collected compelling empirical evidence on physicists' research endeavours. Battiston \textit{et al}.\cite{battiston2019taking} provided a comprehensive census of academic physicists active in recent decades. The authors charted a thorough picture of the evolution of various fields in terms of number of scientists, productivity (including impact and recognitions such as Nobel prizes), team size and role of chaperones -- highlighting a rich heterogeneity among specializations. Moreover, Battiston \textit{et al.}\cite{battiston2019taking} mapped "migration" flows by comparing the field in which a given scientist published her first paper with the one characterizing her later research interests. 

Also Aleta \textit{et al.}\cite{aleta2019explore} mapped flows among physics sub-fields, with the aim of investigating the "essential tension" in the evolution of scholars' research interests.
The authors defined a measure of exploration comparing early- and late-career ranges of actives, and tracked flows using origin-destination matrices among fields. 
Results suggest a preference for exploration over exploitation, but concentrated within the same broad area of research, and non-random transitions among different areas. 
Jia \textit{et al.}\cite{jia2017} observed that the frequency of scientists decays exponentially as one considers increasing degrees of change in interests. 
In order to reconstruct the macroscopic patterns that drive such evolution, the authors proposed a random walk model over a stylized knowledge space, which reproduces empirical observations thanks to the inclusion of key features such as heterogeneity, subject proximity and recency. 
Finally, Zeng \textit{at al.}\cite{zeng2019increasing} analysed the dynamics of "topic switching" by exploring co-citation networks. Results suggest a growing propensity to switch among topics but also that such a strategy might hamper productivity, especially for early-career researchers.

Despite the growing body of evidence and stylized facts provided by this literature, much remains to be done to disentangle and quantify the roles of different contributing factors.
To make progress in this direction, we investigate scientists' research portfolio diversification by quantifying potential drivers of exploration, or, to put it differently, the hurdles faced by scientists when they move out of their immediate specialization.
We use a network approach to compute a measure of similarity among research sub-fields, define a measure of social relatedness and track of scientists' diversification patterns.
We build our empirical strategy upon the intuition of Breschi \textit{et al.}\cite{Breschi2003}, who used patent data to explore the nature and degree of coherence in firms' technological diversification. 

Our analysis proceeds as follows. First, we test and reject the hypothesis that research portfolio diversification is random. Second, we use regression techniques to characterize how subject and social proximity affect diversification, controlling for possible confounding factors. Third, we quantify the relative importance of our relatedness measures.
We provide robust empirical evidence that knowledge and social relatedness are both significant statistical predictors of diversification, as is their interaction -- which corroborates the notion that collaborations modulate knowledge acquisition, especially when scientists move far from their own specialization.
Like many of the articles mentioned above, we analyze data concerning physicists. This focus 
is due in part to the central role of physics among the \textit{hard} sciences, and in part to the reliability of data collected labeling articles through the PACS codes.
Nevertheless, our approach is fully general and could be used in different domains.

\section{Results}

\subsection{Data description}

\noindent
We use the American Physical Society (APS) dataset to reconstruct the activities of 197,682 physicists who published at least one paper in one of the APS outlets in the period ranging from 1977 to 2009 (see section \ref{data} for details). All articles in APS journals are classified according to hierarchical codes that map into physics fields and sub-fields (i.e., PACS codes). 
For our analyses (see section \ref{model}), we filter out authors and sub-fields that appear only sporadically in the data. Specifically, we focus on 105,558 authors who published at least two articles, covering a minimum of two sub-fields over a restricted set of 68 PACS which appear in at least four articles. 

\begin{figure}
\centering
{\includegraphics[scale=1]{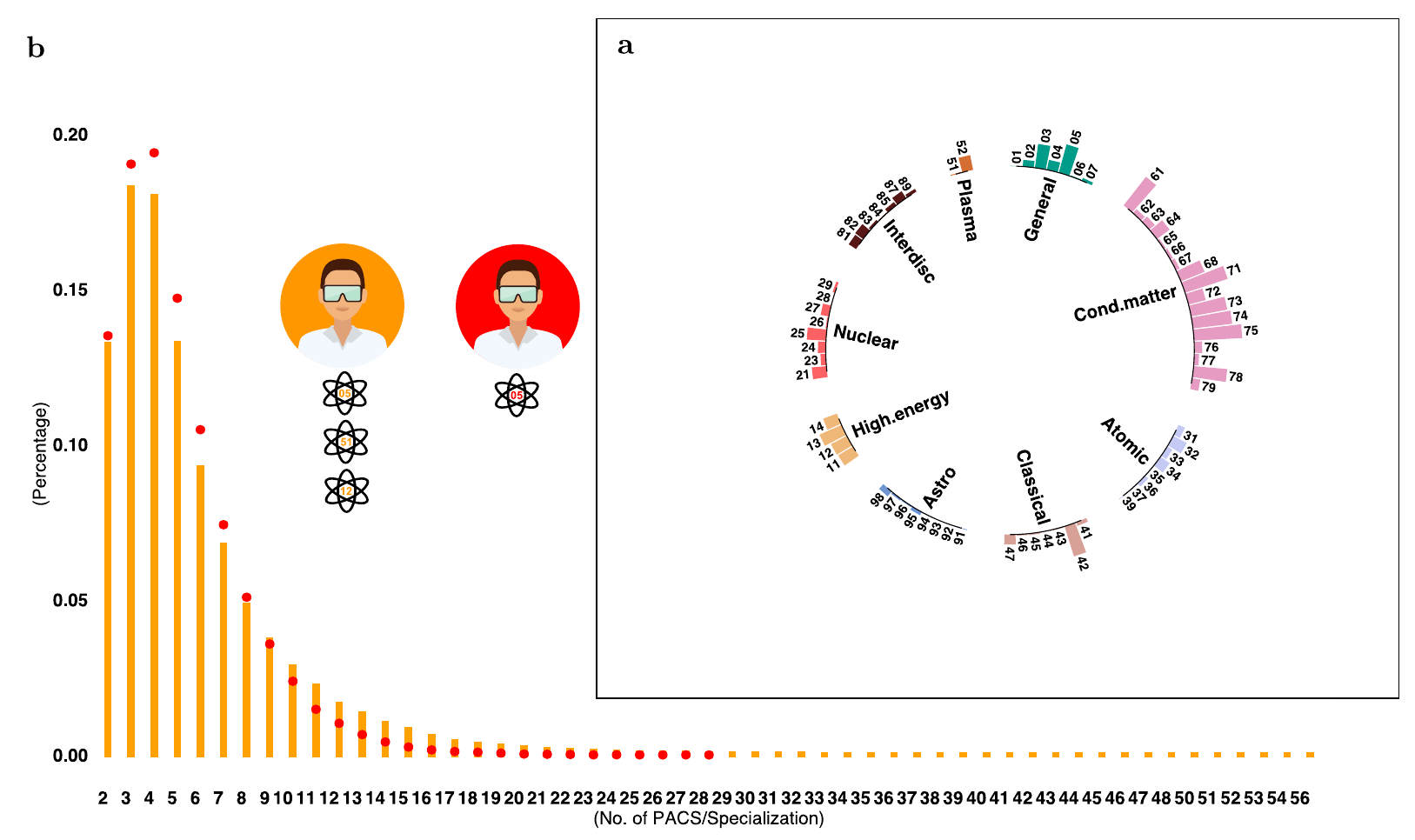}}
\caption{{\small
\textbf{Popularity of fields and scientists' degree of diversification/specialization}. 
\textbf{(a)} Circular bar-chart showing the number of articles assigned to each sub-field in the one-digit PACS codes, taking into account their hierarchical structure. 
The chart highlights the popularity of \textit{Condensed Matter} research in both size and scope.
\textbf{(b)} Distribution of scientists' degree of diversification (the number of sub-field they explored; orange bars) and of their relative specialization (the number of sub-fields in which they have a scientific advantage; red dots).  Scientists explore several sub-fields, but specialize in only a few -- despite the existence of some individuals with a truly interdisciplinary path, by and large research portfolios are fairly limited in scope. Inset: pictorial description of a scientist who explored three sub-fields (orange) but has only one specialization (PACS 05: red).}
\label{popularity}}
\end{figure}

Figure~\ref{popularity} provides a general description of the data and some insights. Figure~\ref{popularity}\textbf{a} shows the popularity, in terms of number of articles, of fields and sub-fields (one- and two-digit level PACS codes, respectively). As expected, PACS popularity is highly heterogeneous and 
reflects the prominence of \textit{condensed matter} research in the last decades. 
Figure~\ref{popularity}\textbf{b} shows scientists' degree of diversification and their relative specialization, as defined in section \ref{core}. The research portfolio of most scholars in our dataset is fairly limited in scope, with a large majority of scientists diversifying in no more than 5 sub-fields. 
The choice of subjects, however, is not random -- as we demonstrate in the next section.

\subsection{Diversification is not random}\label{randomtest}
\noindent 
Do scientists, much like firms \cite{Teece1994, Breschi2003}, shape their research portfolios based on specific strategies and constraints? 
To address this question quantitatively, we draw a parallel with ecology: as species may co-occur in distinct sites, sub-fields may overlap in research portfolios. 
Measuring the relatedness of species based on their geographical co-occurrence is analogous to measuring the relatedness of sub-fields based on their overlap in scientists' ranges of activity. 
Thus, the PACS-Authors binary bipartite network resembles a presence-absence matrix \cite{veech2013probabilistic}. 
The monopartite projection of this bipartite network (see section \ref{infer}) on the PACS layer carries a critical piece of information: for each pair of PACS, it tells us how many scientists are active in both sub-fields irrespective of the number of articles, drawing a diversification network. 

We can assess this network contrasting it against an appropriate null model. 
Which sub-fields overlaps are over- or under-represented relative to what we would expect under the assumption that scientists picked research topics at random, but taking into account the popularity of sub-fields?  
Under a random model, the probability that $x$
scientists are active both in sub-field $a$ and in sub-field $b$, given that $S_a$ and $S_b$ scientists are active in these sub-fields, follows a hypergeometric distribution \cite{tumminello2011statistically}

\begin{equation}
P(X=x) = \frac{\binom{S_a}{x} \binom{S-S_a}{S_b-x}}{\binom{S}{S_b}}
\end{equation} 
where $S$ is the total number of scientists in the sample.

\begin{figure} 
\centering 
{\includegraphics[scale=0.7]{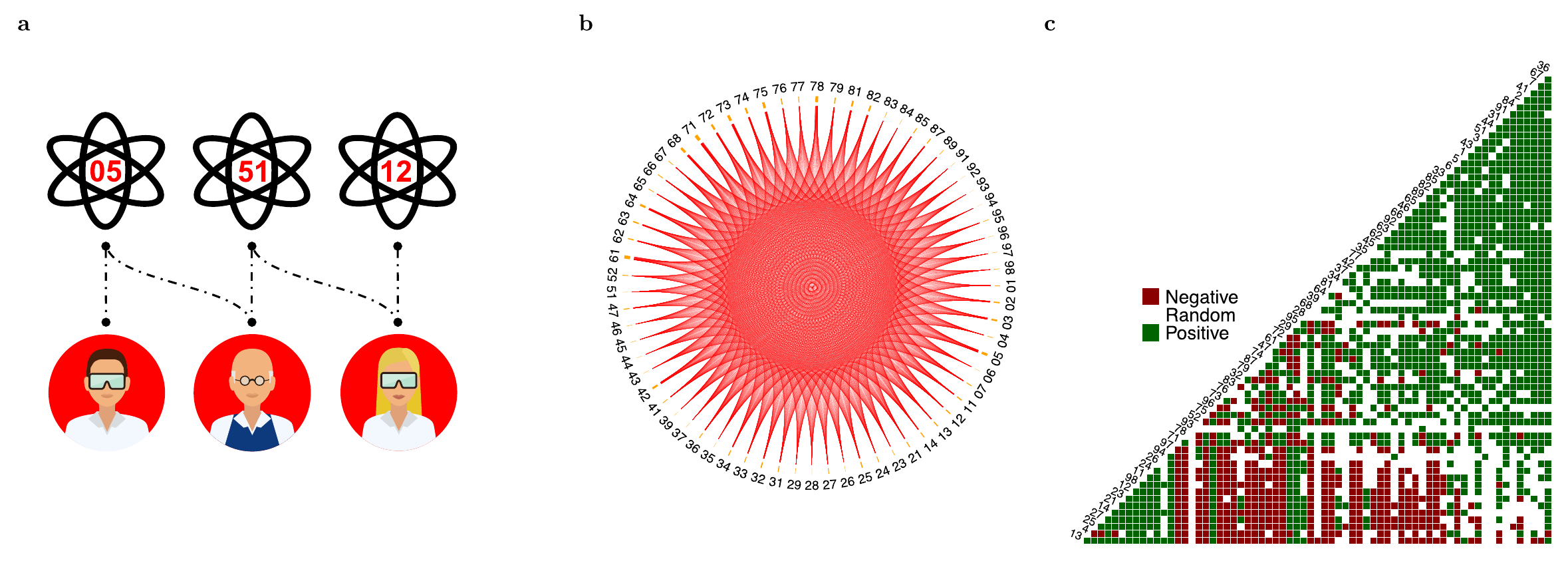}}
\caption{
\textbf{Diversification patterns}.
{\small \textbf{(a)}  A stylized picture of the original PACS-Authors bipartite network representing scientists' diversification patterns. \textbf{(b)} The diversification network (the monopartite projection on PACS): links represents the number of scientists active in each pair of sub-fields. \textbf{(c)} Visual summary of the hypergeometric test, providing evidence of the coherent nature of scientists' diversification choices: 72\% of pairs are classified as non-random ($p<0.05$ after Bonferroni correction).}
\label{patterns}}
\end{figure}

Figure \ref{patterns} describes the steps of our procedure. Starting from the bipartite network (panel \textbf{a}), we derive its monopartite projection (panel \textbf{b}) and 
test whether the resulting structure is non-random, summarizing statistically validated diversification patterns (panel \textbf{c}).  
Out of 2,278 pairs of PACS, 72\% are classified as non-random with a Bonferroni-corrected $p$-value $<0.05$. 
Of these, 1,151 pairs show a positive association and 486 a negative one.
Given the severity of the Bonferroni correction (i.e., power decreases significantly as the number of tests increases) and possible issues related to dependency, we also employ the \textit{False Discovery Rate} (FDR) Benjamini-Hochberg and Benjamini-Yekutieli corrections (see section \ref{multiplecorrection} and Table \ref{rand}).
These results strongly support a coherent nature of scientists' diversification choices, but do not provide a direct quantification of the role played by specific features in shaping such coherence. 
Next, we investigate potential drivers of diversification considering measures of cognitive and social proximity. 

\subsection{Knowledge and social relatedness predict diversification}\label{stime}
\begin{figure} 
\centering 
{\includegraphics[scale=0.7]{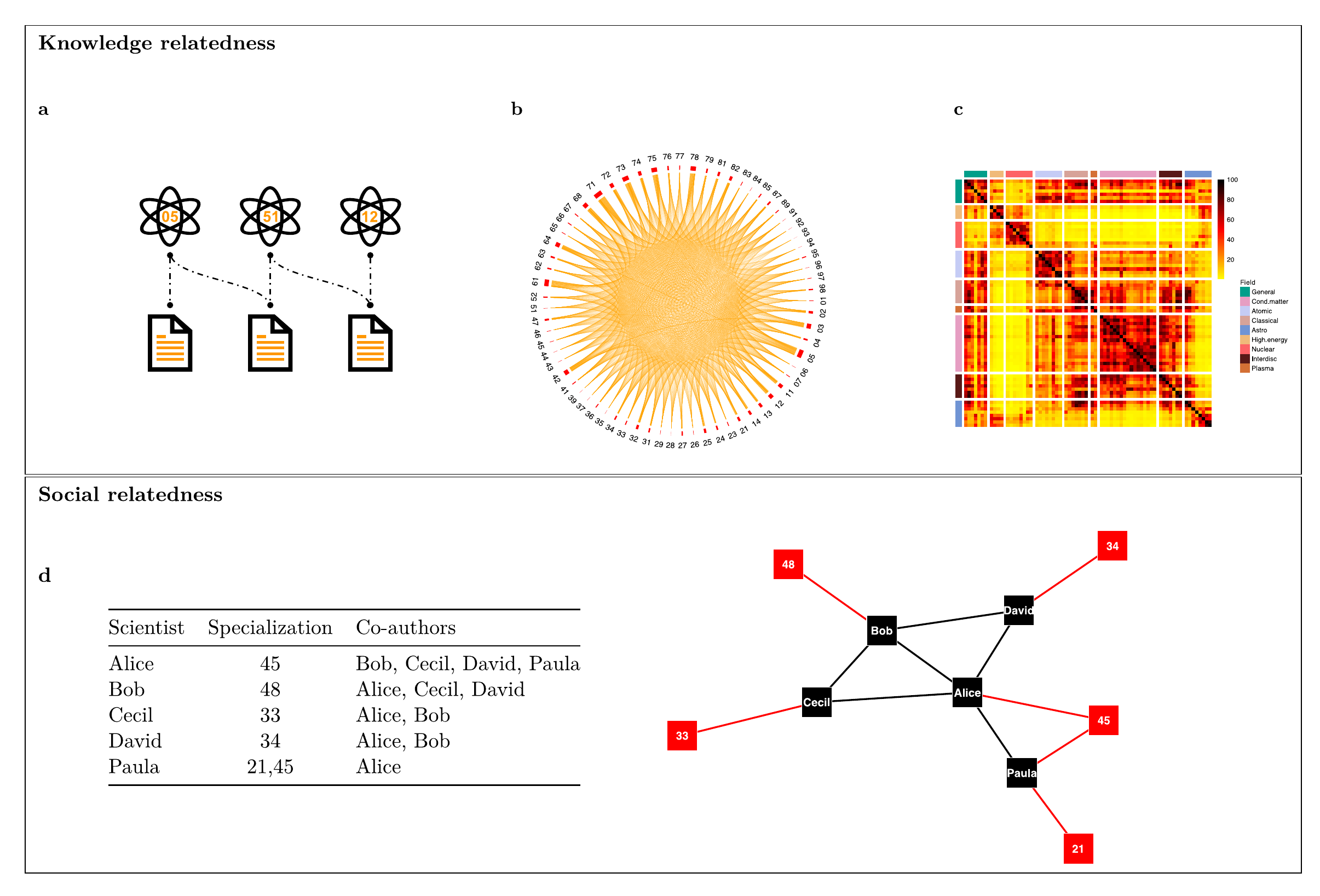}}
\caption{\textbf{Knowledge and social relatedness. \label{measures}}	
{\small \textbf{(a)} A stylized example of the bipartite PACS-Articles network. 
\textbf{(b)}
The PACS co-occurrence network (monopartite projection on PACS codes). 
\textbf{(c)}
The cosine similarity matrix, which "maps" the physics knowledge space and identifies clusters corresponding to fields. 
\textbf{(d)} A table illustrating how co-authorship and specialization information are combined to produce the augmented co-authorship network shown in the figure, which includes nodes attributes (specializations). The nodes represent individual scientists (in black) and specializations (in red). Our measure of social relatedness ($SR_{ib}$) is defined as a dummy that captures whether scholar $i$ can reach a certain sub-field $b$ through social interactions; 
$SR_{ib}$ = 1 if $d(i,b)=2$, where $d(i,b)$ is the geodesic distance between scholar $i$ and sub-field $b$. For instance, $SR_{David,45}=1$ since David could directly exchange knowledge with Alice (specialized in sub-field $45$), while $SR_{David,21}=0$.}}
\end{figure}

\noindent 
The relationships among scientific fields, like those among technologies, can be mapped using network science tools. To chart a knowledge space we need a measure of distance between fields. Several different metrics have been proposed to quantify the relatedness of technologies or scientific domains (see Bowen \textit{et al.}\cite{Bowen2016} for a review). 
When we consider the monopartite projection on the PACS layer of the bipartite PACS-Articles network, counting the co-occurrences of all pairs of PACS produces a first approximation of the relatedness of sub-fields. A similar approach was used by Lamperti \textit{et al.}\cite{doi:10.1080/10438599.2019.1633113} for patent data. However, we need a measure of proximity that: (i) does not depend on the absolute popularity of the fields, and (ii) is symmetric. The most straightforward metric that fulfils both requirements is the cosine similarity (see Figure \ref{measures}\textbf{a}-\textbf{c}, section \ref{prox}). As expected, the proximity matrix has a clear hierarchical block structure, with blocks largely overlapping with fields. Interestingly, several off block elements show the proximity of sub-fields belonging to different PACS fields.

As science becomes an increasingly "social" enterprise, it is also important to capture the relatedness of scholars, which can be done by analysing co-authorships \cite{wuchty2007}. 
Similar to what we did for knowledge relatedness, we construct a measure of social relatedness starting from the bipartite Authors-Articles network. The monopartite projection on the Authors layer defines the co-authorship network from which we compute our desired metric. In addition, to investigate whether diversification is associated with the exploitation of social relationships, we include information on authors' specialization as node attributes in the network and we introduce a dummy $SR_{ib}$ equal to $1$ if scientist $i$ can reach sub-field $b$ through direct social interactions (see Figure \ref{measures}\textbf{d}, section \ref{prox}).

Next, we evaluate the effects of knowledge and social relatedness on diversification with logistic regressions. 
The binary dependent variable encodes whether a scientist is active in a sub-field, the main explanatory variables are our measures of cognitive and social proximity, and a control is introduced for the core field. 
In practice, each scientist is assigned to a core sub-field (specialization) and can possibly diversify in one or more target sub-fields different from her own (see section \ref{core}).
In this first set of regressions, each  scientist  appears 67  times, one for every possible target PACS different from  her  own specialization (see section \ref{model} for more details). 

\begin{figure}[ht]
\centering 
{\includegraphics[scale=0.55]{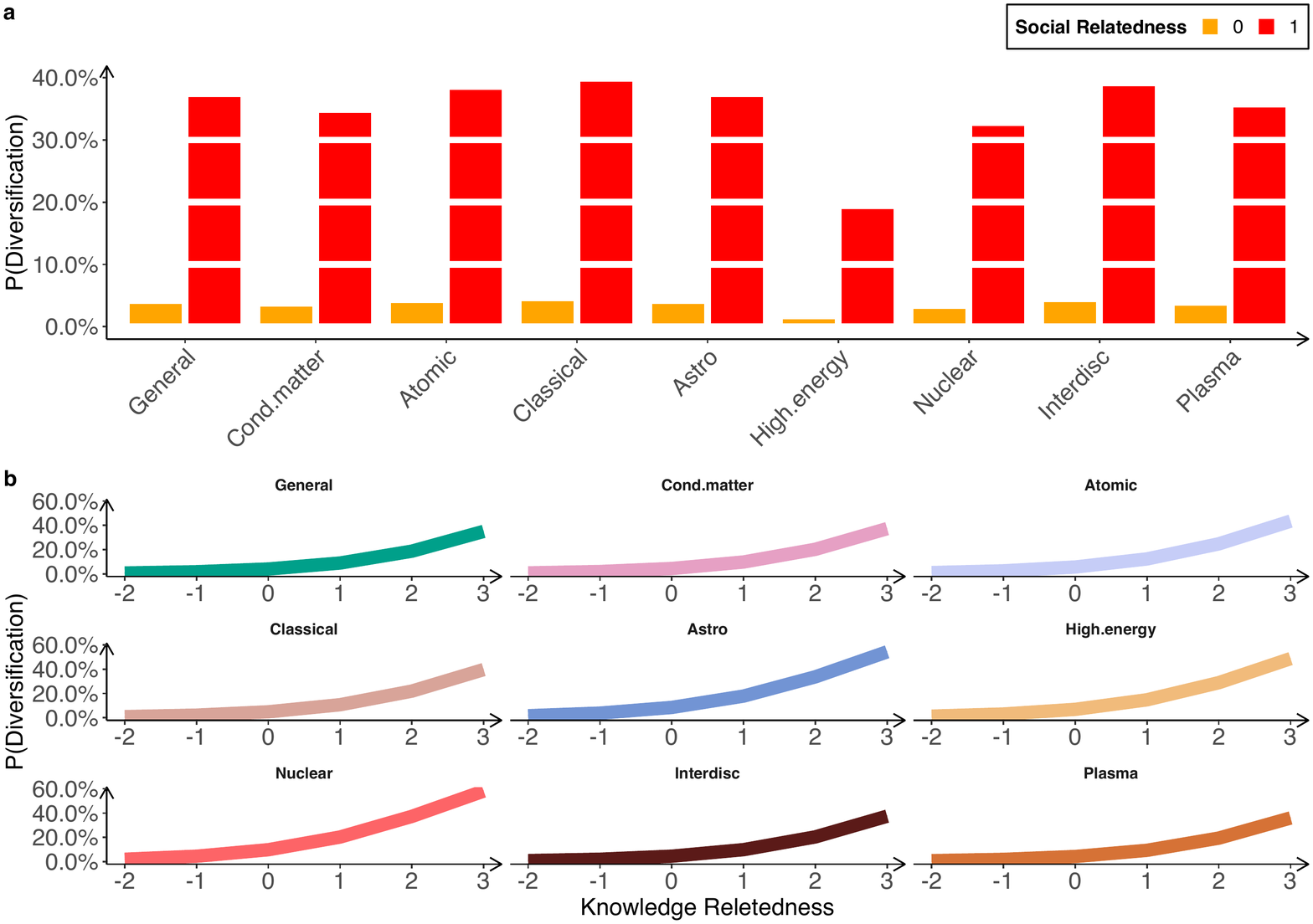}}
\caption{\textbf{
Probabilities of scientists diversifying in a sub-field different from their own specialization}. 
{\small Predicted probabilities of a scientist being active in a sub-field different from her own specialization as a function of \textbf{(a)} (binary) social relatedness, and \textbf{(b)} (standardized) knowledge relatedness. Results are obtained by fitting a logistic regression with only one control variable -- the scientist' core field. All coefficients are statistically significant ($p<0.01$).}
\label{basic}}
\end{figure}

Figure~\ref{basic} provides evidence that both social and knowledge relatedness are associated with scientists' diversification strategies. Social relatedness matters irrespective of the field, as scientists who can acquire new knowledge through social relationships are more likely to be active in a sub-field different form their own specialization (panel \textbf{a}). 
Also knowledge relatedness increases the probability of a scientist being active out of her own specialization, and again this is true for all fields (panel \textbf{b}). 
These results strongly suggest that cognitive and social proximity do contribute to shaping diversification strategies.

\subsection{Model extensions and robustness checks}\label{full}

\noindent
To move further in our investigation of research portfolio diversification, we broaden our analysis in several ways. {\em First}, we expand our logistic regression model 
including a larger set of control variables, such as the number of co-authors or the popularity and citations of the target sub-field (see Table \ref{var} for a complete list). All numerical variables in the expanded model are normalized, and log-transformed to reduce right-skew when necessary (see section \ref{model} for more details). Since the effect of knowledge relatedness on the probability of diversification may be modulated by social relatedness, we also include an interaction term in our analysis.

{\em Second}, we tackle two potential limitations of our original analysis; that is, defining a single specialization for each scientist (while core specializations may actually be multiple), and not separating sub-field movements within and between fields, i.e., one-digit PACS codes (which may be differently affected by various features). We run additional model fits allowing scientists to have multiple specializations (see section \ref{core}) and separating within and between field diversification. Specifically, we perform the following fits:  (i) single specialization with full diversification, (ii) multiple-specialization with full diversification, (iii) single specialization with within field diversification, (iv) multiple specialization with within field diversification, (v) single specialization with between field diversification and (vi) multiple specialization with between field diversification.  

{\em Third}, we account for the fact that the data employed in our fits are "clustered", with several observations associated to each scientist and a potential heteroskedasticity across clusters/scientists. We estimate clustering-robust standard errors using the clustered sandwich estimator from the {\bf R} package \textit{sandwich} \cite{JSSv011i10}.

\begin{figure}[ht]
\centering 
{\includegraphics[scale=0.55]{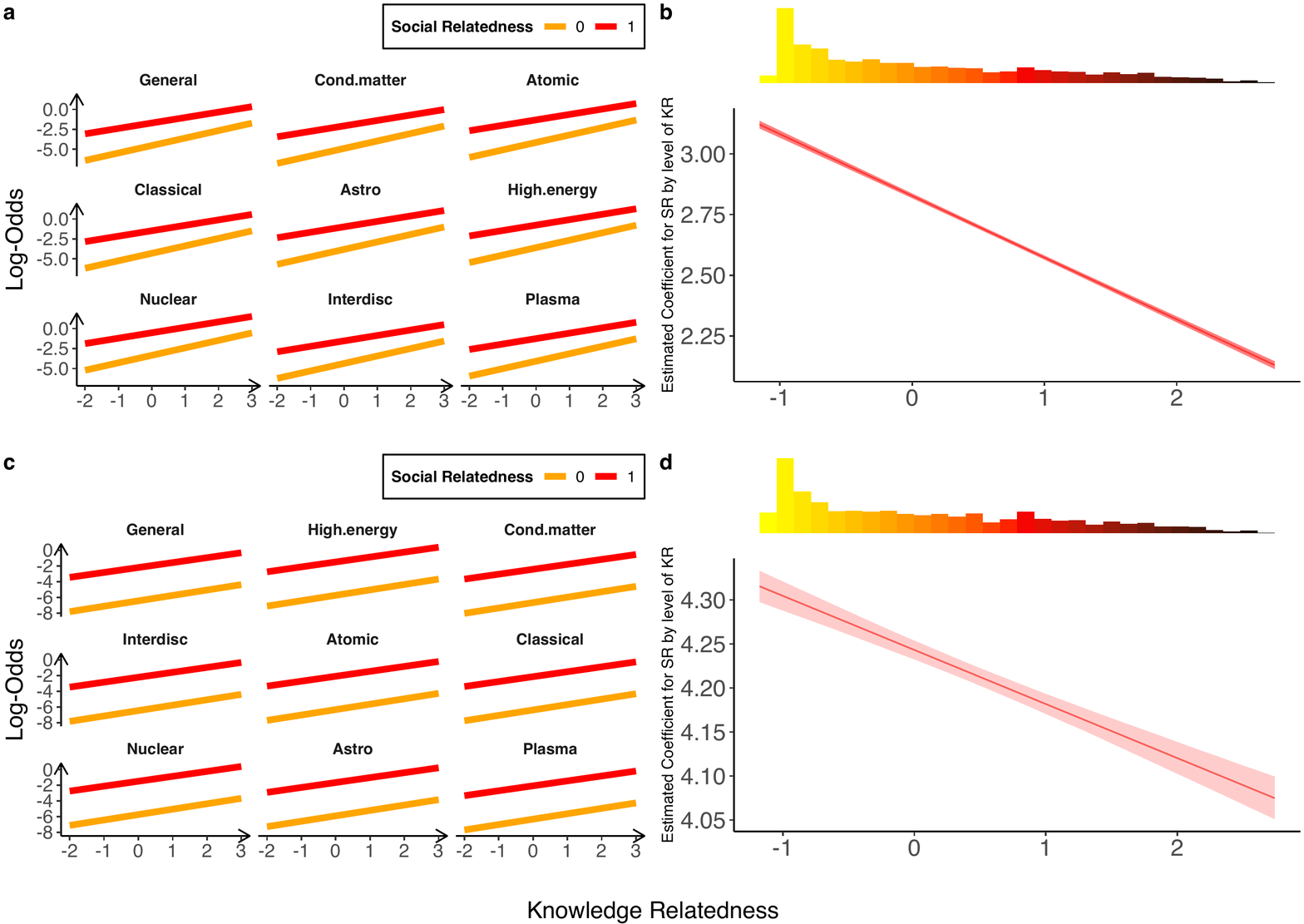}}
\caption{\textbf{Scientists' research portfolio diversification: full diversification, single and multiple specialization}.
{\small \textbf{(a)}
Log-odds as function of (binary) social relatedness and (standardized) knowledge relatedness, accounting for multiple control variables, for the single specialization specification (i).
\textbf{(b)} Estimated coefficient for social relatedness conditional on knowledge relatedness, and distribution of knowledge relatedness (on top, similarity color coded as in Figure \ref{measures}\textbf{c}), for the single specialization specification (i)).
\textbf{(c), (d)} Same as (a) and (b) for the multiple specialization specification (ii).}
\label{single}}
\end{figure}

Fits for specifications (i)-(iv), all including the interaction between knowledge and social relatedness and clustering corrected standard errors, are summarized in Table \ref{table}, confirming the high significance of the relatedness metrics in shaping research diversification. Figure \ref{single} focuses on the full diversification case. Panels \textbf{a} (single specialization, (i)) and \textbf{c} (multiple specialization, (ii)) show the log-odds difference in the probability of diversification as a function of knowledge and social relatedness, accounting for all controls. Social relatedness positively affects the chances of diversification and the effect is moderated by knowledge relatedness in both specifications, though more markedly in (i) than in (ii). Panels \textbf{b} (for (i)) and \textbf{d} (for (ii)) further illustrate this, showing how the estimated coefficient of social relatedness decreases as knowledge relatedness increases. This result indicates that when diversifying toward "close" sub-field, the role of social relatedness becomes less crucial.

\begin{table}[!htbp] \centering 
  \caption{{\bf Regression results.} Coefficients of the logistic regressions of Eq. \ref{eq:reg2}, i.e. the model including the interaction term between knowledge and social relatedness, under different specialization settings. The table reports clustering corrected standard errors (in parenthesis) and significance level:{$^{*}$p$<$0.1; $^{**}$p$<$0.05; $^{***}$p$<$0.01}. \xmark  \ variable not included in the model.}
  \label{table} 
                                 \fontsize{6}{8}\selectfont

\begin{tabular}{@{\extracolsep{5pt}}lcccccc} 
\\[-1.8ex]\hline 
\hline \\[-1.8ex] 
 & \multicolumn{6}{c}{\textit{Dependent variable: Prob(diversification)}} \\ 
\cline{2-7} \\
& \multicolumn{2}{c}{Full Diversification} & \multicolumn{2}{c}{Within Field Diversfication} & \multicolumn{2}{c}{Between Field Diversfication} 
\\ \cmidrule(lr){2-3} \cmidrule(lr){4-5} \cmidrule(lr){6-7} \\ 
 & single & multiple & single & multiple & single & multiple \\ 
\\[-1.8ex] & (i) & (ii) & (iii) & (iv) & (v) & (vi)\\ 
\hline \\[-1.8ex] 
  Knowledge Relatedness & 0.936$^{***}$ & 0.688$^{***}$ & 0.184$^{***}$ & 0.121$^{***}$ & 0.702$^{***}$ & 0.511$^{***}$ \\ 
  & (0.003) & (0.009) & (0.005) & (0.013) & (0.003) & (0.011) \\  \hline
  
  Social Relatedness & 2.827$^{***}$ & 4.243$^{***}$ & 2.272$^{***}$ & 3.968$^{***}$ & 2.914$^{***}$ & 4.284$^{***}$ \\ 
  & (0.006) & (0.019) & (0.008) & (0.021) & (0.008) & (0.021) \\  \hline
  
  field core-Atomic & $-$0.332$^{***}$ & $-$0.428$^{***}$ & 0.056$^{**}$ & $-$0.276$^{***}$ & $-$0.303$^{***}$ & $-$0.385$^{***}$ \\ 
  & (0.010) & (0.007) & (0.025) & (0.021) & (0.010) & (0.008) \\ 
  
  field core-Classical & $-$0.490$^{***}$ & $-$0.477$^{***}$ & $-$1.001$^{***}$ & $-$0.932$^{***}$ & $-$0.313$^{***}$ & $-$0.328$^{***}$ \\ 
  & (0.010) & (0.007) & (0.029) & (0.023) & (0.010) & (0.008) \\ 
  
  field core-Cond.matter & $-$1.088$^{***}$ & $-$0.761$^{***}$ & $-$1.110$^{***}$ & $-$0.892$^{***}$ & $-$1.263$^{***}$ & $-$0.903$^{***}$ \\ 
  & (0.012) & (0.009) & (0.024) & (0.020) & (0.017) & (0.013) \\ 
  
  field core-General & $-$0.722$^{***}$ & $-$0.537$^{***}$ & $-$0.927$^{***}$ & $-$0.823$^{***}$ & $-$0.632$^{***}$ & $-$0.422$^{***}$ \\ 
  & (0.011) & (0.007) & (0.028) & (0.021) & (0.012) & (0.008) \\ 
  
  field core-High.energy & 0.219$^{***}$ & 0.168$^{***}$ & 1.806$^{***}$ & 1.176$^{***}$ & $-$0.360$^{***}$ & $-$0.060$^{***}$ \\ 
  & (0.010) & (0.006) & (0.027) & (0.023) & (0.013) & (0.008) \\ 
  
  field core-Interdisc & $-$0.557$^{***}$ & $-$0.553$^{***}$ & $-$0.357$^{***}$ & $-$0.724$^{***}$ & $-$0.365$^{***}$ & $-$0.367$^{***}$ \\ 
  & (0.010) & (0.007) & (0.026) & (0.021) & (0.011) & (0.008) \\ 
  
  field core-Nuclear & 0.463$^{***}$ & 0.164$^{***}$ & 0.969$^{***}$ & 0.692$^{***}$ & 0.068$^{***}$ & $-$0.161$^{***}$ \\ 
  & (0.010) & (0.006) & (0.024) & (0.021) & (0.011) & (0.009) \\ 
  
  field core-Plasma & $-$0.269$^{***}$ & $-$0.419$^{***}$ & $-$0.155$^{**}$ & $-$0.361$^{***}$ & $-$0.074$^{***}$ & $-$0.256$^{***}$ \\ 
  & (0.013) & (0.008) & (0.068) & (0.058) & (0.015) & (0.009) \\  \hline
  
  {\#}\ of\ PACS & 0.882$^{***}$ & 0.806$^{***}$ & 0.769$^{***}$ & 0.497$^{***}$ & 1.003$^{***}$ & 0.944$^{***}$ \\ 
  & (0.002) & (0.003) & (0.005) & (0.004) & (0.004) & (0.004) \\ 
  
  {\#}\ of\ papers & 0.010$^{***}$ & 0.113$^{***}$ & 0.065$^{***}$ & 0.252$^{***}$ & $-$0.032$^{***}$ & 0.050$^{***}$ \\ 
  & (0.002) & (0.003) & (0.005) & (0.004) & (0.004) & (0.004) \\ 
  
  {\#}\ of\ co-authors & $-$0.406$^{***}$ & $-$0.347$^{***}$  & $-$0.240$^{***}$ & $-$0.145$^{***}$  & $-$0.488$^{***}$ &  $-$0.444$^{***}$ \\ 
  & (0.002) & (0.004)  & (0.004) & (0.004)  & (0.003) & (0.006)  \\  \hline
  
  PACS target popularity & 1.130$^{***}$ & 0.611$^{***}$ & 1.370$^{***}$ & 0.774$^{***}$ & 1.108$^{***}$ & 0.559$^{***}$ \\ 
  & (0.002) & (0.002) & (0.005) & (0.003) & (0.003) & (0.002) \\ 
  
  $\Delta$ crowd & 0.239$^{***}$ & 0.358$^{***}$ & 0.131$^{***}$ & 0.345$^{***}$ & 0.320$^{***}$ & 0.393$^{***}$ \\ 
  & (0.002) & (0.003) & (0.003) & (0.003) & (0.003) & (0.003) \\  \hline
  
  $\Delta$ PACS citations & $-$0.273$^{***}$ & $-$0.332$^{***}$ & $-$0.208$^{***}$ & $-$0.313$^{***}$ & $-$0.369$^{***}$ & $-$0.354$^{***}$ \\ 
  & (0.002) & (0.003) & (0.004) & (0.003) & (0.004) & (0.003) \\ 
  
  $\Delta$ field citations & $-$0.156$^{***}$ & $-$0.070$^{***}$ & \xmark  & \xmark & $-$0.196$^{***}$ & $-$0.143$^{***}$ \\ 
  & (0.004) & (0.004) &  &  & (0.006) & (0.005) \\  \hline
  
  KR:SR & $-$0.255$^{***}$ & $-$0.061$^{***}$ & $-$0.047$^{***}$ & $-$0.001 & $-$0.234$^{***}$ & $-$0.067$^{***}$ \\ 
  & (0.004) & (0.010) & (0.007) & (0.013) & (0.005) & (0.011) \\ 
  
  Constant & $-$3.812$^{***}$ & $-$5.903$^{***}$ & $-$1.882$^{***}$ & $-$4.250$^{***}$ & $-$4.168$^{***}$ & $-$6.165$^{***}$ \\ 
  & (0.010) & (0.020) & (0.022) & (0.028) & (0.010) & (0.022) \\ 
 \hline \\[-1.8ex] 
Observations & 7,072,386 & 35,968,615 & 1,000,230 & 5,407,404 & 6,072,156 & 30,154,990 \\ 
Log Likelihood & $-$1,086,281.000 & $-$7,303,198.000 & $-$334,697.300 & $-$2,166,803.000 & $-$716,398.900 & $-$4,971,497.000 \\ 
Akaike Inf. Crit. & 2,172,600.000 & 14,606,434.000 & 669,430.600 & 4,333,642.000 & 1,432,836.000 & 9,943,033.000 \\ 
\hline 
\hline \\[-1.8ex] 
\end{tabular} 
\end{table} 

Next, we contrast scientists moving within their specialization field (between two sub-fields, i.e.~two-digit PACS codes, belonging to the same field, i.e.~one-digit PACS code; e.g.~PACS 12 \textit{Specific theories and interaction models; particle systematics} and PACS 13 \textit{Specific reactions and phenomenology}, both belonging to PACS 1 \textit{High Energy} physics) and scientists moving out of their field and towards a completely different subject (i.e.~a different one-digit PACS code). These choices may be driven by different factors. 
Scientists moving within their field may be less dependent on external collaborations, since such a diversification strategy requires a smaller learning effort. Our estimates do highlight differences. Looking at the within field diversification case, single specialization (Table \ref{table}, (iii)), we see that knowledge and social relatedness, as well as their interaction, are still significant -- but the magnitude of the coefficients is smaller with respect to the full diversification case. When we consider multiple specialization (Table \ref{table}, (iv)), coefficients shrink even further and the interaction is no longer significant (see also Figure \ref{within}). 
On the contrary, looking at the between field diversification case, the general trends outlined for the full diversification case are confirmed -- including the negative interaction term remaining sizeable and significant for both single and multiple specialization (see Table \ref{table}, (v) and (vi), and Figure \ref{bet}).These results are in line with expectations: while having a co-author in a different sub-field may well be useful, knowledge is not a barrier to entry when scientists move within the same general area of inquiry. This explains why the interaction between social and knowledge relatedness becomes less prominent or non-significant in our estimates. 

\subsection{Quantifying the relative importance of knowledge and social relatedness}

\noindent
Can we quantify the (relative) role of knowledge and social relatedness in explaining research portfolio diversification? How important are these quantities when evaluated in the presence of several control covariates, and under a range of model specifications? To answer these questions we follow two approaches.

\begin{figure}[ht]
\centering 
{\includegraphics[scale=1]{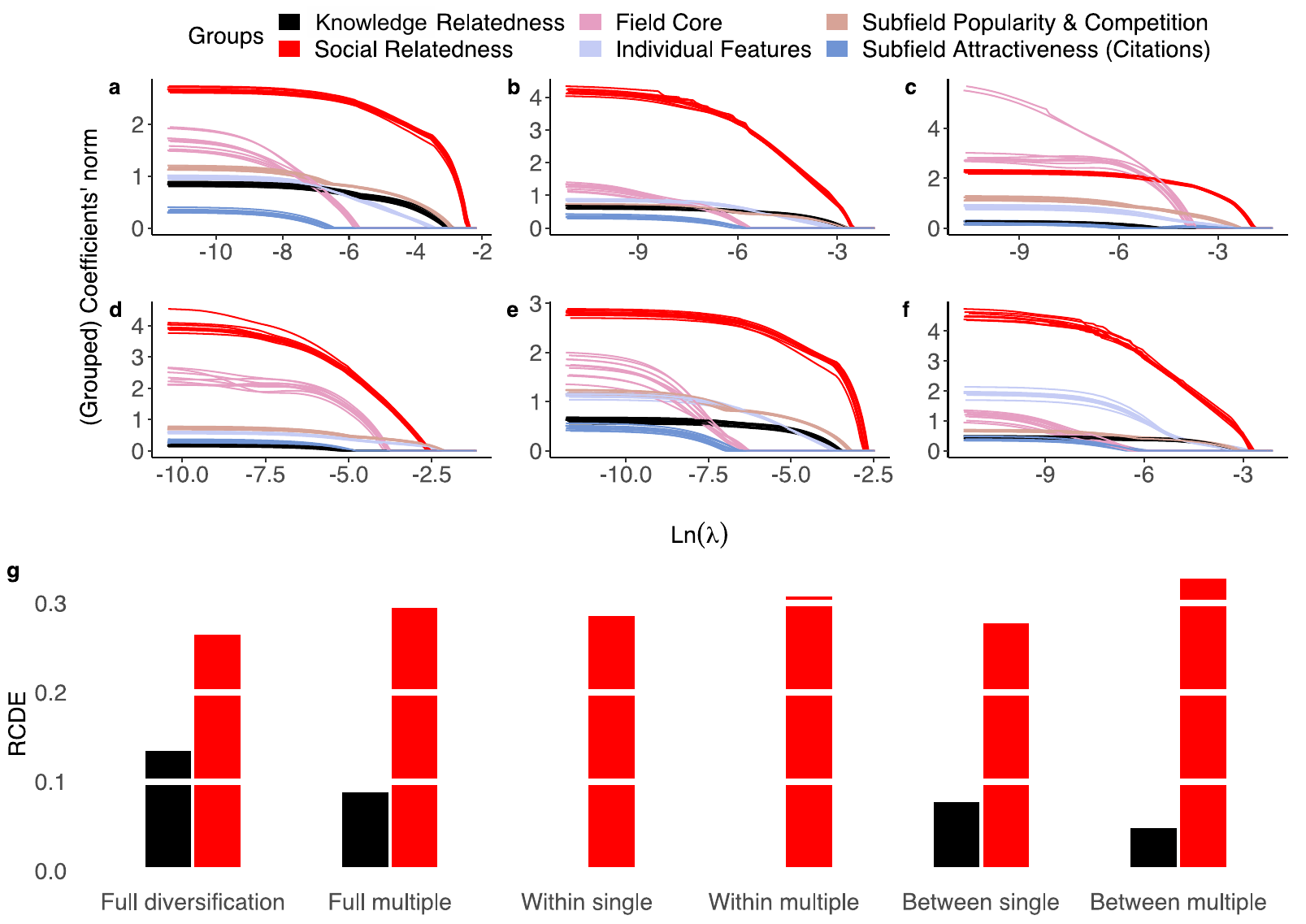}}%
\caption{\textbf{Relative importance of predictors.} 
{\small \textbf{(a)-(f)} Group LASSO paths for \textbf{(a)} full diversification, single specialization; \textbf{(b)} full diversification, multiple specialization; \textbf{(c)} within-field diversification, single specialization; \textbf{(d)} within-field diversification, multiple specialization; \textbf{(e)} between-field diversification, single specialization; \textbf{(f)} between-field diversification, multiple specialization. In each panel, variables in the same group are color coded, and their average coefficient norm is plotted (as a single path) against the penalty parameter (log $\lambda$). The multiple paths for each color correspond to separate group LASSO runs on 10 random sub-samples of 1,000 scientists. 
\textbf{(g)} Relative Contributions to Deviance Explained for knowledge relatedness (black) and social relatedness (red) across all fits.}
\label{lasso}}
\end{figure}

{\em First}, we run a LASSO feature selection procedure to gauge the relative importance and role of different predictors by tracking how they are excluded/included in a model as one varies the regularization penalty. Since our predictors include categorical variables (i.e., groups of dummies), as well as naturally grouped variables (e.g., scientists' individual characteristics, sub-fields' popularity and competition, etc.) we run a {\em group} LASSO algorithm \cite{yuan2006model} with features grouped as shown in Table~\ref{var}.  
Moreover, to counteract collinearity and finite sample issues which can render the LASSO unstable \cite{mullainathan2017machine}, we split our data forming ten random subsamples of 1,000 scientists each, and repeat the group LASSO fit on each of the subsamples for all the considered model specifications.  Figure \ref{lasso}\textbf{a-f} show the (grouped) coefficient norms as a function of the penalization parameter $\lambda$. 
Results clearly demonstrate the crucial role played by social and knowledge relatedness. They also confirm that the role of knowledge relatedness weakens markedly in the case of within-field diversification (panels \textbf{c} and \textbf{d}).

{\em Second}, we compute the {\em Relative Contributions to Deviance Explained} (RCDEs; see section \ref{model} for details). This index captures what percentage of the logistic regression deviance is captured by a predictor. Figure \ref{lasso}\textbf{g} strongly supports a prominent role for social relatedness, with RCDEs around or above 30\%  across all 
specifications. 
The RCDEs of knowledge relatedness are smaller, around 5-10\%, and again become negligible in the case of within-field diversification. 
In summary, our results provide additional evidence that both social and knowledge proximity shape scientists' diversification strategies, but highlight social interactions as the dominant channel through which knowledge is exchanged and acquired.

\subsection{Digging deeper: multidisciplinarity  and time}

\noindent
Next, we tackle two additional potential limitations of our original analysis, which might overestimate the probability of diversification for truly multidisciplinary scientists and suffer from reverse causality issues. To investigate diversification into truly unexplored sub-fields, we fitted the model specification (i) (see section \ref{full}) considering scientists' specialization (see section \ref{core}) and limiting their diversification choices to sub-fields in which they have no revealed scientific advantage (see section {\ref{constrained}}). 
To at least partially address causality in the effects of knowledge and social relatedness on diversification, we included a temporal dimension: we split the original dataset in three time periods, re-computed our measures of relatedness in each, and used them to predict scientists' diversification introducing time lags (see section \ref{timestep}). In both exercises, results confirmed our previous findings: social relatedness shapes scientists' diversification strategies more than knowledge relatedness.

Finally, and again related to time, our findings may be influenced by underlying trends in the temporal evolution of PACS co-occurrence networks -- and thus knowledge proximity. A detailed study of the evolution of relationships among sub-fields, which is of course of interest {\em per se}, is beyond the scope of the present article.
Nevertheless, to gather at least some approximate sense of its potential impact, we recomputed our measure of knowledge relatedness separately for each of the different decades in the original dataset. Based on results shown in section \ref{temporal}, the physics knowledge space remained rather stable over the time span considered. 
A valuable alternative approach to take into account the temporal evolution of the physics knowledge space is provided by Chinazzi \textit{et al.}\cite{chinazzi2019mapping}

\section{Discussion}

\noindent
Scientists try to balance the "tension" between exploitation and exploration, but the exploration phase is, to some extent, constrained by the "burden of knowledge". To tackle the rising complexity of producing new knowledge, scientists adapt their diversification strategies leveraging social interactions; that is, proximity to other scientists. Our analysis attempts to identify and quantify drivers of research portfolio diversification. Based on data concerning a very large sample of physicists we find that, while knowledge relatedness 
plays a role, contemporary science is a profoundly social enterprise.
When scientists move out of their specialization, they do so through collaborations. And the further the move, the more these collaborations matter.

Limitations in the methodology we employed for this study point towards needed future developments. First and foremost, we are not assessing causal effects; we analyse research diversification patterns irrespective of the mechanisms which determine the similarity among sub-fields and the co-authorship network. Indeed, knowledge relatedness and collaborations may themselves be affected by scientists' diversification strategies. 
We believe that the observed negative interaction between knowledge and social relatedness helps us rule out, at least partially, the contingency of reverse causality for social relatedness: if diversification were causally driving the link, we would expect a positive interaction. There is no reason to believe that new collaborators are easier to find in sub-fields far from a scientist's own specialization; in fact,  the opposite may be more likely -- the closer the sub-fields, the higher the chances to collaborate. 
Moreover, since the structure of the knowledge space appears fairly stable over time, the direction of causality is more likely from subject proximity to diversification -- not the other way around. 
Additional analyses with methods that fully exploit the temporal trajectories of scientists' activities
will be instrumental to elucidating the causal interplay between individual strategies and collaborations. In the Supplement we do provide results for the checks we were able to run based on the data and methods at our disposal. 

Another critical development will be expanding the investigation to scientific and/or technological domains beyond physics -- shedding further light on behaviours and potential sources of heterogeneity. 
Our initial focus on physics was due to its central role in the natural sciences and to the availability of reliable and abundant data. Nevertheless, the approach used in this study is fully applicable to different domains. Patents and publications records would both be useful grounds to validate and extend our results -- thus providing 
a quantitative benchmark to inform science and technology policy. 

From a policy perspective, our current results already provide some insights. They support the notion that social interactions constitute the core medium to foster new scientific venues, allowing scientists to overcome knowledge barriers.
Thus, social interactions should be a focus of efforts aimed at improving cross-disciplinary team formation. Institutions should strive to create environments that favor social proximity and collaboration, and funding for interdisciplinary research should reward matches among scholars specialized in very distant domains.  

\section{Methods}
\begin{small}

\subsection{Data}\label{data}
\noindent 
We use the American Physical Society (henceforth APS) dataset, which is maintained by the APS and publicly available for research purposes upon request (see \href{https://journals.aps.org/datasets}{APS website}). Each article in the dataset is labeled with up to 5 PACS codes.
As an example, the PACS code \textit{42.65.-k} refers to \textit{nonlinear optics}; the first digit represents a broad field (Classical Physics), and the second a more specific sub-field (Optics).
A brief description of the one-digit level fields is provided in Table \ref{PACS}. 
In our analyses, we work at the level of sub-fields; our measure of knowledge relatedness is based on similarity of PACS at two-digit level.
Based on our aims (analysing research diversification strategies), we created a dataset based on two requirements: (i) the ability to reconstruct the career of each individual, and (ii) a standardized classification system for each article. (i) poses several issues related to name disambiguation, which have been successfully investigated in previous studies. We rely on the disambiguated dataset made available by Sinatra \textit{et al.}\cite{Sinatra2016}. (ii) concerns the classification scheme applied to physics articles. The PACS classification has been broadly employed from 1970 to 2016, but then the APS adopted a different labelling procedure (Physics Subject Headings; PhySH). We limit our analysis to a period entirely covered by the PACS system.
Our final dataset includes information regarding 197,682 scholars that published at least one article in one of the 9 APS journals in the period ranging from 1977 to 2009.  Figure \ref{date} shows the number of papers (panel \textbf{a}) and the number of papers per author (panel \textbf{b}) over time.  
\end{small}

\begin{small}
\subsection{Monopartite projections of bipartite networks} \label{infer}

\noindent 
A bipartite network is a graph whose nodes can be divided into two distinct sets (layers) such that no edge connects a pair of nodes belonging to the same set.
A binary undirected bipartite network is identified by a rectangular biadjacency matrix \textbf{b} of dimensions $N_R \times N_C$. The number of rows $N_R$ is the number of nodes in layer $R$, and the number of columns $N_C$ is the number of nodes in layer $C$ \cite{saracco2017inferring}. Being binary simply means that the elements of the matrix are 
\begin{linenomath*}
\begin{equation}
b_{_{rc}}=\begin{cases}
1 & \textrm{if node \textit{r} $\in$ R and \textit{c} $\in$ C are linked}\\
0 & \textrm{otherwise}
\end{cases}
\end{equation}
\end{linenomath*}

\noindent
The weighted monopartite projection on one of the layers is constructed counting so-called V-motifis: we draw a link in the projected network if two nodes share a neighbour in the bipartite network. For instance, to derive the weighted monopartite projection on layer R, we count co-occurences in the bipartite network and construct the square $N_R \times N_R$ matrix \textbf{M} with elements

\begin{linenomath*}
\begin{equation}
m_{rr'}=\sum_{c=1}^{N_C}b_{rc}b_{r'c}
\end{equation}
\end{linenomath*}

\noindent
For our analyses, we derive weighted monopartite projections from three binary bipartite networks; namely, Subfields-Articles, Authors-Articles and Subfields-Authors.

\end{small}

\begin{small}
\subsection{Scientists' specializations}
\label{core}

\noindent
Our analyses require us to assign specializations (single or multiple) to individuals. Unfortunately, there is no standard way to approach 
this problem -- in part because, unlike articles or patents which can often be unambiguously linked to a limited number of classes, scientists can explore the knowledge space quite extensively. 
For our purposes, a suitable assignment should take into account both the relative specialization of a scientist and the distribution of publications across areas. 
Share-based metrics can be used to construct effective assignments. An instance is the Revealed Scientific Advantage (RSA) recently used by Battiston \textit{et al.}\cite{battiston2019taking}, which is akin to a metric originally used by Balassa\cite{balassa1965trade} to analyse comparative international trade advantages among countries. 
We consider the normalized metric; for each author $i$ and sub-field (two-digit PACS) $s$ this is defined as
\begin{linenomath*}
\begin{equation}
RSA_{is} = \frac{\frac{w_{i,s}}{\sum_s w_{i,s}} }{\frac{\sum_i w_{i,s}}{\sum_{i,s} w_{i,s}}},
\end{equation}
\end{linenomath*}
where $w_{i,s}$ is the number of articles author $i $ has published in sub-field $s$. By construction, $RSA_{is} \in [-1,1]$, and a positive value indicates an advantage for author $i$ in sub-field $s$. To assign a single specialization to $i$, we simply take $s(i) = argmax_s \lbrace RSA_{is} \rbrace$.
\end{small}

\begin{small}
To assign multiple specializations to $i$, we take $S(i) = \{s\ s.t.\ RSA_{is}>0\}$.  
In this case we actually create a fictitious "copy" of $i$ for each of the sub-fields in $S(i)$ -- keeping all individual characteristics but the specialization for each copy. This overcomes possible biases stemming from classification errors or marked heterogeneity in the distribution of articles across sub-fields. 
\end{small}

\begin{small}
\subsection{Measures of knowledge and social relatedness}\label{prox}

\noindent 
We define knowledge relatedness among sub-fields (two-digit PACS) from the bipartite network PACS-Articles. Specifically, we derive the monopartite projection on the PACS layer (a $68 \times  68$ co-occurrence matrix) and then apply the cosine similarity to construct a knowledge relatedness matrix. 
The procedure is illustrated in Figure \ref{measures}: panel \textbf{a} shows a stylized example of the bipartite network PACS-Articles, panel \textbf{b} shows the network of co-occurrences of all pairs of PACS (the monopartite projection on the PACS layer), and panel \textbf{c} shows the cosine similarity matrix describing proximity among physics sub-fields.
\end{small}
\begin{small} 

We define social relatedness from the initial co-authorship network $G(V,E)$. Specifically, we build an augmented graph $G'(V',E')$ to integrate scientists' specializations: for each node (author) $V \in G$, we create an {\em individual} node in $G'$ and for each edge $E\in G$ we draw the corresponding edge in $G'$. 
Then for each PACS $s$, we create an {\em attribute} node in $G'$. 
Next, we add further edges to $G'$ considering the specialization(s) of each scientist and creating an edge between her individual node and her specialization(s)'s attribute node(s) (Figure \ref{measures}\textbf{d} provides a simple example).
Finally, we capture social relatedness with a binary variable based on whether an author has at least one coauthor specialized in a sub-field different from her own; that is
\begin{linenomath*}
\begin{equation}
SR_{is}=  \begin{cases} 
1 & \text{if}\ d(i,s)=2\\ 
0 & \text{otherwise}
\end{cases}
\end{equation}
\end{linenomath*}
where $d(i,s)$ is the geodesic distance between scientist $i$ and sub-field $s$ in the augmented graph.
\end{small}

\begin{small}
\subsection{Modeling and assessment of predictors' contributions} 
\label{model}

\noindent
Consider an author $i$ specialized in the sub-field $a$. The probability that she is also active in sub-field $b\ne a$ is modeled as
\begin{linenomath*}
\begin{equation}
p:= f(KR_{ab},SR_{ib},\mathbf{IF_i},\mathbf{SC_b},\mathbf{Cit_b})
\end{equation}
\end{linenomath*}
where $KR_{ab}$ is the knowledge relatedness between the two sub-fields, $SR_{ib}$ is the social relatedness between the author and the sub-field $b$, $\mathbf{IF_i}$ is a vector of author's characteristics, $\mathbf{SC_b}$ is a vector of variables capturing the sub-field popularity and competition (i.e., for each sub-field,  number of papers and number of specialized scientists), and $\mathbf{Cit_b}$ is a vector of variables capturing the relative attractiveness of the sub-field.  
A full list of the variables comprised in these vectors is provided in Table \ref{var}. 
We reformulate the model as a
logistic regression and consider two baseline specifications, with and without the interaction term between knowledge and social relatedness:
\begin{linenomath*}
\begin{equation}\label{eq:reg1}
ln(\frac{p}{1-p}) = \alpha + \beta KR_{ab} + \gamma SR_{ib} + \bm{\theta} \cdot \mathbf{IF_i} + \bm{\eta} \cdot \mathbf{SC_b} + \bm{\phi} \cdot \mathbf{Cit_b} 
\end{equation}
\end{linenomath*}

\begin{linenomath*}
\begin{equation}\label{eq:reg2}
ln(\frac{p}{1-p}) = \alpha + \beta KR_{ab} + \gamma SR_{ib} + \zeta (KR_{ab} \times SR_{ib}) + \bm{\theta} \cdot \mathbf{IF_i} + \bm{\eta} \cdot \mathbf{SC_b} + \bm{\phi} \cdot \mathbf{Cit_b} 
\end{equation}
\end{linenomath*}

For both the single-and multiple-specialization settings, we fit these logistic regressions in three scenarios; namely, {\em full} (no constraint on sub-fields $a$ and $b$), {\em within field} ($a$ and $b$ in the same field; i.e.~one-digit PACS code) and {\em between field} ($a$ and $b$ in different fields) diversification. 

In order to quantify the roles of knowledge and social relatedness, we compute the {\it Relative Contribution to Deviance Explained} (RCDE) for each of these variables \cite{10.1371/journal.pcbi.1004956}. For a generic predictor $X$ this is defined as  
\begin{linenomath*}
\begin{equation}
RCDE_{X} = \frac{(D_{null}-D_{full}) - (D_{null}-D_{full \setminus X})}{(D_{null}-D_{full})} 
\end{equation}
\end{linenomath*}
where $D_{null}$ is the null deviance, $D_{full}$ is the residual deviance of the full model (including all predictors) and $D_{full \setminus X}$ is the residual deviance of the model obtained by removing $X$ (in our case $KR$ or $SR$). The RCDE thus quantifies the percentage of the total logistic deviance attributable $X$.
\end{small}

\section*{Data availability}
\noindent All data used in this study are publicly available. The APS data can be downloaded at \href{https://journals.aps.org/datasets}{https://journals.aps.org/datasets}. The list of disambiguated authors’ names is available at \href{https://science.sciencemag.org/content/354/6312/aaf5239}{https://science.sciencemag.org/content/354/6312/aaf5239}.

\section*{Code availability}
\noindent Code and functions for regression and other statistical analyses/tests are available in standard R libraries and packages including cooccur, stats (glm), interplot, sandwich. Other code used in this study (e.g., for data cleaning and results visualization, based on Python and R packages) is available from the corresponding authors upon request.

\bibliography{bibRP}

\section*{Acknowledgements}
\noindent We thank Francesco Lamperti, Andrea Mina, Fabio Saracco, Gianmichele Blasi and Dino Pedreschi for useful discussions. This work was partially supported by the Huck Institutes of the Life Sciences of the Pennsylvania State University (FC).

\section*{Authors’contributions}
\noindent GT conceived the study, and designed it together with FC and FL. GT performed all analyses. All authors wrote the manuscript.

\section*{Competing interests}
\noindent The authors declare no competing interests.

\newpage
\beginsupplement
\section*{Supplementary Information}

\subsection{Data}

\begin{figure}[b]
\centering 
\subcaptionbox*{\textbf{a}}
{\includegraphics[scale=0.5]{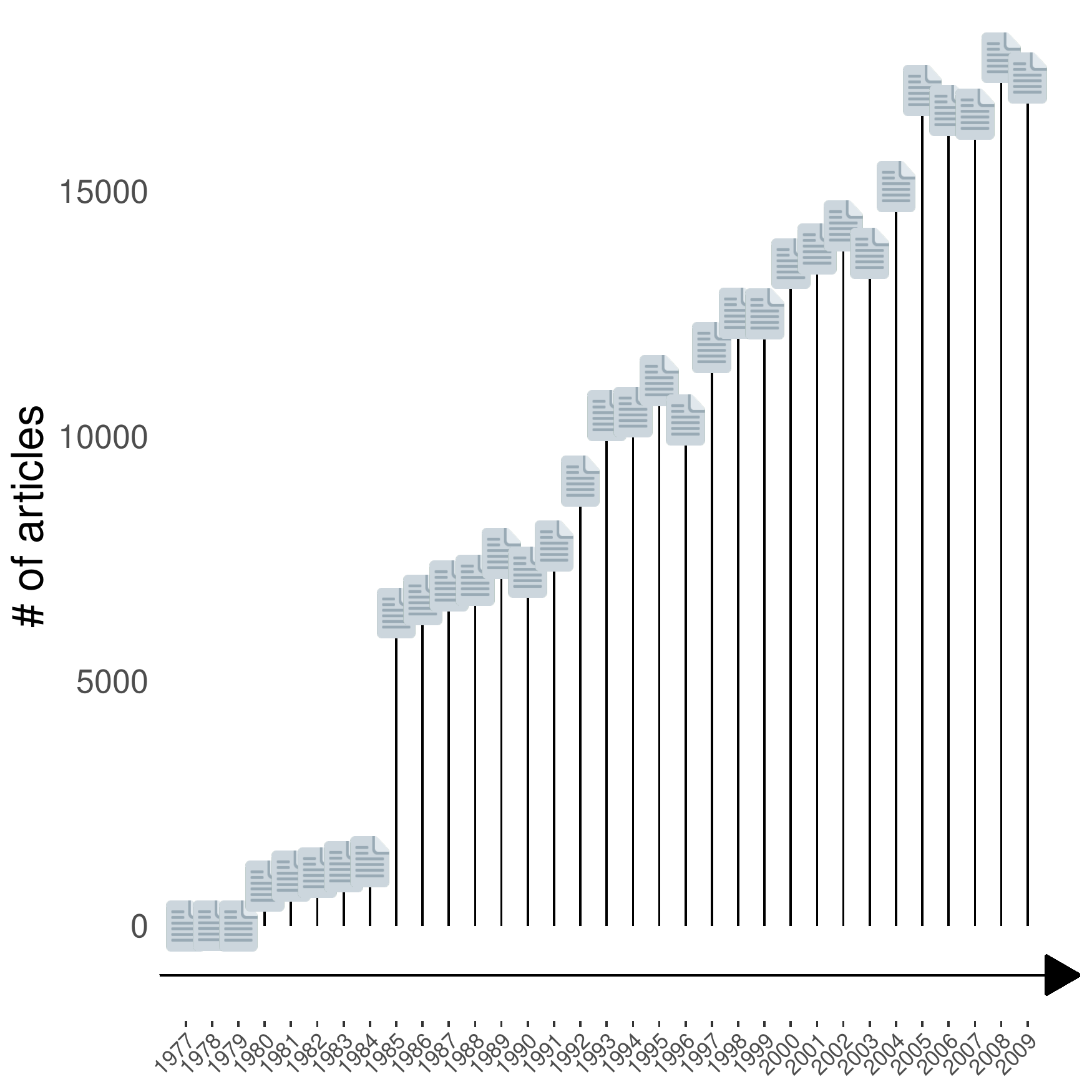}}
\subcaptionbox*{\textbf{b}}
{\includegraphics[scale=0.5]{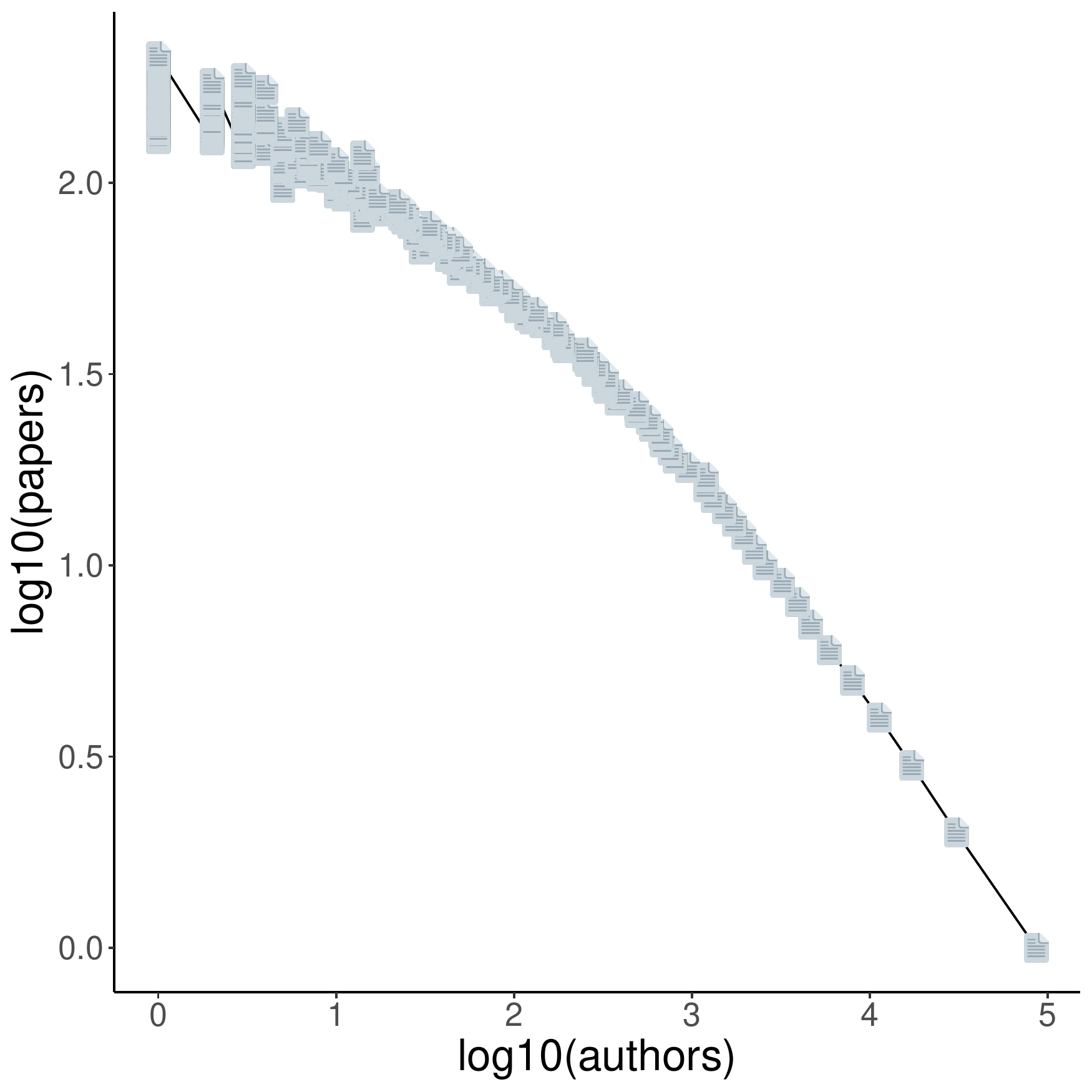}}
\caption{\textbf{Statistical properties of the APS data}. {\small \textbf{a}, The time series of papers over time shows that the number of papers published in APS outlets increased substantially from 1977 to 2009. \textbf{b}, The distribution of the number of papers per author is fat-tailed: the large majority of authors published just few articles while some authors have been extremely productive. } \label{date}}
\end{figure}

The American Physical Society (APS) grants access to data containing information about papers published in 9 journals: Physical Review A, B, C, D, E, I, L, ST and Review of Modern Physics. 
The APS makes available, under request, two datasets including over 450,000 articles metadata and citations from 1893 onwards. Each article has a unique identifier and most of them contain reference codes that map into physics sub-fields (PACS codes). As mentioned in section \ref{data}, we make use of such a classification to keep track of scientists' diversification patterns. Moreover, we use a disambiguated list of authors made available by \cite{Sinatra2016}.
As a result, we analyse a sub-sample for which we have access to all the necessary information: it includes more than 300,000 articles published by 197,682 authors over the period 1977-2009. Figure \ref{date} provides simple statistical properties of the dataset.

\begin{table}[b]
\tiny \centering
\begin{threeparttable} 
\caption{One-digit PACS codes \label{PACS}}\fontsize{5}{18}\selectfont
\begin{tabular}{lcc} 
\toprule PACS & Field & Description \\ \midrule 
0  & General & Mathematical Methods, Quantum Mechanics, Relativity, Nonlinear Dynamics and Metrolog \\ 
1 & High-energy & Physics of Elementary Particles and Fields\\ 
2  &  Nuclear & Nuclear Structure and Reactions\\ 
3 & Atomic & Atomic and Molecular Physics \\ 
4 &  Classical & Electromagnetism, Optics, Acoustics, Heat Transfer, Classical Mechanics and Fluid Dynamics\\ 
5 &  Plasma & Physics of Gases, Plasmas and Electric Discharges
\\ 
6 - 7 &  Condensed Matter &  Structural, Mechanical and Thermal Properties, Electronic Structure and Electrical, Magnetic and Optical Properties \\ 
8 &   Interdisc & Interdisciplinary Physics and Related Areas of Science and Technology
\\  
9 &  Astro & Astrophysics, Astronomy and Geophysics \\  \bottomrule 
\end{tabular} 
\end{threeparttable} 
\end{table}

\subsection{Test of randomness - multiple hypothesis correction}\label{multiplecorrection}
As stated in section \ref{randomtest}, under the hypothesis that scientists diversify their research portfolio at random, the probability that exactly $x$ authors are active in two sub-fields follows a hypergeometric distribution \cite{tumminello2011statistically}. 
A clear advantage of such a formulation is that we can easily associate a p-value to each element (i.e., link in the projected network) and evaluate significance. However, since we are performing hypothesis testing, we need to set a level of statistical significance accordingly. One of the most commonly used method to deal with multiple hypothesis testing is the \textit{Bonferroni} correction. Such a well known method controls for \textit{family-wise error rate} (FWER) in a very stringent fashion, computing the adjusted p-value by directly multiplying the number of simultaneously tested hypotheses. In many multiple testing settings, the Bonferroni correction might result too strict, and not appropriate in dealing with dependence. Thus, many less restrictive and more flexible alternatives such as FDR have been developed  \cite{benjamini1995controlling,benjamini2001control}.  Table \ref{rand} reports results for non-corrected p-values and after several correction methods: Bonferroni, Benjamini-Hochberg (BH), and Benjamini-Yekutieli (BY).  Results confirm that scientists' diversification choices can be hardly seen as a random phenomenon, irrespective of the correction method employed.

\begin{table}[b]
\centering \begin{threeparttable}
\caption{\label{rand}Test of randomness in scientists' research portfolio diversification}\fontsize{15}{25}\selectfont
\begin{tabular}{@{}cccc@{}}
\toprule
                       & \textbf{Positive} & \textbf{Negative} & \textbf{\% Non-Random} \\ \midrule
\textbf{No correction} & 1361              & 616               & 86.8                   \\
\textbf{Bonferroni}    & 1151              & 486               & 71.8                   \\
\textbf{BH}            & 1339              & 580               & 84.2                   \\
\textbf{BY}            & 1264              & 547               & 79.4                   \\ \bottomrule
\end{tabular}
{\footnotesize \textit{Note:}   2,278 pairs analyzed, 68 sub-fields, 197,682 scientists. Analysis performed employing the \textbf{R} package \textit{cooccur}.}\\ 
\end{threeparttable} 
\end{table}

\subsection{Additional estimation results}
As mentioned in section \ref{model} and \ref{stime}, we use a multivariate logistic regression to estimate the probability that a scientist diversifies in a sub-field different from her own specialization. Table \ref{var} summarizes our independent variables and includes information about our grouping strategy. Here, we provide results for each and every specification: (i) single specialization (full diversification), (ii) multiple-specialization (full diversification), (iii) single specialization (within field diversification), (iv) multiple specialization (within field diversification), (v) single specialization (between field diversification) and (vi) multiple specialization (between field diversification).

\begin{table}[b]
\tiny \centering
\begin{threeparttable}
\caption{\label{var}Variables and grouping strategy}\fontsize{7}{25}\selectfont
\begin{tabular}{lcc} 
\toprule Name & Group & Description \\ \midrule 
Knowledge relatedness   & 1 - KR  & Cosine similarity among sub-fields\\ 
Social relatedness & 2 - SR & Scientist' co-authors specialized in the sub-field different from her core one (dummy) \\ 
Field core &  3 - IF  &    macro-field specialization (categorical) \\ 
{\#}\ of\ PACS & 4 - IF & Number of PACS explored \\ 
{\#}\ of\ papers & 4  - IF &  Number of papers published\\ 
{\#}\ of\ co-authors & 4 - IF & Number of co-authors \\ 
PACS target popularity & 5 - SC &  Number of articles assigned to the target sub-field \\ 
$\Delta$ crowd &  5 - SC & Difference in the number of specialized scientists between core and target sub-field  \\  
$\Delta$ PACS citations &  6 - Cit &  Difference in the number citations between core and target sub-field \\  
$\Delta$ field citations &  6 - Cit  &  Difference in the number citations between core and target macro-field \\  
\bottomrule 
\end{tabular} 
\end{threeparttable} 
\end{table}

\paragraph*{Full diversification - Specification (i) and (ii)}
Results are summarized in Table \ref{tablesingle} and \ref{tablemultiple}, where the first column refers to the baseline model (without the interaction term between social and knowledge relatedness), column (2) refers to the model including the interaction term while column (3) presents the same results with clustering corrected standard errors. 
Figure \ref{single}-\textbf{a}/\textbf{c} show the differences in the probability of diversification as a function of knowledge and social relatedness, taking into account all the control variables. Figure \ref{single}-\textbf{b}/\textbf{d} provide evidence of the moderating role played by the similarity across sub-fields on the estimated coefficient of social relatedness

\begin{table}[b] \centering 
  \caption{(i) Single specialization - full diversification.} 
  \label{tablesingle} 
                     \fontsize{11}{12}\selectfont

\begin{tabular}{@{\extracolsep{5pt}}lccc} 
\\[-1.8ex]\hline 
\hline \\[-1.8ex] 
 & \multicolumn{3}{c}{\textit{Dependent variable: P(diversification)}} \\ 
\cline{2-4}  \\ 
 & Baseline & Interactions & Robust SE \\ 
\\[-1.8ex] & (1) & (2) & (3)\\ 
\hline \\[-1.8ex] 
  Knowledge Relatedness& 0.846$^{***}$ & 0.936$^{***}$ & 0.936$^{***}$ \\ 
  & (0.002) & (0.002) & (0.003) \\ 
  Social Relatedness & 2.647$^{***}$ & 2.827$^{***}$ & 2.827$^{***}$ \\ 
  & (0.004) & (0.005) & (0.006) \\ 
  field core-Atomic & $-$0.332$^{***}$ & $-$0.332$^{***}$ & $-$0.332$^{***}$ \\ 
  & (0.009) & (0.009) & (0.010) \\ 
  field core-Classical & $-$0.480$^{***}$ & $-$0.490$^{***}$ & $-$0.490$^{***}$ \\ 
  & (0.010) & (0.010) & (0.010) \\ 
  field core-Cond.matter & $-$1.094$^{***}$ & $-$1.088$^{***}$ & $-$1.088$^{***}$ \\ 
  & (0.010) & (0.010) & (0.012) \\ 
  field core-General & $-$0.710$^{***}$ & $-$0.722$^{***}$ & $-$0.722$^{***}$ \\ 
  & (0.010) & (0.010) & (0.011) \\ 
  field core-High.energy & 0.221$^{***}$ & 0.219$^{***}$ & 0.219$^{***}$ \\ 
  & (0.011) & (0.011) & (0.010) \\ 
  field core-Interdisc & $-$0.546$^{***}$ & $-$0.557$^{***}$ & $-$0.557$^{***}$ \\ 
  & (0.009) & (0.009) & (0.010) \\ 
  field core-Nuclear & 0.438$^{***}$ & 0.463$^{***}$ & 0.463$^{***}$ \\ 
  & (0.009) & (0.009) & (0.010) \\ 
  field core-Plasma & $-$0.258$^{***}$ & $-$0.269$^{***}$ & $-$0.269$^{***}$ \\ 
  & (0.014) & (0.014) & (0.013) \\ 
  {\#}\ of\ PACS & 0.891$^{***}$ & 0.882$^{***}$ & 0.882$^{***}$ \\ 
  & (0.003) & (0.003) & (0.002) \\ 
  {\#}\ of\ papers & $-$0.007$^{**}$ & 0.010$^{***}$ & 0.010$^{***}$ \\ 
  & (0.003) & (0.003) & (0.002) \\ 
  PACS target popularity & 1.130$^{***}$ & 1.130$^{***}$ & 1.130$^{***}$ \\ 
  & (0.003) & (0.003) & (0.002) \\ 
  $\Delta$ crowd & 0.239$^{***}$ & 0.239$^{***}$ & 0.239$^{***}$ \\ 
  & (0.002) & (0.002) & (0.002) \\ 
  {\#}\ of\ co-authors & $-$0.382$^{***}$ & $-$0.406$^{***}$ & $-$0.406$^{***}$ \\ 
  & (0.003) & (0.003) & (0.002) \\ 
  $\Delta$ PACS citations & $-$0.272$^{***}$ & $-$0.273$^{***}$ & $-$0.273$^{***}$ \\ 
  & (0.003) & (0.003) & (0.002) \\ 
  $\Delta$ field citations & $-$0.167$^{***}$ & $-$0.156$^{***}$ & $-$0.156$^{***}$ \\ 
  & (0.003) & (0.003) & (0.004) \\ 
  KR:SR &  & $-$0.255$^{***}$ & $-$0.255$^{***}$ \\ 
  &  & (0.004) & (0.004) \\ 
  Constant & $-$3.749$^{***}$ & $-$3.812$^{***}$ & $-$3.812$^{***}$ \\ 
  & (0.008) & (0.008) & (0.010) \\ 
 \hline \\[-1.8ex] 
Observations & 7,072,386 & 7,072,386 & 7,072,386 \\ 
Log Likelihood & $-$1,088,731.000 & $-$1,086,281.000 & $-$1,086,281.000 \\ 
Akaike Inf. Crit. & 2,177,498.000 & 2,172,600.000 & 2,172,600.000 \\ 
\hline 
\hline \\[-1.8ex] 
\textit{Note:}  & \multicolumn{3}{r}{$^{*}$p$<$0.1; $^{**}$p$<$0.05; $^{***}$p$<$0.01} \\ 
\end{tabular} 
\end{table}

\begin{table}[b] \centering 
  \caption{(ii) Multiple-specialization - full diversification} 
  \label{tablemultiple}
                       \fontsize{11}{12}\selectfont

\begin{tabular}{@{\extracolsep{5pt}}lccc} 
\\[-1.8ex]\hline 
\hline \\[-1.8ex] 
 & \multicolumn{3}{c}{\textit{Dependent variable: P(diversification)}} \\ 
\cline{2-4} 
\\[-1.8ex] & \multicolumn{3}{c}{} \\ 
 & Baseline & Interactions & Robust SE \\ 
\\[-1.8ex] & (1) & (2) & (3)\\ 
\hline \\[-1.8ex] 
 Knowledge Relatedness& 0.628$^{***}$ & 0.689$^{***}$ & 0.689$^{***}$ \\ 
  & (0.001) & (0.005) & (0.009) \\ 
  Social Relatedness& 4.221$^{***}$ & 4.243$^{***}$ & 4.243$^{***}$ \\ 
  & (0.005) & (0.005) & (0.019) \\ 
  field core-Atomic & $-$0.427$^{***}$ & $-$0.427$^{***}$ & $-$0.427$^{***}$ \\ 
  & (0.004) & (0.004) & (0.007) \\ 
  field core-Classical & $-$0.475$^{***}$ & $-$0.475$^{***}$ & $-$0.475$^{***}$ \\ 
  & (0.005) & (0.005) & (0.007) \\ 
  field core-Cond.matter & $-$0.761$^{***}$ & $-$0.760$^{***}$ & $-$0.760$^{***}$ \\ 
  & (0.004) & (0.004) & (0.009) \\ 
  field core-General & $-$0.537$^{***}$ & $-$0.537$^{***}$ & $-$0.537$^{***}$ \\ 
  & (0.004) & (0.004) & (0.007) \\ 
  field core-High.energy & 0.165$^{***}$ & 0.165$^{***}$ & 0.165$^{***}$ \\ 
  & (0.005) & (0.005) & (0.006) \\ 
  field core-Interdisc & $-$0.552$^{***}$ & $-$0.552$^{***}$ & $-$0.552$^{***}$ \\ 
  & (0.004) & (0.004) & (0.007) \\ 
  field core-Nuclear & 0.163$^{***}$ & 0.163$^{***}$ & 0.163$^{***}$ \\ 
  & (0.004) & (0.004) & (0.006) \\ 
  field core-Plasma & $-$0.409$^{***}$ & $-$0.409$^{***}$ & $-$0.409$^{***}$ \\ 
  & (0.007) & (0.007) & (0.008) \\ 
  {\#}\ of\ PACS & 0.768$^{***}$ & 0.768$^{***}$ & 0.768$^{***}$ \\ 
  & (0.001) & (0.001) & (0.003) \\ 
  {\#}\ of\ papers & 0.117$^{***}$ & 0.117$^{***}$ & 0.117$^{***}$ \\ 
  & (0.001) & (0.001) & (0.003) \\ 
  PACS target popularity & 0.611$^{***}$ & 0.611$^{***}$ & 0.611$^{***}$ \\ 
  & (0.001) & (0.001) & (0.002) \\ 
  $\Delta$ crowd & 0.359$^{***}$ & 0.359$^{***}$ & 0.359$^{***}$ \\ 
  & (0.001) & (0.001) & (0.003) \\ 
  {\#}\ of\ co-authors & $-$0.346$^{***}$ & $-$0.346$^{***}$ & $-$0.346$^{***}$ \\ 
  & (0.001) & (0.001) & (0.004) \\ 
  $\Delta$ PACS citations & $-$0.333$^{***}$ & $-$0.333$^{***}$ & $-$0.333$^{***}$ \\ 
  & (0.001) & (0.001) & (0.003) \\ 
  $\Delta$ field citations& $-$0.071$^{***}$ & $-$0.070$^{***}$ & $-$0.070$^{***}$ \\ 
  & (0.001) & (0.001) & (0.004) \\ 
  KR:SR&  & $-$0.062$^{***}$ & $-$0.062$^{***}$ \\ 
  &  & (0.005) & (0.010) \\ 
  Constant & $-$5.855$^{***}$ & $-$5.877$^{***}$ & $-$5.877$^{***}$ \\ 
  & (0.006) & (0.007) & (0.020) \\ 
 \hline \\[-1.8ex] 
Observations & 35,562,394 & 35,562,394 & 35,562,394 \\ 
Log Likelihood & $-$7,299,777.000 & $-$7,299,692.000 & $-$7,299,692.000 \\ 
Akaike Inf. Crit. & 14,599,590.000 & 14,599,421.000 & 14,599,421.000 \\ 
\hline 
\hline \\[-1.8ex] 
\textit{Note:}  & \multicolumn{3}{r}{$^{*}$p$<$0.1; $^{**}$p$<$0.05; $^{***}$p$<$0.01} \\ 
\end{tabular} 
\end{table}

\begin{figure}[b]
\centering 
{\includegraphics[scale=0.5]{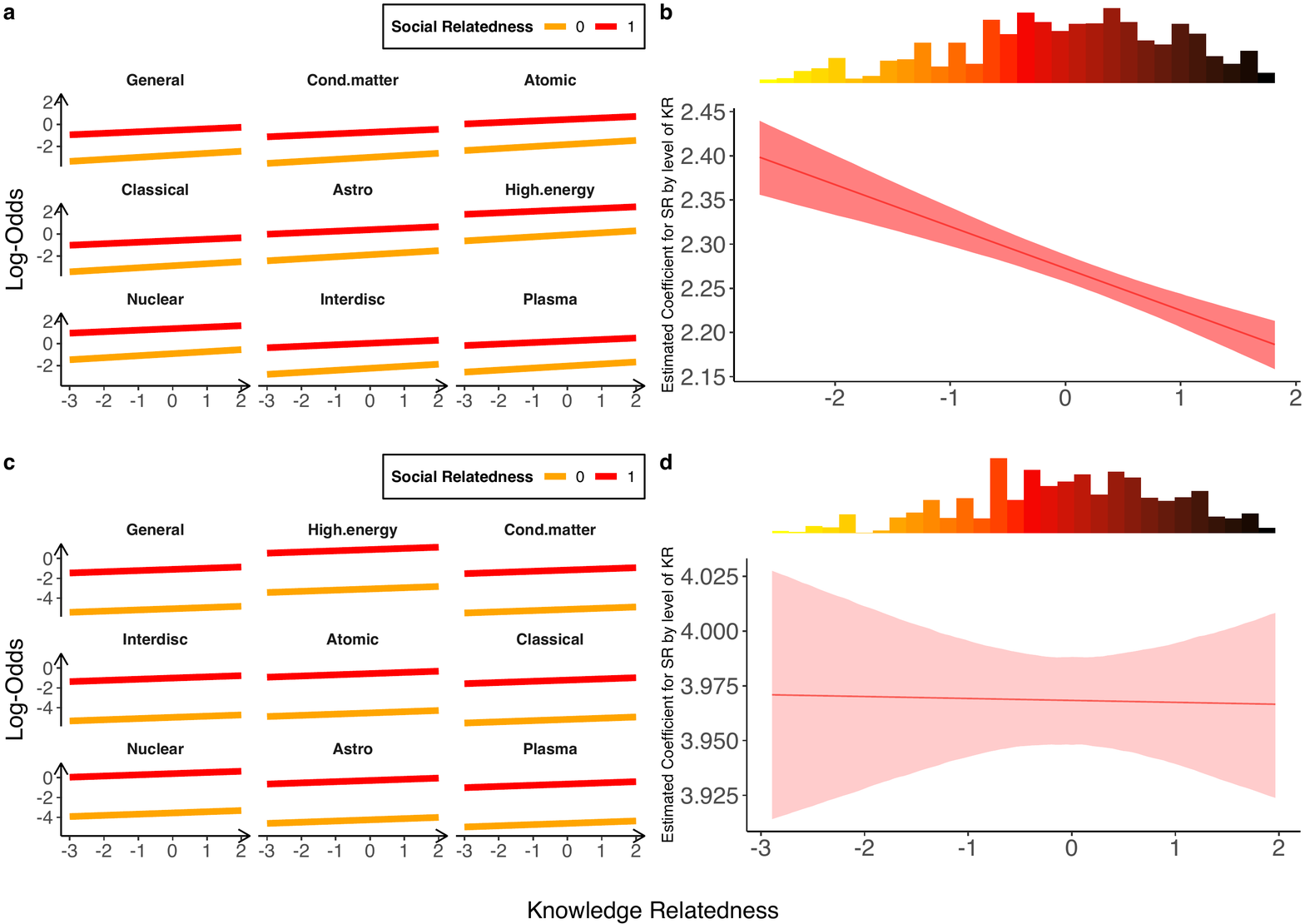}}
\caption{\textbf{Scientists' research portfolio diversification - (within field diversification) single and multiple specialization}.
{\small \textbf{a}, Log-odds as a function of social and (standardized) knowledge relatedness, controlling for all the confounding variables - specification (iii). \textbf{b}, Estimated coefficient for social relatedness conditional on (standardized) knowledge relatedness - specification (iii). \textbf{c}, Log-odds as a function of social and (standardized) knowledge relatedness, controlling for the all confounding variables - specification (iv). \textbf{d}, Estimated coefficient for social relatedness conditional on (standardized) knowledge relatedness - specification (iv). \textbf{b} and \textbf{d} include the distribution of the conditional variable (i.e., knowledge relatedness). The color palette is in accordance with the similarity matrix (Figure \ref{measures}-\textbf{c})}.
\label{within}}
\end{figure}

\paragraph*{Within field diversification - Specification (iii) and (iv).}
Figure \ref{within}-\textbf{a}/\textbf{b} plots the results for the single specialization case: knowledge and  social relatedness are still significant as well as their interaction, but the magnitude of the coefficients is smaller with respect to the full diversification case. In addition, when we consider the multiple specialization case (Figure \ref{within}-\textbf{c}/\textbf{d}), coefficients shrink  further and the interaction term between social and knowledge relatedness is no longer significant (see Table \ref{withinsingle} and \ref{withinmultiple} for details).  

\begin{table}[b] \centering 
  \caption{(iii) Single specialization - within field diversification} 
  \label{withinsingle} 
                         \fontsize{11}{12}\selectfont

\begin{tabular}{@{\extracolsep{5pt}}lccc} 
\\[-1.8ex]\hline 
\hline \\[-1.8ex] 
 & \multicolumn{3}{c}{\textit{Dependent variable: P(diversification)}} \\ 
\cline{2-4} 
\\[-1.8ex] & \multicolumn{3}{c}{} \\ 
 & Baseline & Interactions & Robust SE \\ 
\\[-1.8ex] & (1) & (2) & (3)\\ 
\hline \\[-1.8ex] 
 Knowledge Relatedness& 0.166$^{***}$ & 0.184$^{***}$ & 0.184$^{***}$ \\ 
  & (0.004) & (0.004) & (0.005) \\ 
  Social Relatedness& 2.265$^{***}$ & 2.272$^{***}$ & 2.272$^{***}$ \\ 
  & (0.008) & (0.008) & (0.008) \\ 
  field core-Atomic & 0.059$^{***}$ & 0.056$^{**}$ & 0.056$^{**}$ \\ 
  & (0.022) & (0.022) & (0.025) \\ 
  field core-Classical & $-$1.003$^{***}$ & $-$1.001$^{***}$ & $-$1.001$^{***}$ \\ 
  & (0.026) & (0.026) & (0.029) \\ 
  field core-Cond.matter & $-$1.108$^{***}$ & $-$1.110$^{***}$ & $-$1.110$^{***}$ \\ 
  & (0.021) & (0.021) & (0.024) \\ 
  field core-General & $-$0.931$^{***}$ & $-$0.927$^{***}$ & $-$0.927$^{***}$ \\ 
  & (0.026) & (0.026) & (0.028) \\ 
  field core-High.energy & 1.809$^{***}$ & 1.806$^{***}$ & 1.806$^{***}$ \\ 
  & (0.025) & (0.025) & (0.027) \\ 
  field core-Interdisc & $-$0.353$^{***}$ & $-$0.357$^{***}$ & $-$0.357$^{***}$ \\ 
  & (0.024) & (0.024) & (0.026) \\ 
  field core-Nuclear & 0.978$^{***}$ & 0.969$^{***}$ & 0.969$^{***}$ \\ 
  & (0.021) & (0.021) & (0.024) \\ 
  field core-Plasma & $-$0.149$^{**}$ & $-$0.155$^{**}$ & $-$0.155$^{**}$ \\ 
  & (0.068) & (0.068) & (0.068) \\ 
  {\#}\ of\ PACS & 0.769$^{***}$ & 0.769$^{***}$ & 0.769$^{***}$ \\ 
  & (0.005) & (0.005) & (0.005) \\ 
  {\#}\ of\ papers & 0.064$^{***}$ & 0.065$^{***}$ & 0.065$^{***}$ \\ 
  & (0.006) & (0.006) & (0.005) \\ 
  PACS target popularity & 1.372$^{***}$ & 1.370$^{***}$ & 1.370$^{***}$ \\ 
  & (0.006) & (0.006) & (0.005) \\ 
  $\Delta$ crowd & 0.130$^{***}$ & 0.131$^{***}$ & 0.131$^{***}$ \\ 
  & (0.004) & (0.004) & (0.003) \\ 
  {\#}\ of\ co-authors & $-$0.239$^{***}$ & $-$0.240$^{***}$ & $-$0.240$^{***}$ \\ 
  & (0.005) & (0.005) & (0.004) \\ 
  $\Delta$ PACS citations & $-$0.209$^{***}$ & $-$0.208$^{***}$ & $-$0.208$^{***}$ \\ 
  & (0.005) & (0.005) & (0.004) \\ 
  KR:SR&  & $-$0.047$^{***}$ & $-$0.047$^{***}$ \\ 
  &  & (0.007) & (0.007) \\ 
  Constant & $-$1.883$^{***}$ & $-$1.882$^{***}$ & $-$1.882$^{***}$ \\ 
  & (0.020) & (0.020) & (0.022) \\ 
 \hline \\[-1.8ex] 
Observations & 1,000,230 & 1,000,230 & 1,000,230 \\ 
Log Likelihood & $-$334,720.800 & $-$334,697.300 & $-$334,697.300 \\ 
Akaike Inf. Crit. & 669,475.700 & 669,430.600 & 669,430.600 \\ 
\hline 
\hline \\[-1.8ex] 
\textit{Note:}  & \multicolumn{3}{r}{$^{*}$p$<$0.1; $^{**}$p$<$0.05; $^{***}$p$<$0.01} \\ 
\end{tabular} 
\end{table}

\begin{table}[b] \centering 
  \caption{(iv) Multiple specialization - within field diversification} 
  \label{withinmultiple} 
                           \fontsize{11}{12}\selectfont

\begin{tabular}{@{\extracolsep{5pt}}lccc} 
\\[-1.8ex]\hline 
\hline \\[-1.8ex] 
 & \multicolumn{3}{c}{\textit{Dependent variable: P(diversification)}} \\ 
\cline{2-4} 
\\[-1.8ex] & \multicolumn{3}{c}{} \\ 
 & Baseline & Interactions & Robust SE \\ 
\\[-1.8ex] & (1) & (2) & (3)\\ 
\hline \\[-1.8ex] 
 Knowledge Relatedness& 0.121$^{***}$ & 0.121$^{***}$ & 0.121$^{***}$ \\ 
  & (0.009) & (0.009) & (0.013) \\ 
  Social Relatedness& 3.968$^{***}$ & 3.968$^{***}$ & 3.968$^{***}$ \\ 
  & (0.010) & (0.010) & (0.021) \\ 
  field core-Atomic & $-$0.276$^{***}$ & $-$0.276$^{***}$ & $-$0.276$^{***}$ \\ 
  & (0.011) & (0.011) & (0.021) \\ 
  field core-Classical & $-$0.932$^{***}$ & $-$0.932$^{***}$ & $-$0.932$^{***}$ \\ 
  & (0.013) & (0.013) & (0.023) \\ 
  field core-Cond.matter & $-$0.892$^{***}$ & $-$0.892$^{***}$ & $-$0.892$^{***}$ \\ 
  & (0.011) & (0.011) & (0.020) \\ 
  field core-General & $-$0.823$^{***}$ & $-$0.823$^{***}$ & $-$0.823$^{***}$ \\ 
  & (0.012) & (0.012) & (0.021) \\ 
  field core-High.energy & 1.176$^{***}$ & 1.176$^{***}$ & 1.176$^{***}$ \\ 
  & (0.012) & (0.012) & (0.023) \\ 
  field core-Interdisc & $-$0.724$^{***}$ & $-$0.724$^{***}$ & $-$0.724$^{***}$ \\ 
  & (0.012) & (0.012) & (0.021) \\ 
  field core-Nuclear & 0.692$^{***}$ & 0.692$^{***}$ & 0.692$^{***}$ \\ 
  & (0.011) & (0.011) & (0.021) \\ 
  field core-Plasma & $-$0.361$^{***}$ & $-$0.361$^{***}$ & $-$0.361$^{***}$ \\ 
  & (0.041) & (0.041) & (0.058) \\ 
  {\#}\ of\ PACS & 0.497$^{***}$ & 0.497$^{***}$ & 0.497$^{***}$ \\ 
  & (0.002) & (0.002) & (0.004) \\ 
  {\#}\ of\ papers & 0.252$^{***}$ & 0.252$^{***}$ & 0.252$^{***}$ \\ 
  & (0.002) & (0.002) & (0.004) \\ 
  PACS target popularity & 0.774$^{***}$ & 0.774$^{***}$ & 0.774$^{***}$ \\ 
  & (0.002) & (0.002) & (0.003) \\ 
  $\Delta$ crowd & 0.345$^{***}$ & 0.345$^{***}$ & 0.345$^{***}$ \\ 
  & (0.002) & (0.002) & (0.003) \\ 
  {\#}\ of\ co-authors & $-$0.145$^{***}$ & $-$0.145$^{***}$ & $-$0.145$^{***}$ \\ 
  & (0.002) & (0.002) & (0.004) \\ 
  $\Delta$ PACS citations & $-$0.313$^{***}$ & $-$0.313$^{***}$ & $-$0.313$^{***}$ \\ 
  & (0.002) & (0.002) & (0.003) \\ 
  KR:SR& $-$0.001 & $-$0.001 & $-$0.001 \\ 
  & (0.009) & (0.009) & (0.013) \\ 
  Constant & $-$4.250$^{***}$ & $-$4.250$^{***}$ & $-$4.250$^{***}$ \\ 
  & (0.014) & (0.014) & (0.028) \\ 
 \hline \\[-1.8ex] 
Observations & 5,407,404 & 5,407,404 & 5,407,404 \\ 
Log Likelihood & $-$2,166,803.000 & $-$2,166,803.000 & $-$2,166,803.000 \\ 
Akaike Inf. Crit. & 4,333,642.000 & 4,333,642.000 & 4,333,642.000 \\ 
\hline 
\hline \\[-1.8ex] 
\textit{Note:}  & \multicolumn{3}{r}{$^{*}$p$<$0.1; $^{**}$p$<$0.05; $^{***}$p$<$0.01} \\ 
\end{tabular} 
\end{table}

\begin{figure}[b]
\centering 
{\includegraphics[scale=0.5]{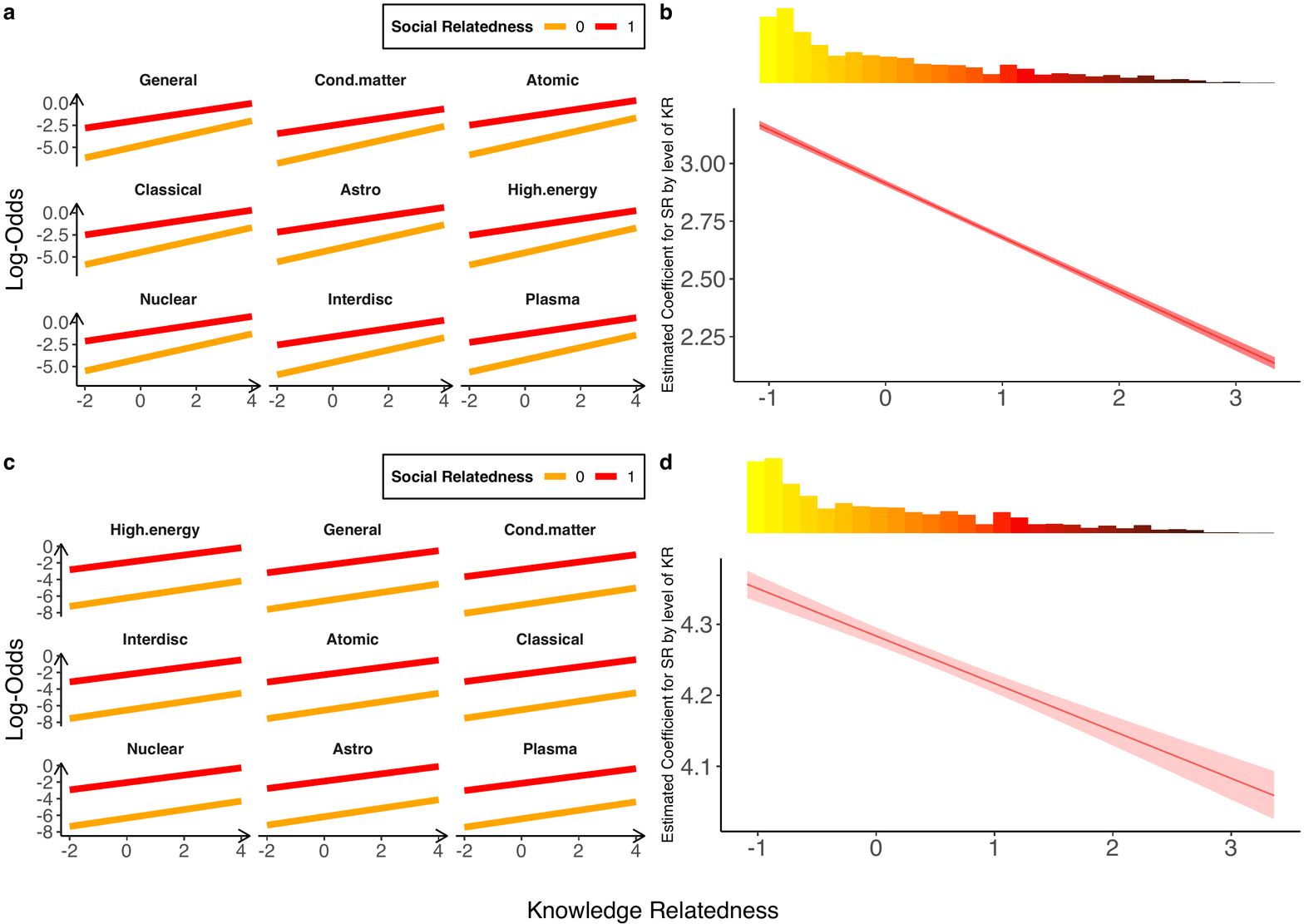}}
\caption{\textbf{Scientists' research portfolio diversification - (between field diversification) single and multiple specialization}.
{\small \textbf{a}, Log-odds as a function of social and (standardized) knowledge relatedness, controlling for all the confounding variables - specification (v). \textbf{b}, Estimated coefficient for social relatedness conditional on (standardized) knowledge relatedness - specification (v). \textbf{c}, Log-odds as a function of social and (standardized) knowledge relatedness, controlling for all the confounding variables - specification (vi). \textbf{d}, Estimated coefficient for social relatedness conditional on (standardized) knowledge relatedness - specification (vi). \textbf{b} and \textbf{d} include the distribution of the conditional variable (i.e., knowledge relatedness). The color palette is in accordance with the similarity matrix (Figure \ref{measures}-\textbf{c}).}
\label{bet} }
\end{figure}

\paragraph*{Between field diversification - Specification (v) and (vi).}
As far as the between field diversification is concerned, the general trends in terms of social and cognitive proximity are confirmed. Moreover, the negative interaction term remains statistically significant and not negligible in magnitude for both model specifications (single and multiple specialization). Figure \ref{bet}, Table \ref{betsingle} and Table \ref{betmultiple}  summarize the results.

\begin{table}[b] \centering 
  \caption{(v) Single specialization - between field diversification} 
  \label{betsingle} 
                             \fontsize{11}{12}\selectfont

\begin{tabular}{@{\extracolsep{5pt}}lccc} 
\\[-1.8ex]\hline 
\hline \\[-1.8ex] 
 & \multicolumn{3}{c}{\textit{Dependent variable: P(diversification)}} \\ 
\cline{2-4} 
\\[-1.8ex] & \multicolumn{3}{c}{} \\ 
 & Baseline & Interactions & Robust SE \\ 
\\[-1.8ex] & (1) & (2) & (3)\\ 
\hline \\[-1.8ex] 
 Knowledge Relatedness& 0.622$^{***}$ & 0.702$^{***}$ & 0.702$^{***}$ \\ 
  & (0.002) & (0.003) & (0.003) \\ 
  Social Relatedness& 2.768$^{***}$ & 2.914$^{***}$ & 2.914$^{***}$ \\ 
  & (0.006) & (0.006) & (0.008) \\ 
  field core-Atomic & $-$0.292$^{***}$ & $-$0.303$^{***}$ & $-$0.303$^{***}$ \\ 
  & (0.010) & (0.010) & (0.010) \\ 
  field core-Classical & $-$0.304$^{***}$ & $-$0.313$^{***}$ & $-$0.313$^{***}$ \\ 
  & (0.010) & (0.010) & (0.010) \\ 
  field core-Cond.matter & $-$1.294$^{***}$ & $-$1.263$^{***}$ & $-$1.263$^{***}$ \\ 
  & (0.013) & (0.013) & (0.017) \\ 
  field core-General & $-$0.628$^{***}$ & $-$0.632$^{***}$ & $-$0.632$^{***}$ \\ 
  & (0.011) & (0.011) & (0.012) \\ 
  field core-High.energy & $-$0.352$^{***}$ & $-$0.360$^{***}$ & $-$0.360$^{***}$ \\ 
  & (0.013) & (0.014) & (0.013) \\ 
  field core-Interdisc & $-$0.356$^{***}$ & $-$0.365$^{***}$ & $-$0.365$^{***}$ \\ 
  & (0.010) & (0.010) & (0.011) \\ 
  field core-Nuclear & 0.060$^{***}$ & 0.068$^{***}$ & 0.068$^{***}$ \\ 
  & (0.011) & (0.011) & (0.011) \\ 
  field core-Plasma & $-$0.062$^{***}$ & $-$0.074$^{***}$ & $-$0.074$^{***}$ \\ 
  & (0.014) & (0.014) & (0.015) \\ 
  {\#}\ of\ PACS & 1.010$^{***}$ & 1.003$^{***}$ & 1.003$^{***}$ \\ 
  & (0.004) & (0.004) & (0.004) \\ 
  {\#}\ of\ papers & $-$0.050$^{***}$ & $-$0.032$^{***}$ & $-$0.032$^{***}$ \\ 
  & (0.004) & (0.004) & (0.004) \\ 
  PACS target popularity & 1.114$^{***}$ & 1.108$^{***}$ & 1.108$^{***}$ \\ 
  & (0.003) & (0.003) & (0.003) \\ 
  $\Delta$ crowd & 0.322$^{***}$ & 0.320$^{***}$ & 0.320$^{***}$ \\ 
  & (0.003) & (0.003) & (0.003) \\ 
  {\#}\ of\ co-authors & $-$0.461$^{***}$ & $-$0.488$^{***}$ & $-$0.488$^{***}$ \\ 
  & (0.003) & (0.003) & (0.003) \\ 
  $\Delta$ PACS citations & $-$0.375$^{***}$ & $-$0.369$^{***}$ & $-$0.369$^{***}$ \\ 
  & (0.004) & (0.004) & (0.004) \\ 
  $\Delta$ field citations& $-$0.209$^{***}$ & $-$0.196$^{***}$ & $-$0.196$^{***}$ \\ 
  & (0.004) & (0.004) & (0.006) \\ 
  KR:SR&  & $-$0.234$^{***}$ & $-$0.234$^{***}$ \\ 
  &  & (0.004) & (0.005) \\ 
  Constant & $-$4.115$^{***}$ & $-$4.168$^{***}$ & $-$4.168$^{***}$ \\ 
  & (0.009) & (0.009) & (0.010) \\ 
 \hline \\[-1.8ex] 
Observations & 6,072,156 & 6,072,156 & 6,072,156 \\ 
Log Likelihood & $-$717,839.000 & $-$716,398.900 & $-$716,398.900 \\ 
Akaike Inf. Crit. & 1,435,714.000 & 1,432,836.000 & 1,432,836.000 \\ 
\hline 
\hline \\[-1.8ex] 
\textit{Note:}  & \multicolumn{3}{r}{$^{*}$p$<$0.1; $^{**}$p$<$0.05; $^{***}$p$<$0.01} \\ 
\end{tabular} 
\end{table} 

\begin{table}[b] \centering 
  \caption{(vi) Multiple specialization - between field diversification} 
  \label{betmultiple} 
                               \fontsize{11}{12}\selectfont
\begin{tabular}{@{\extracolsep{5pt}}lccc} 
\\[-1.8ex]\hline 
\hline \\[-1.8ex] 
 & \multicolumn{3}{c}{\textit{Dependent variable: P(diversification)}} \\ 
\cline{2-4} 
\\[-1.8ex] & \multicolumn{3}{c}{} \\ 
 & Baseline & Interactions & Robust SE \\ 
\\[-1.8ex] & (1) & (2) & (3)\\ 
\hline \\[-1.8ex] 
 cos & 0.446$^{***}$ & 0.511$^{***}$ & 0.511$^{***}$ \\ 
  & (0.001) & (0.005) & (0.011) \\ 
  Social Relatedness& 4.261$^{***}$ & 4.284$^{***}$ & 4.284$^{***}$ \\ 
  & (0.006) & (0.006) & (0.021) \\ 
  field core-Atomic & $-$0.384$^{***}$ & $-$0.385$^{***}$ & $-$0.385$^{***}$ \\ 
  & (0.005) & (0.005) & (0.008) \\ 
  field core-Classical & $-$0.328$^{***}$ & $-$0.328$^{***}$ & $-$0.328$^{***}$ \\ 
  & (0.005) & (0.005) & (0.008) \\ 
  field core-Cond.matter & $-$0.904$^{***}$ & $-$0.903$^{***}$ & $-$0.903$^{***}$ \\ 
  & (0.005) & (0.005) & (0.013) \\ 
  field core-General & $-$0.421$^{***}$ & $-$0.422$^{***}$ & $-$0.422$^{***}$ \\ 
  & (0.005) & (0.005) & (0.008) \\ 
  field core-High.energy & $-$0.060$^{***}$ & $-$0.060$^{***}$ & $-$0.060$^{***}$ \\ 
  & (0.005) & (0.005) & (0.008) \\ 
  field core-Interdisc & $-$0.367$^{***}$ & $-$0.367$^{***}$ & $-$0.367$^{***}$ \\ 
  & (0.005) & (0.005) & (0.008) \\ 
  field core-Nuclear & $-$0.160$^{***}$ & $-$0.161$^{***}$ & $-$0.161$^{***}$ \\ 
  & (0.005) & (0.005) & (0.009) \\ 
  field core-Plasma & $-$0.255$^{***}$ & $-$0.256$^{***}$ & $-$0.256$^{***}$ \\ 
  & (0.007) & (0.007) & (0.009) \\ 
  {\#}\ of\ PACS & 0.944$^{***}$ & 0.944$^{***}$ & 0.944$^{***}$ \\ 
  & (0.001) & (0.001) & (0.004) \\ 
  {\#}\ of\ papers & 0.050$^{***}$ & 0.050$^{***}$ & 0.050$^{***}$ \\ 
  & (0.001) & (0.001) & (0.004) \\ 
  PACS target popularity & 0.559$^{***}$ & 0.559$^{***}$ & 0.559$^{***}$ \\ 
  & (0.001) & (0.001) & (0.002) \\ 
  $\Delta$ crowd & 0.392$^{***}$ & 0.393$^{***}$ & 0.393$^{***}$ \\ 
  & (0.002) & (0.002) & (0.003) \\ 
  {\#}\ of\ co-authors & $-$0.444$^{***}$ & $-$0.444$^{***}$ & $-$0.444$^{***}$ \\ 
  & (0.001) & (0.001) & (0.006) \\ 
  $\Delta$ PACS citations & $-$0.354$^{***}$ & $-$0.354$^{***}$ & $-$0.354$^{***}$ \\ 
  & (0.002) & (0.002) & (0.003) \\ 
  $\Delta$ field citations& $-$0.143$^{***}$ & $-$0.143$^{***}$ & $-$0.143$^{***}$ \\ 
  & (0.002) & (0.002) & (0.005) \\ 
  KR:SR&  & $-$0.067$^{***}$ & $-$0.067$^{***}$ \\ 
  &  & (0.005) & (0.011) \\ 
  Constant & $-$6.142$^{***}$ & $-$6.165$^{***}$ & $-$6.165$^{***}$ \\ 
  & (0.007) & (0.008) & (0.022) \\ 
 \hline \\[-1.8ex] 
Observations & 30,154,990 & 30,154,990 & 30,154,990 \\ 
Log Likelihood & $-$4,971,576.000 & $-$4,971,497.000 & $-$4,971,497.000 \\ 
Akaike Inf. Crit. & 9,943,188.000 & 9,943,033.000 & 9,943,033.000 \\ 
\hline 
\hline \\[-1.8ex] 
\textit{Note:}  & \multicolumn{3}{r}{$^{*}$p$<$0.1; $^{**}$p$<$0.05; $^{***}$p$<$0.01} \\ 
\end{tabular} 
\end{table} 

\subsection{Temporal evolution of the physics knowledge space}\label{temporal}
\begin{figure}[b]
\centering 
\subcaptionbox{Full data}
{\includegraphics[scale=0.5]{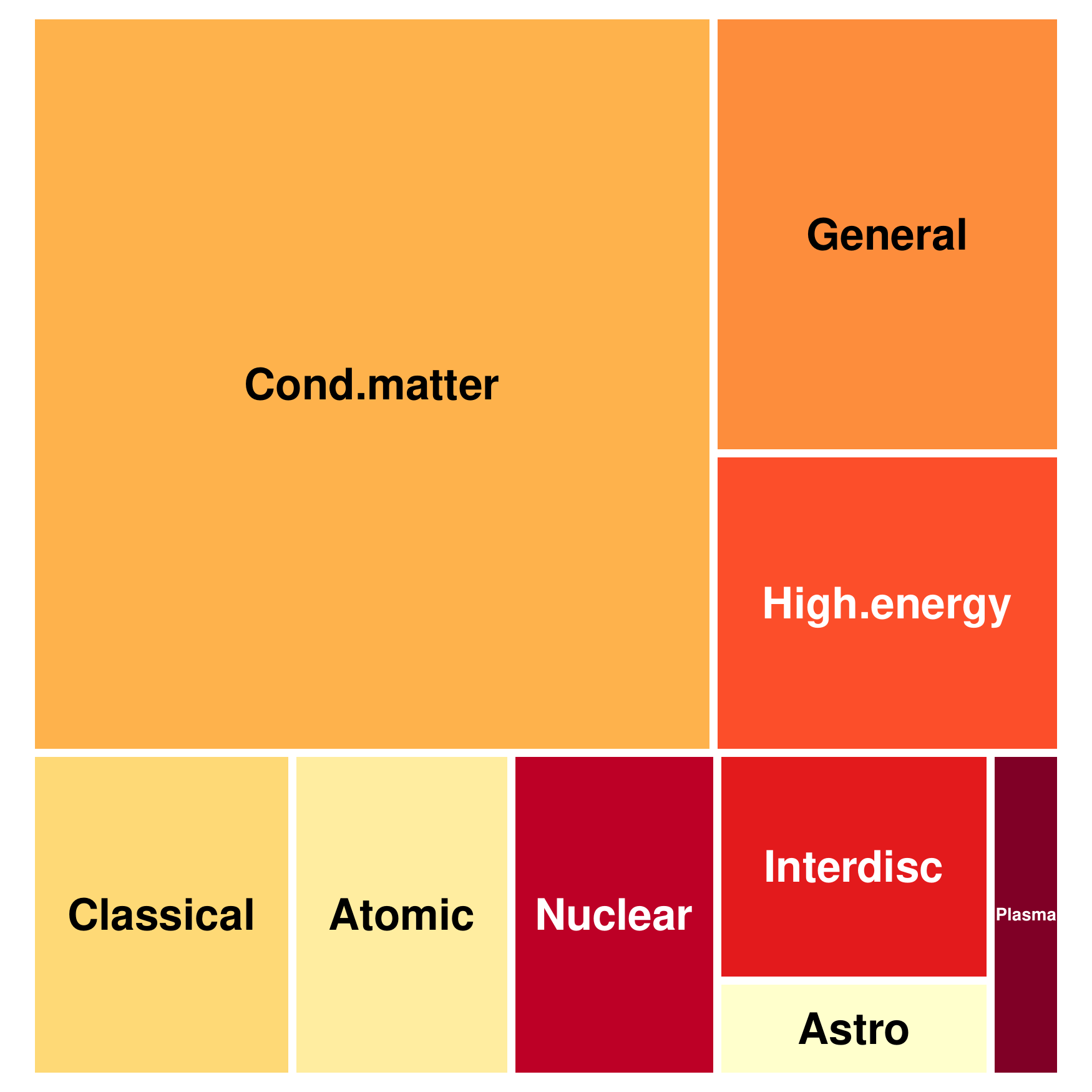}}%
\subcaptionbox{Last 10 years}
{\includegraphics[scale=0.5]{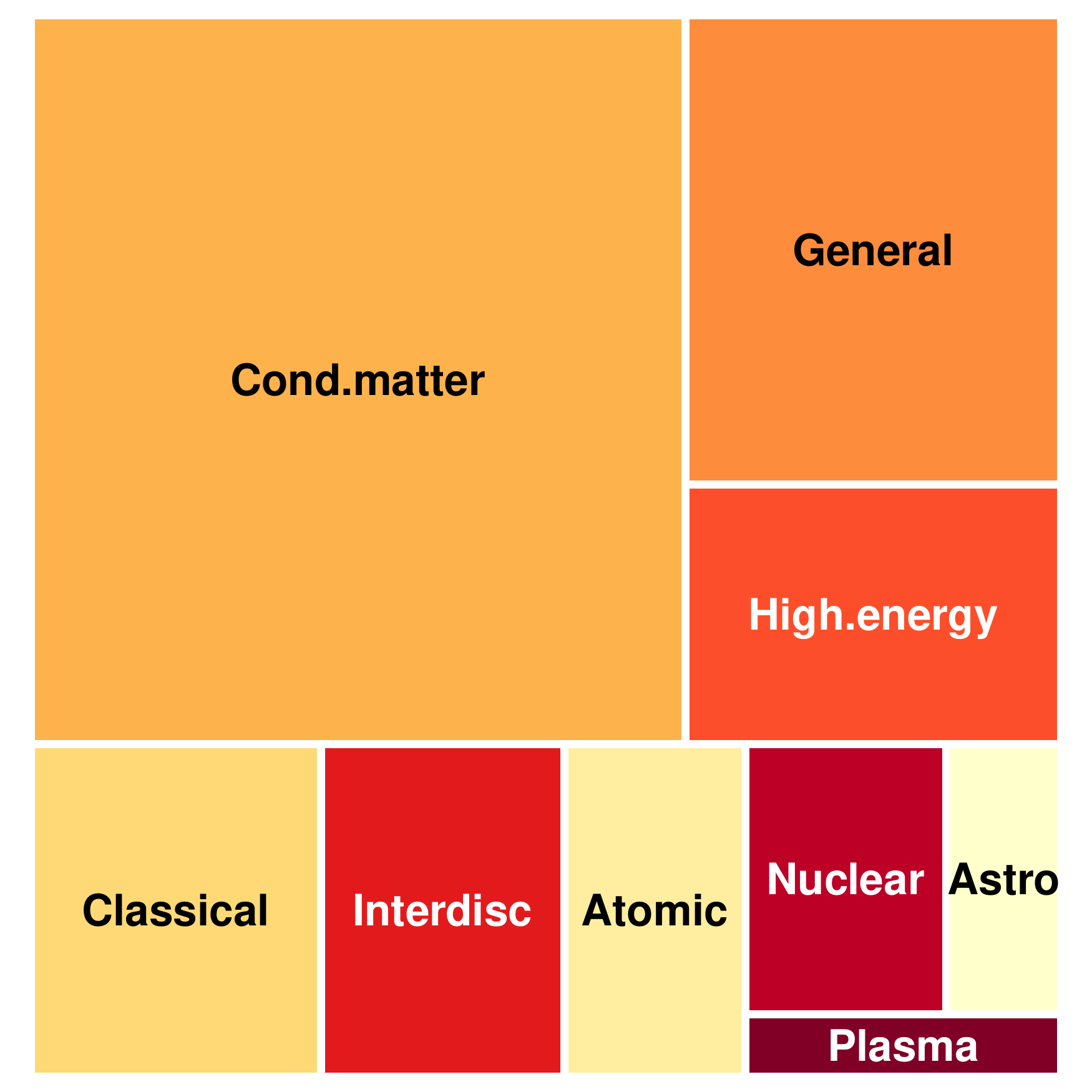}}
\subcaptionbox{Full data}
{\includegraphics[scale=0.5]{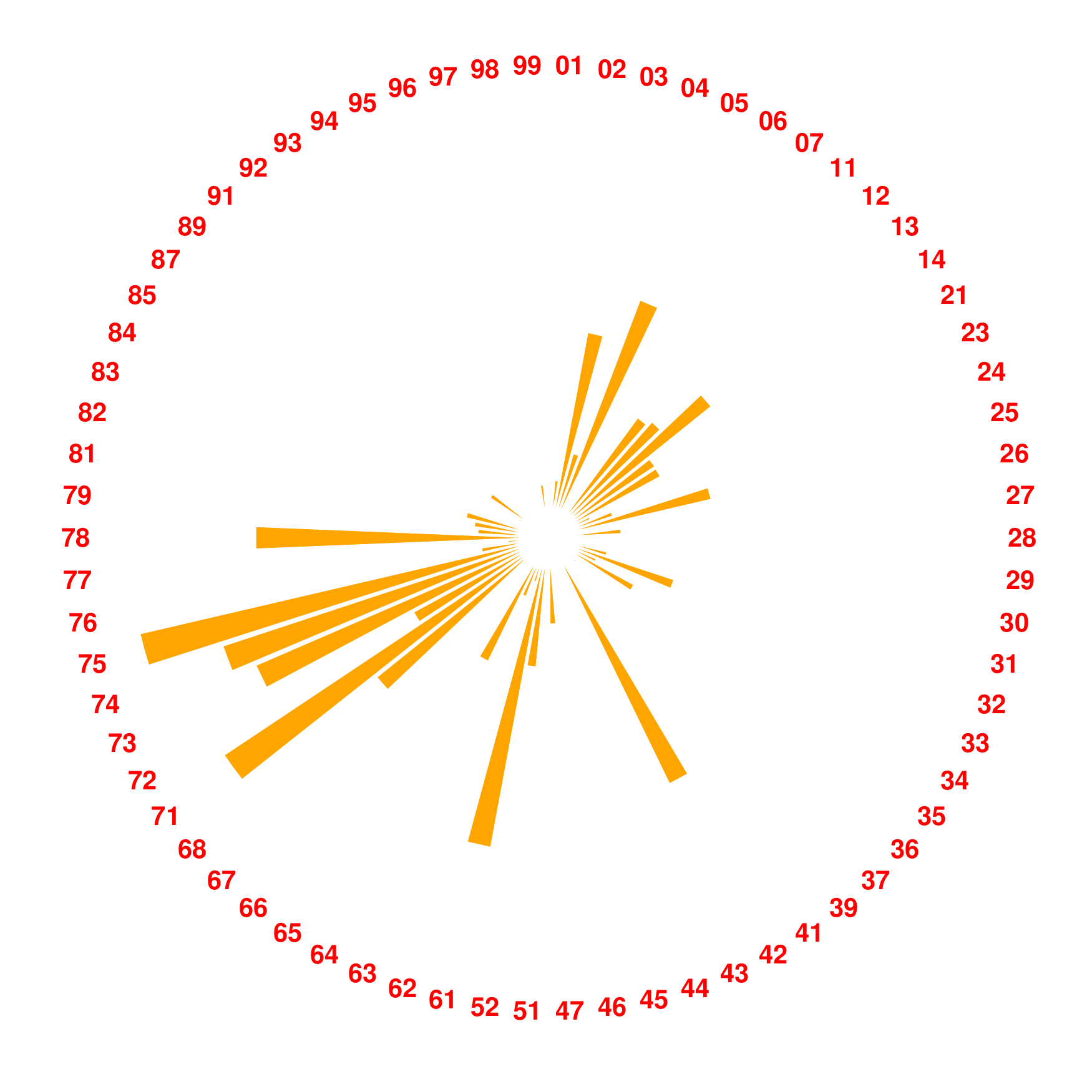}}%
\subcaptionbox{Last 10 years}
{\includegraphics[scale=0.5]{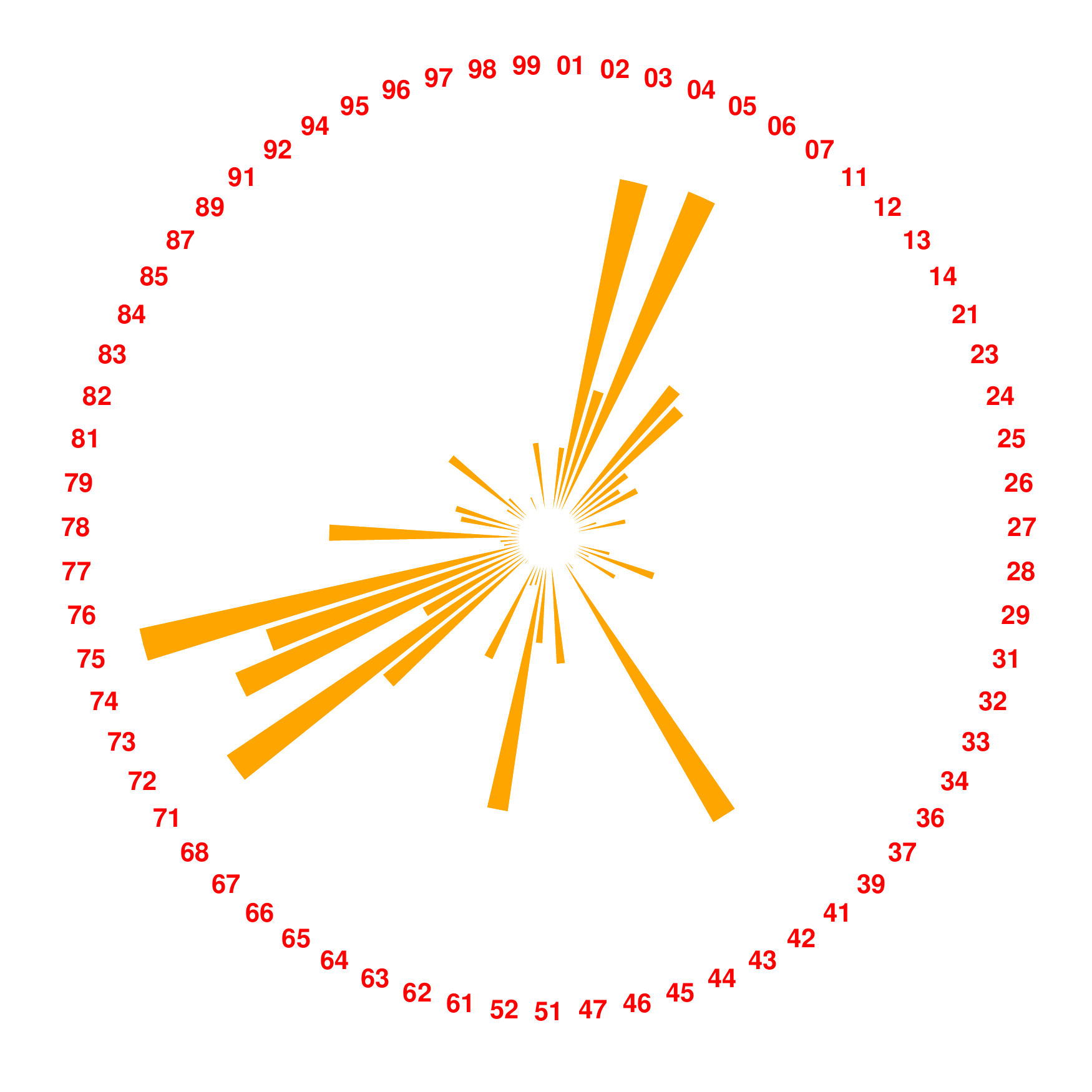}}
\caption{\textbf{Popularity of fields and sub-fields over time.}
{\small We focus on a subset including articles published from 2000 to 2009 (last 10 years in our data) to compare the popularity of physics fields and sub-fields over time (i.e., number of articles assigned to a given field/sub-field). The distribution of topics remains fairly stable, except for the rise of interdisciplinary physics.}
\label{pacs_time}}
\end{figure}

\begin{figure}[b]
\centering 
\subcaptionbox{Full Data}
{\includegraphics[scale=0.5]{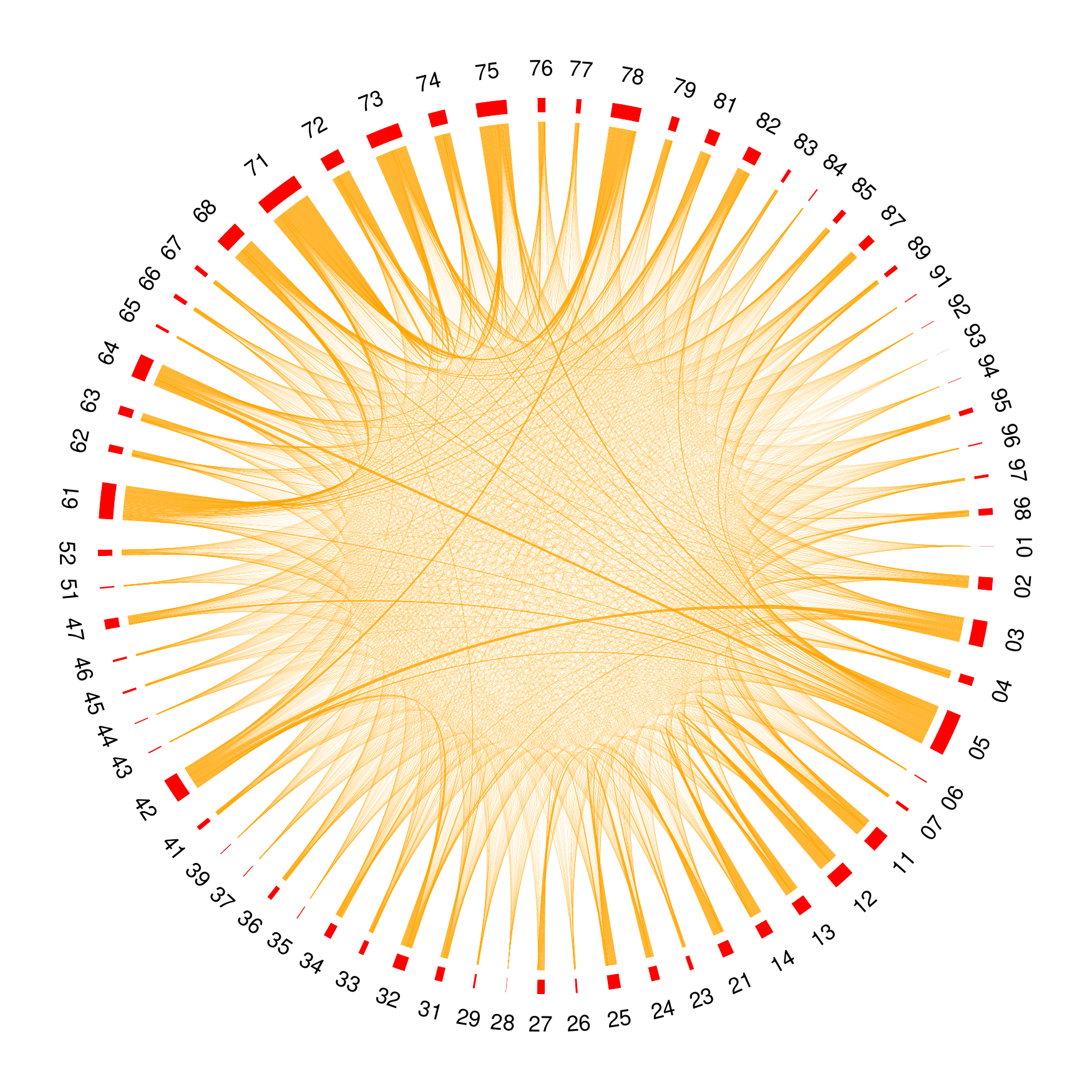}}%
\subcaptionbox{Last 10 Years}
{\includegraphics[scale=0.5]{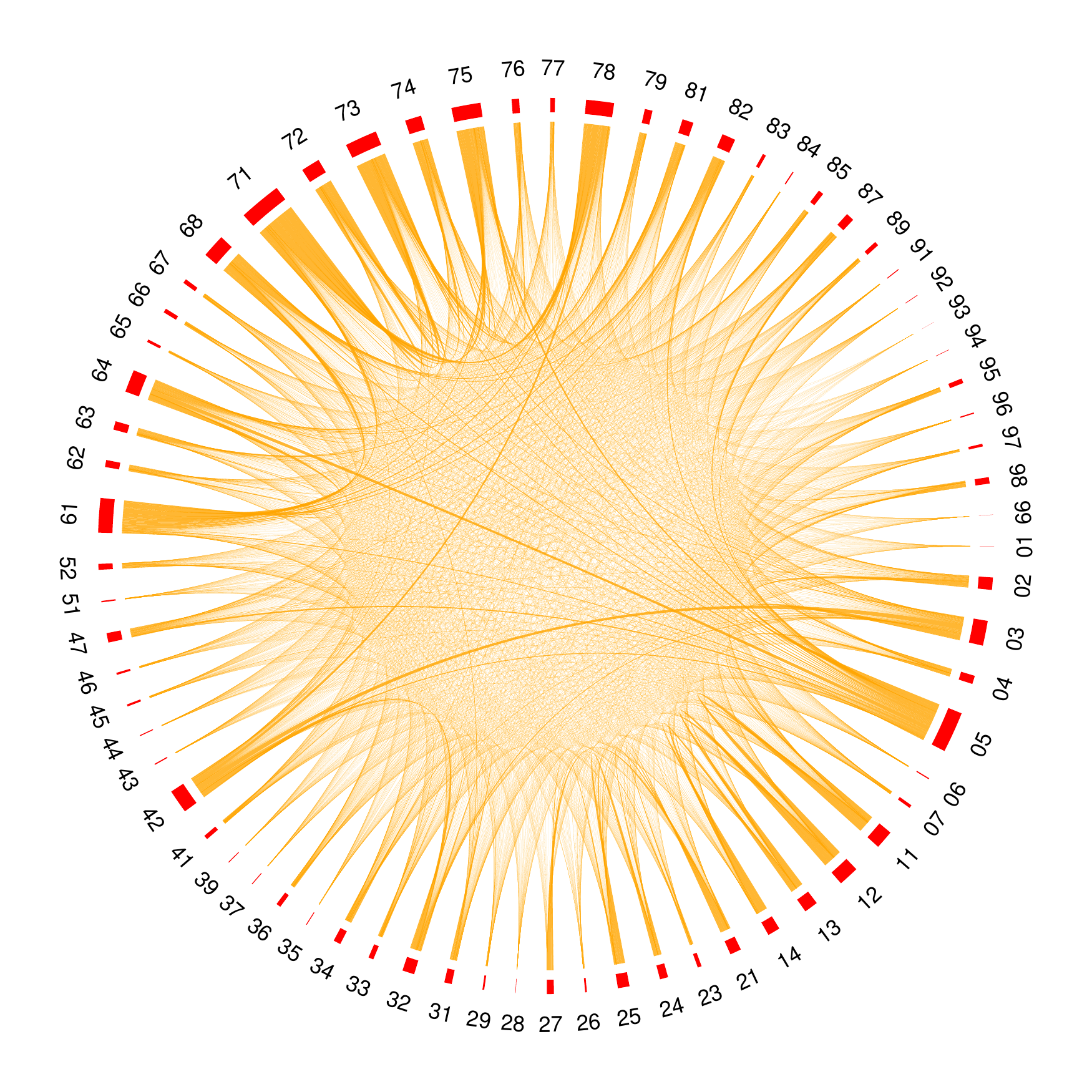}}
\subcaptionbox{Full Data}
{\includegraphics[scale=0.5]{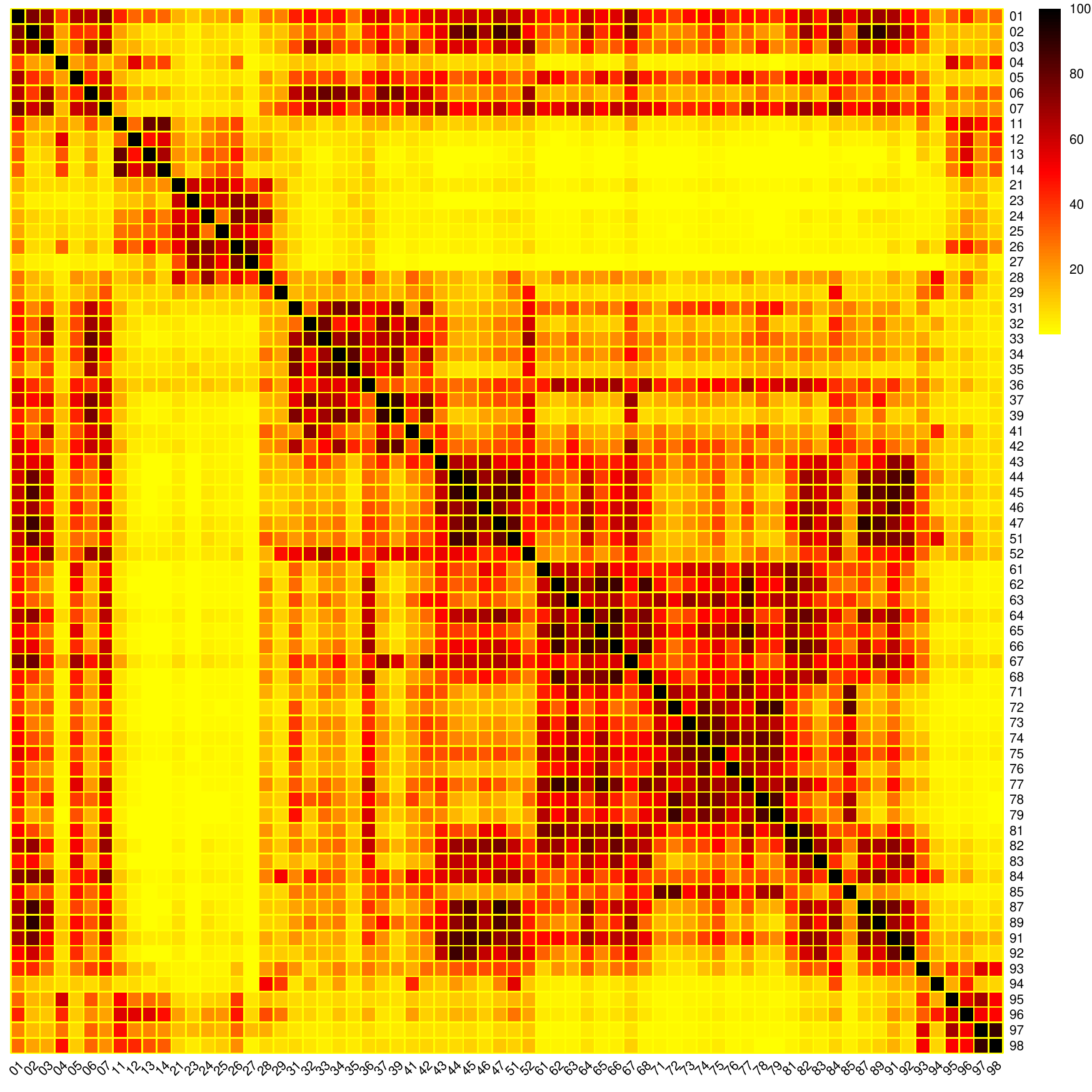}}%
\subcaptionbox{Last 10 Years}
{\includegraphics[scale=0.5]{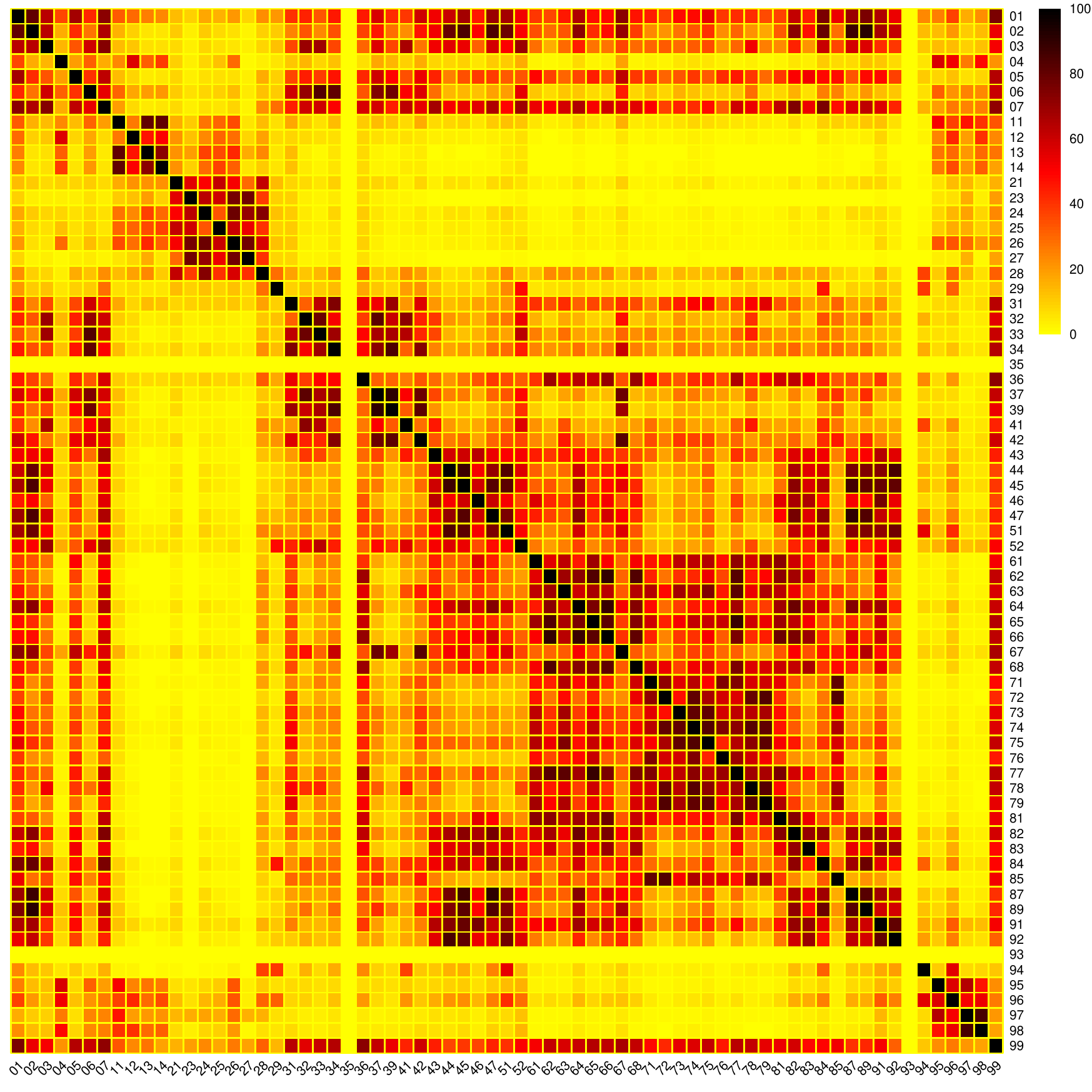}}
\caption{\textbf{Knowledge relatedness over time.}
{\small We focus on a subset including articles published from 2000 to 2009 (last 10 years in our data) to evaluate the evolution of the physics knowledge space over time. Despite a slightly general increase of interdisciplinarity, subject proximity indicates a stable structure among sub-fields.}
\label{knowledge_time}}
\end{figure}

\begin{figure}[b]
\centering 
{
\subcaption{1980 - 1989}

\includegraphics[scale=0.35]{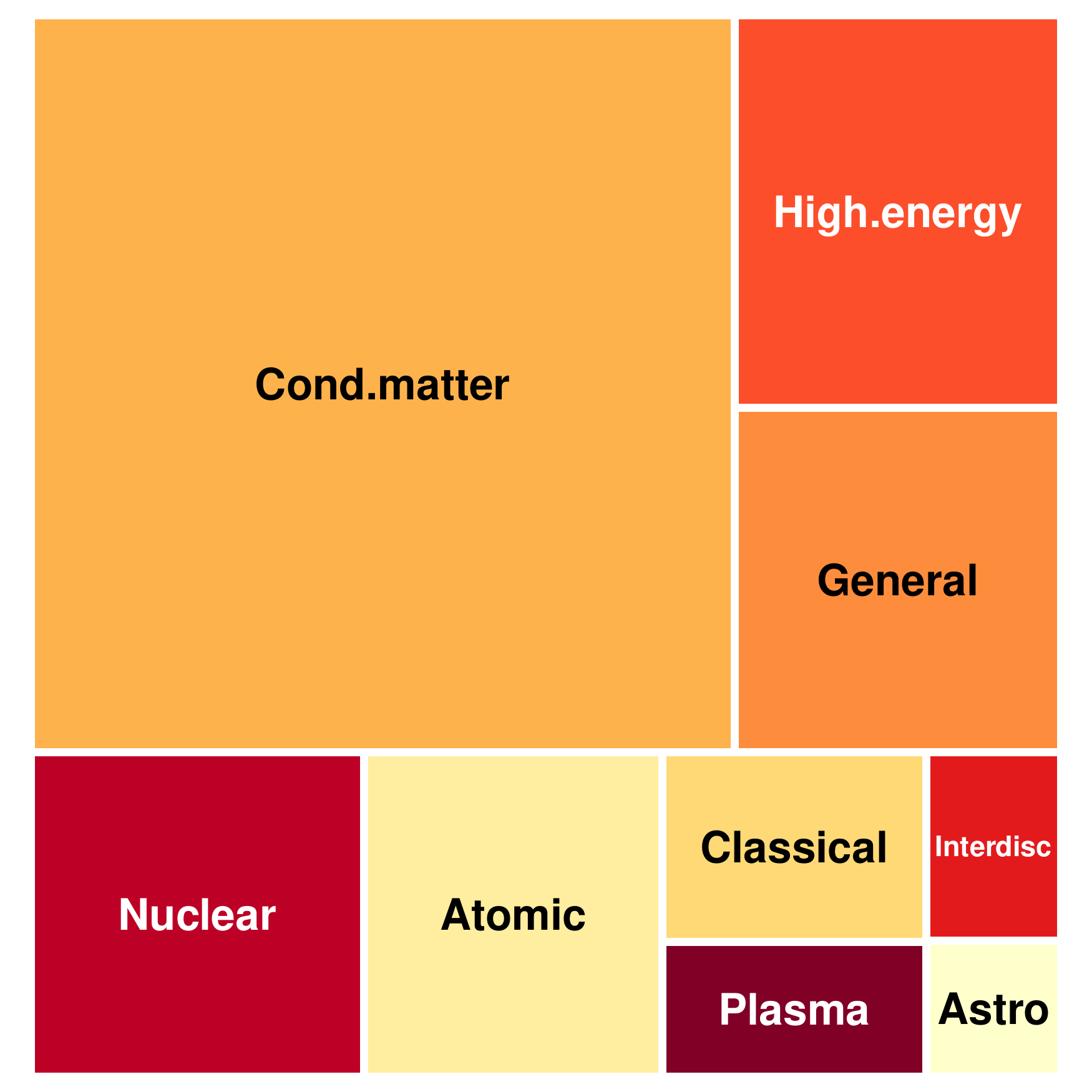}%
\includegraphics[scale=0.35]{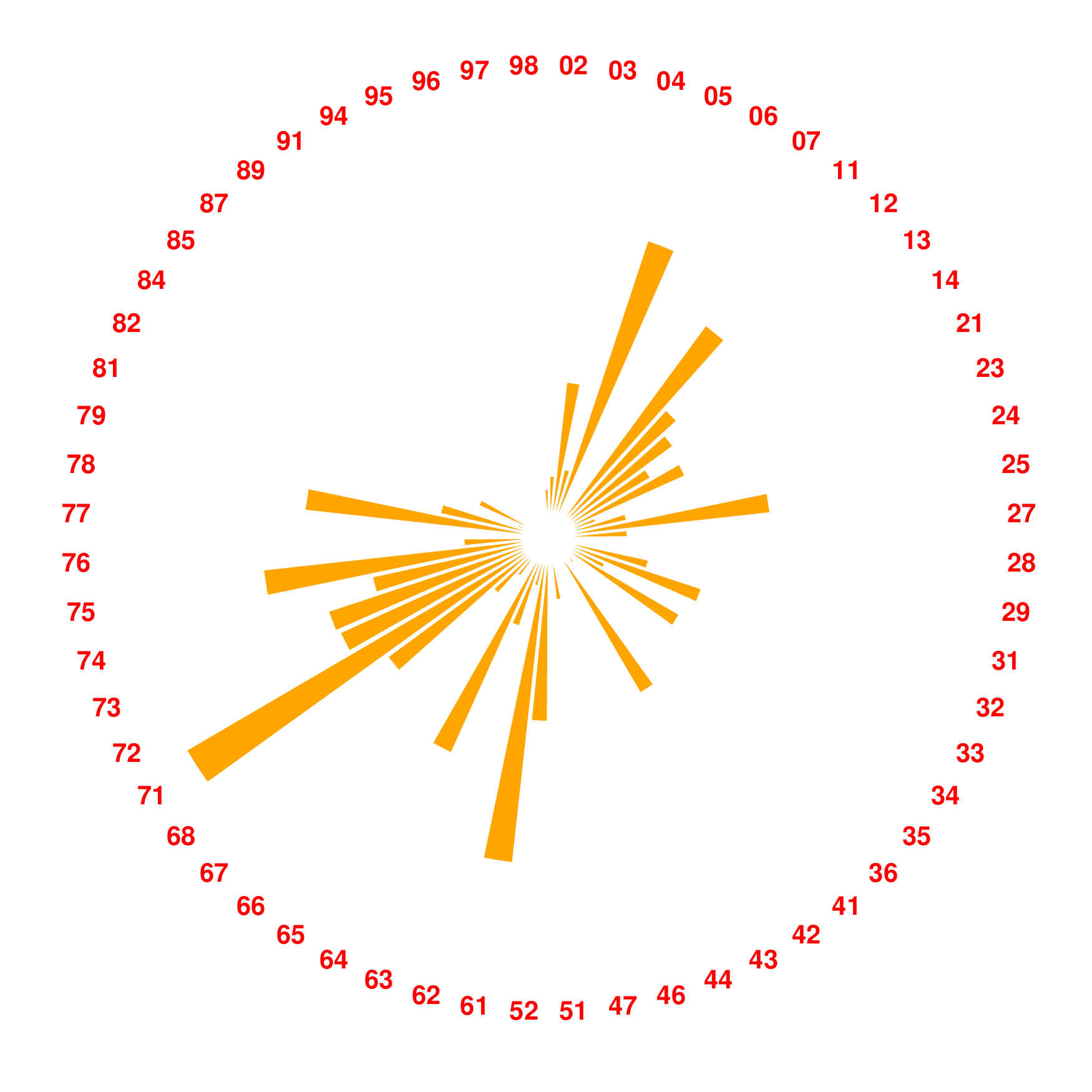}

\subcaption{1990 - 1999}
\includegraphics[scale=0.35]{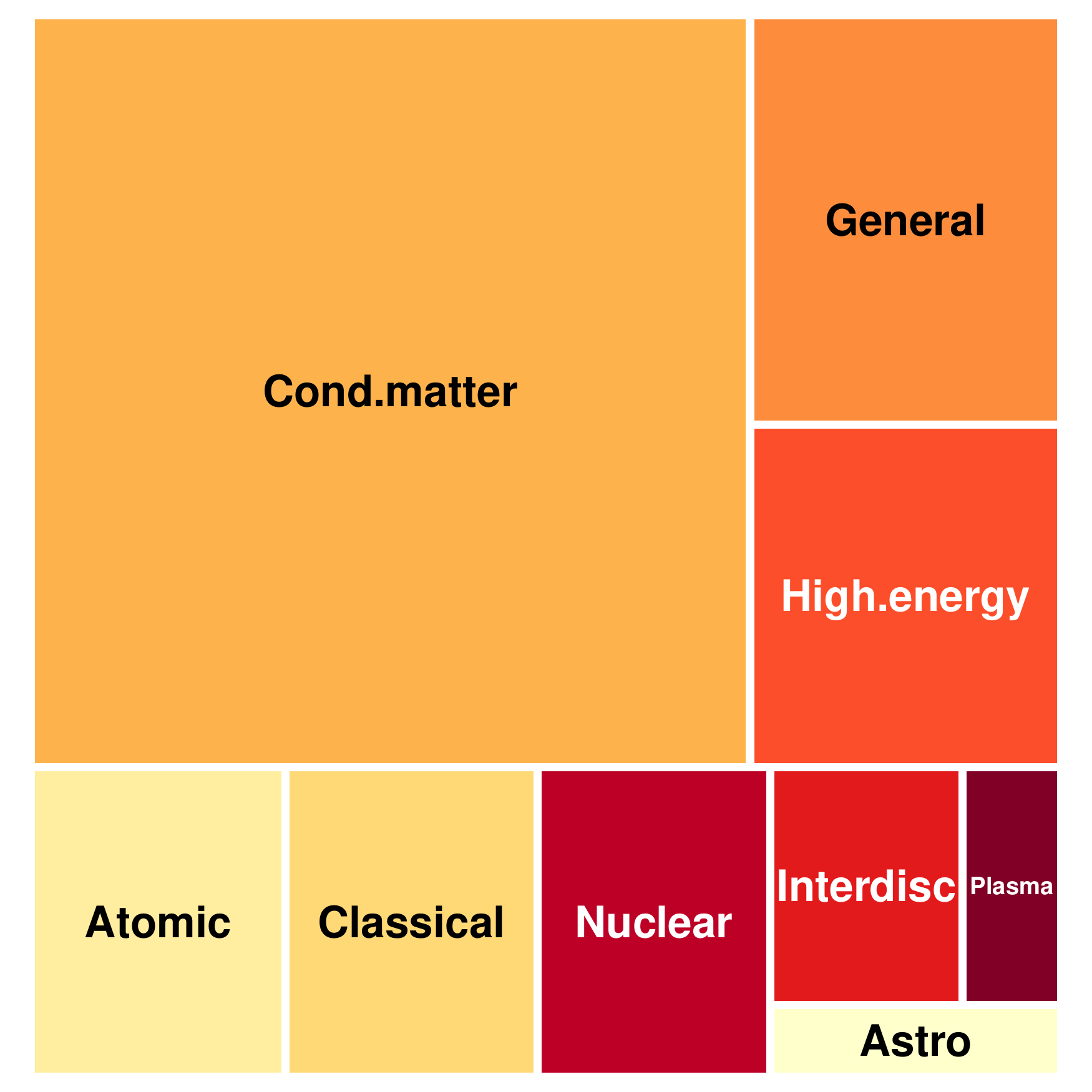}%
\includegraphics[scale=0.35]{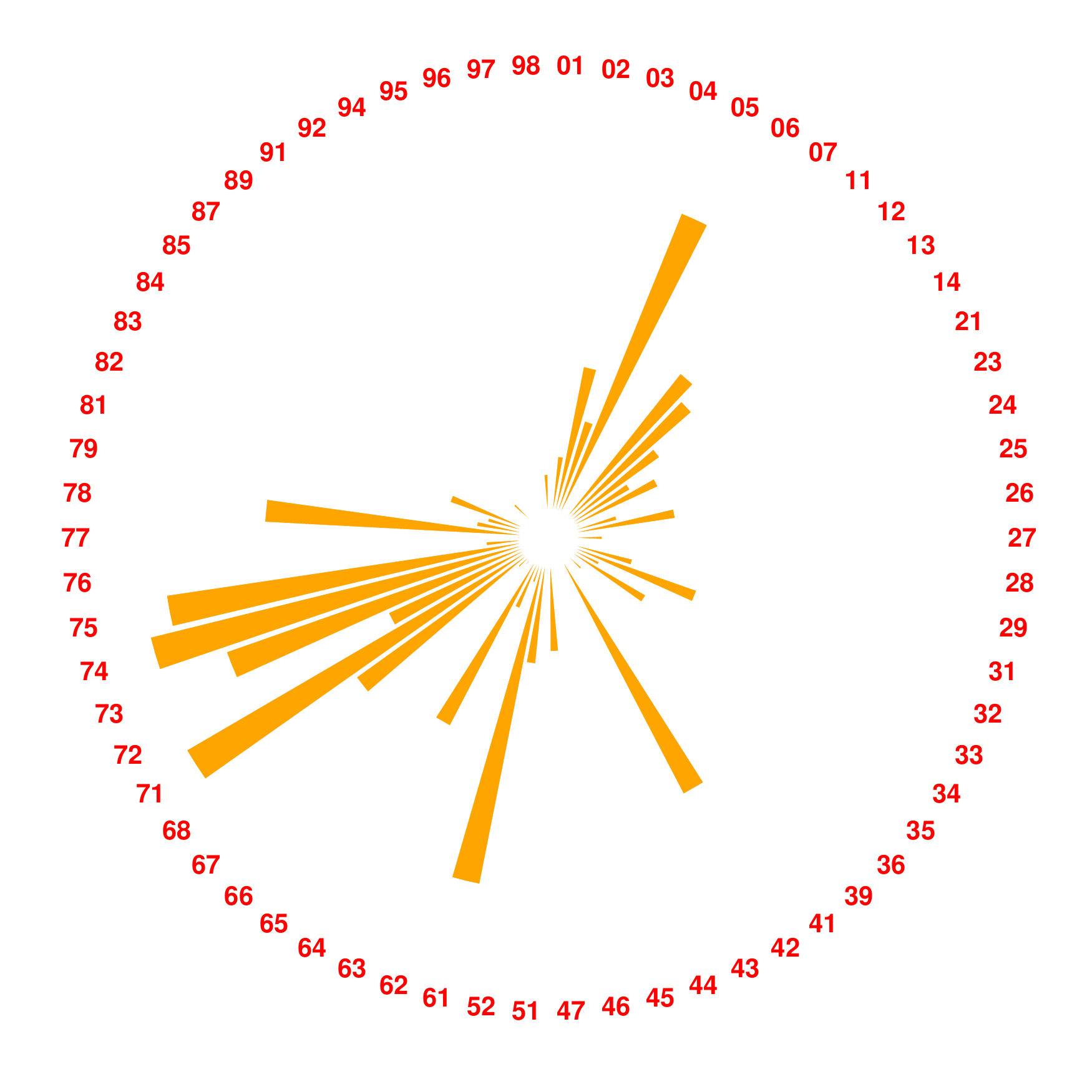}

\subcaption{2000 - 2009}
\includegraphics[scale=0.35]{treemap_field_00_09}%
\includegraphics[scale=0.35]{circle_pacs__00_09}
}
\caption{\textbf{Popularity of fields and sub-fields through decades}. {\small The plots compare the popularity of physics fields and sub-fields over time (i.e., number of articles assigned to a given field/sub-field).} \label{decades}}
\end{figure}

The structure of the  knowledge space can evolve over time, and sharp differences might undermine our strategy. To check whether such changes are significant, we split our initial dataset into three subsets, one for each decade: 1980-1989, 1990-1990, 2000-2009.  
We compare the structure of the physics knowledge space in the last decade of our sample with the one referring to the entire period. Figure \ref{pacs_time} compares popularity of one- and two-digit PACS in the last decade with the one for the full sample. Figure \ref{knowledge_time} shows how the network and, as a consequence, the cosine similarity matrix have  changed in the last ten years. Figure \ref{decades} shows the popularity of one- and two-digit PACS in the three decades. Data confirm the rise of interdisciplinary physics within an otherwise stable distribution of interests, as much as it was observed in previous works \cite{pan2012evolution}.

\begin{figure}[b]
\centering 
\subcaptionbox{\tiny{1980-1989}}
{\includegraphics[scale=0.5]{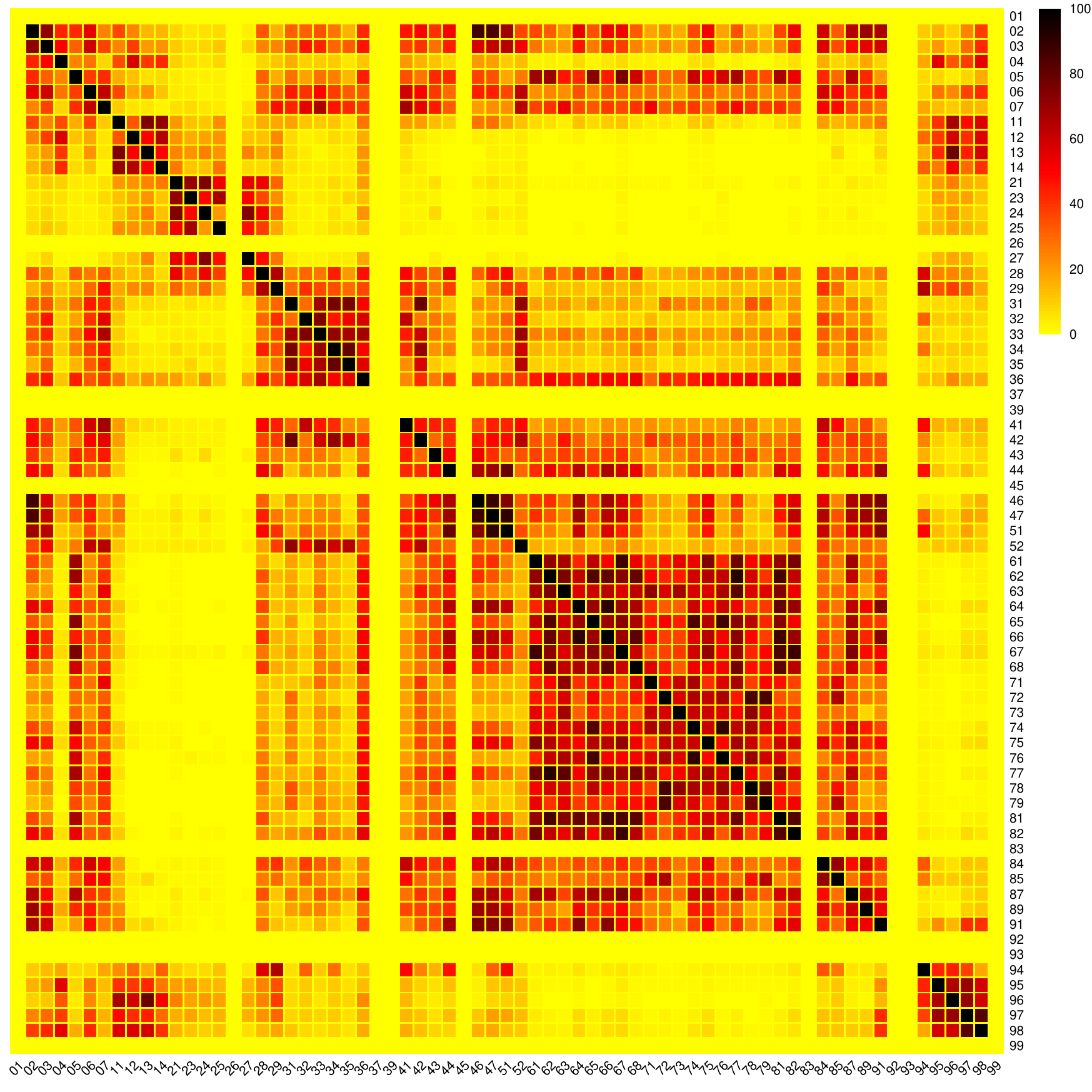}}%
\subcaptionbox{\tiny{1990-1999}}
{\includegraphics[scale=0.5]{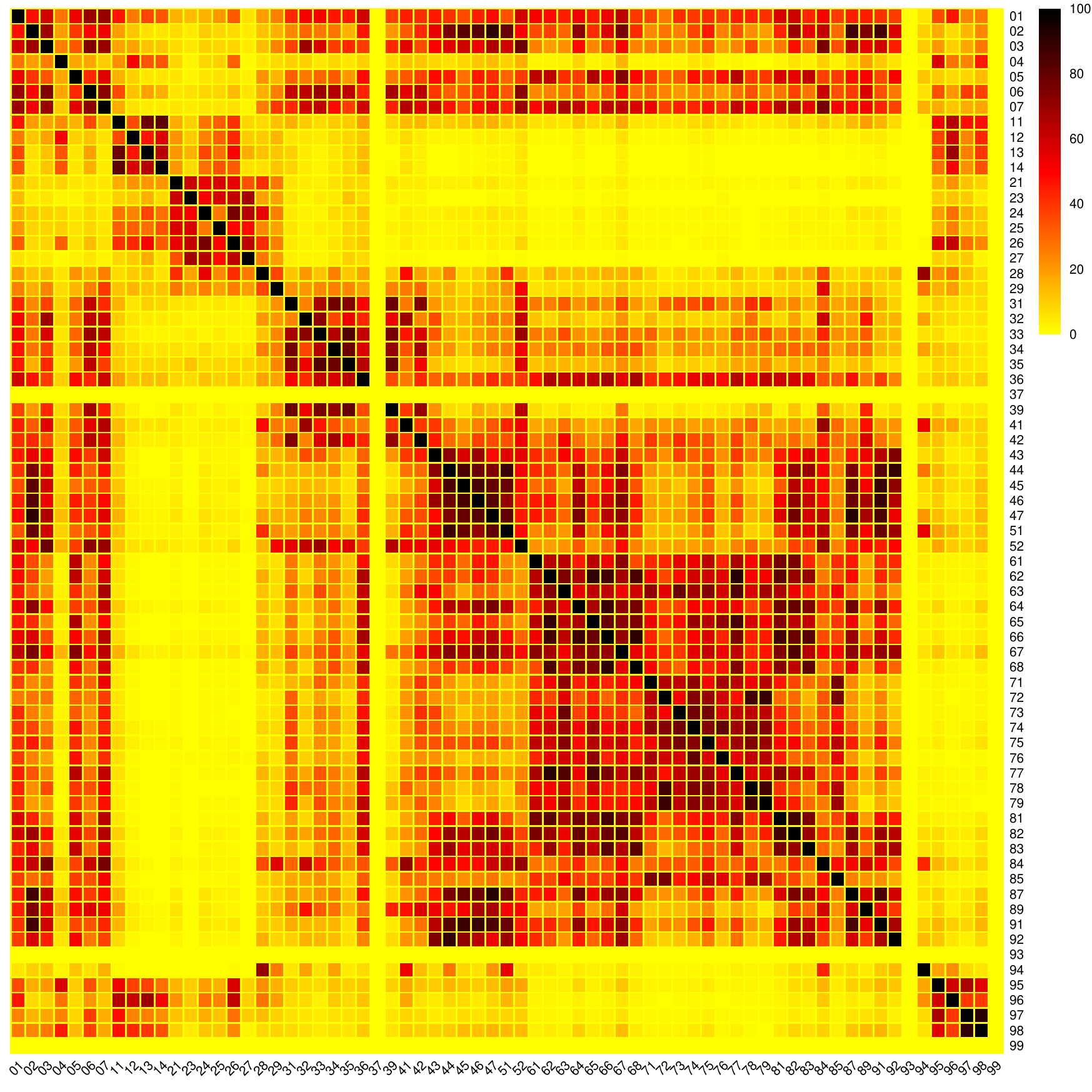}}
\subcaptionbox{\tiny{Difference}}
{\includegraphics[scale=0.5]{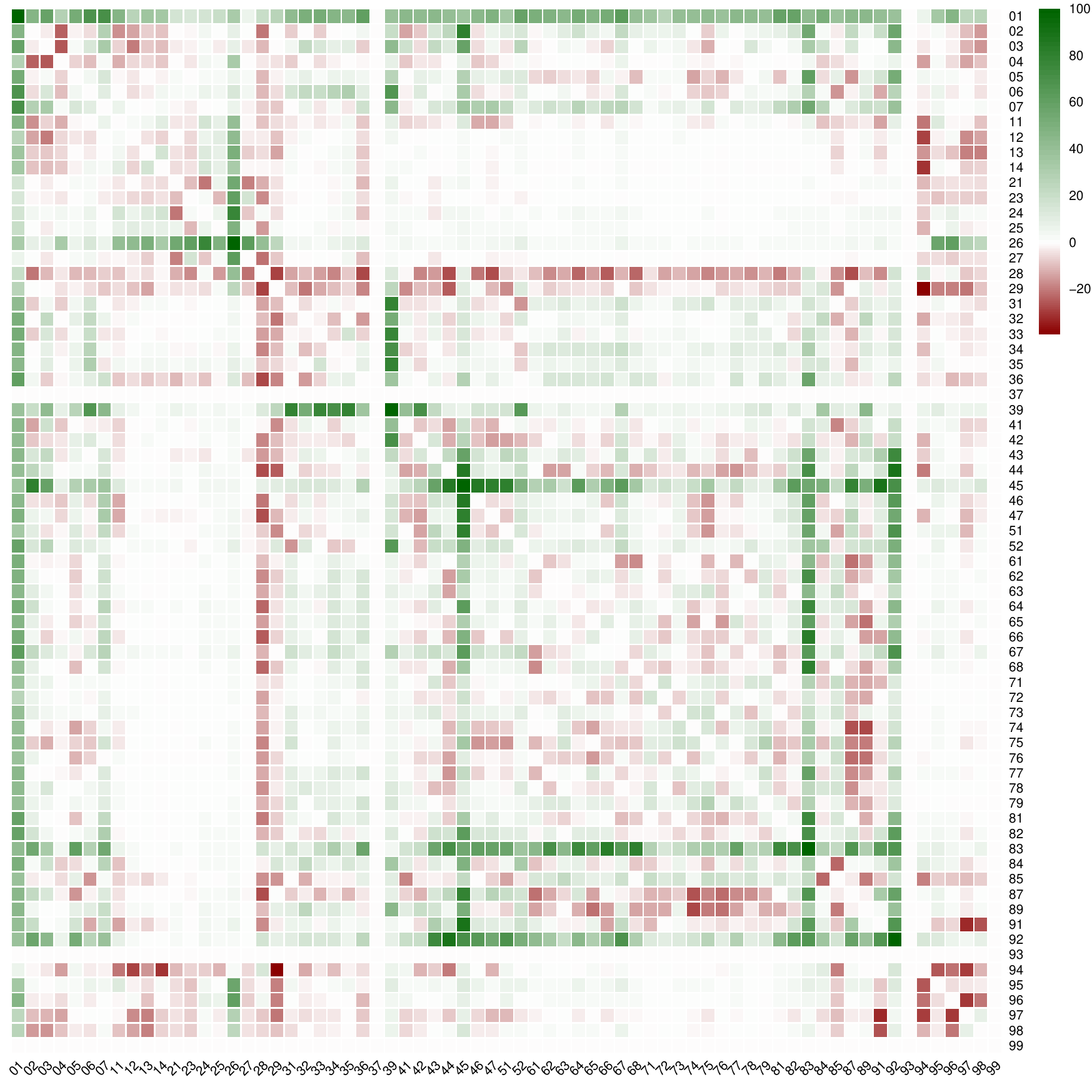}}%
\caption{\textbf{Knowledge relatedness evolution over the first two decades. } The top panels show the cosine similarity matrix between two-digit PACS in two decades, while the bottom panel shows their difference. \label{diff1}}
\end{figure}

\begin{figure}[b]
\centering 
\subcaptionbox{\tiny{1980-1989}}
{\includegraphics[scale=0.5]{sim_80_89}}%
\subcaptionbox{\tiny{2000-2009}}
{\includegraphics[scale=0.5]{sim_00_09}}
\subcaptionbox{\tiny{Difference}}
{\includegraphics[scale=0.5]{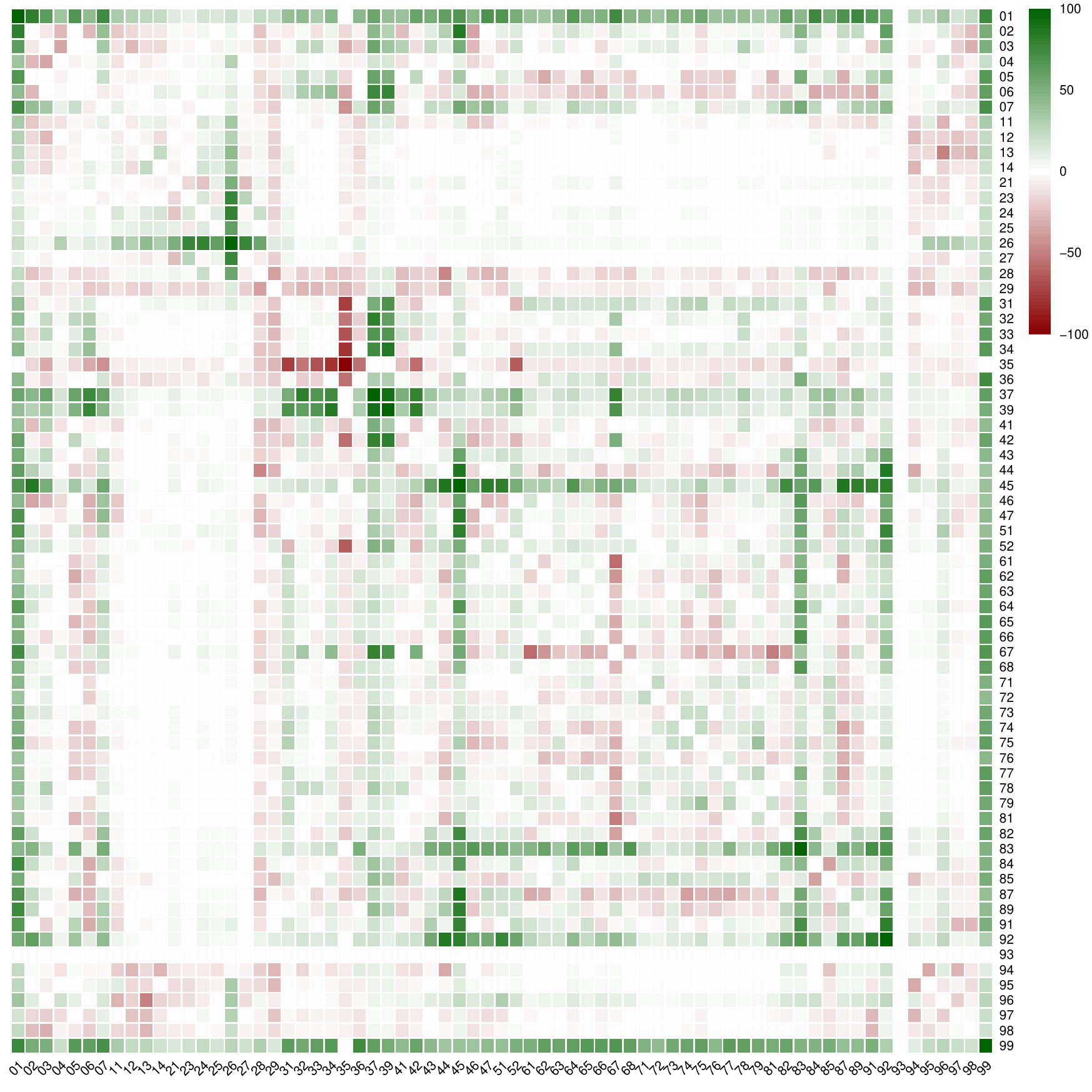}}%
\caption{\textbf{Knowledge relatedness evolution over three decades.} The top panels show the cosine similarity matrix between two-digit PACS in two decades, while the bottom panel shows their difference.\label{diff3}}
\end{figure}

\begin{figure}[b]
\centering 
\subcaptionbox{\tiny{90/99-80/89}}
{\includegraphics[scale=0.25]{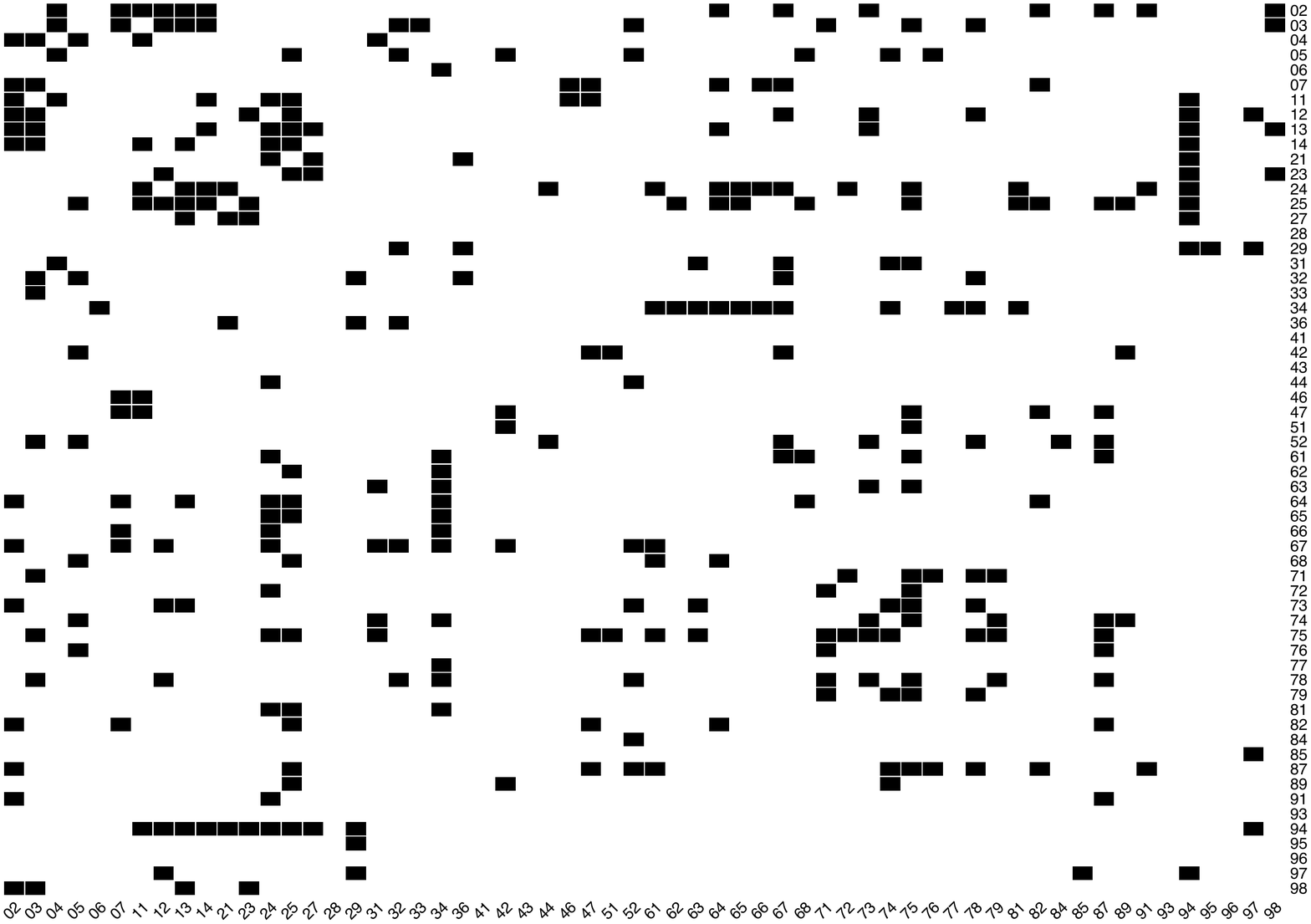}}

\subcaptionbox{\tiny{00/09-90/99}}
{\includegraphics[scale=0.25]{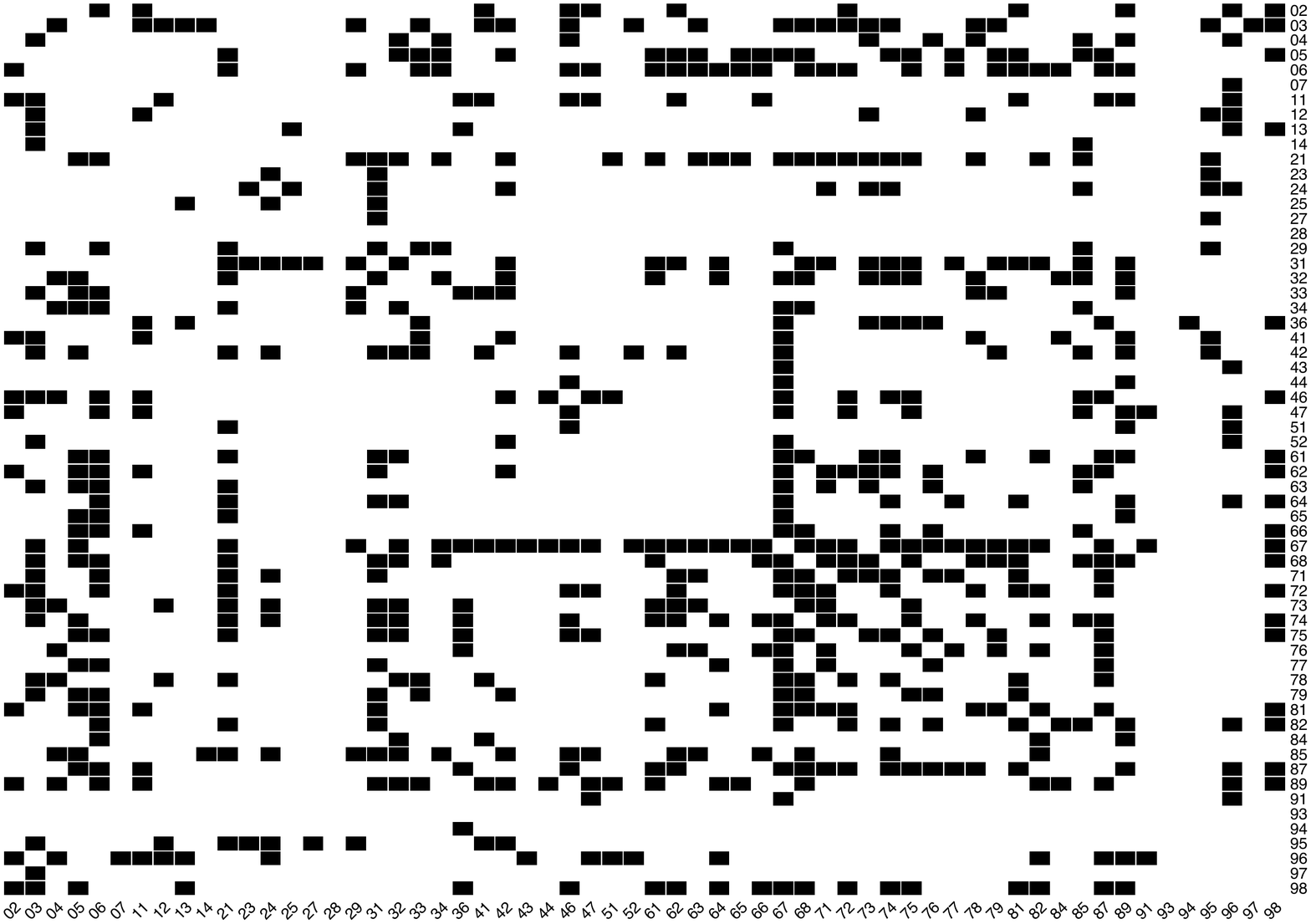}}

\subcaptionbox{\tiny{00/09-80/89}}
{\includegraphics[scale=0.25]{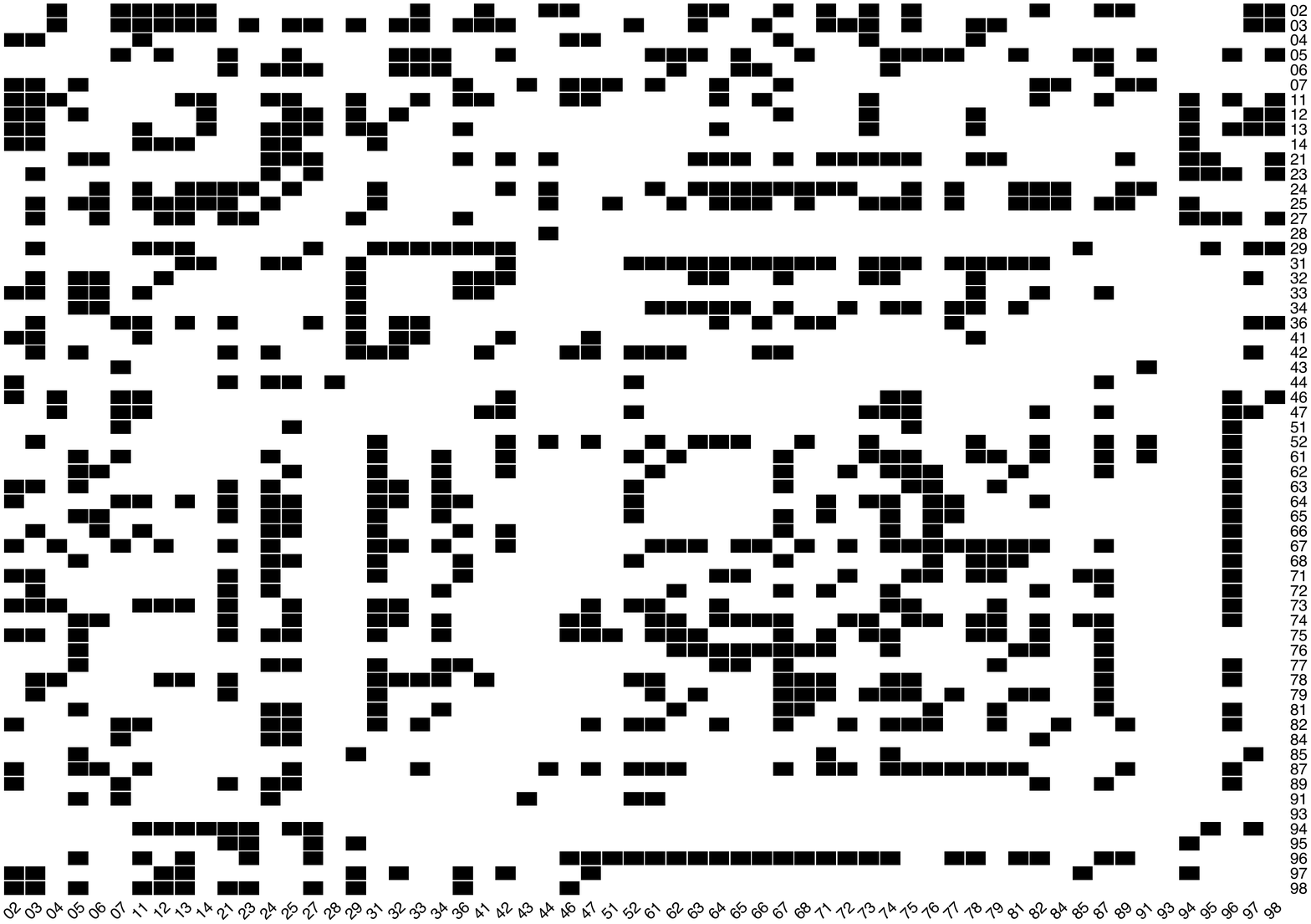}}

\caption{\textbf{Bootstrap validation.} 
{\small Bootstrap validation of the difference matrices computed over decades by sampling with replacement and generating 1,000 additional difference matrices (showing only PACS codes present in each decade). The confidence interval ($\alpha=0.05$) for each element of the matrix assesses the statistical significance (elements in black), taking into account multiple hypothesis testing correction (Bonferroni correction). }
\label{boot4}}
\end{figure}

Since our measure of knowledge relatedness depends on PACS co-occurrences in research articles, we provide a more robust quantitative test to check whether the relationships among sub-fields have changed significantly over time. To do so, we
first construct the difference between the cosine similarity matrix in two decades (see Figure \ref{diff1} and \ref{diff3}). Then we validate the resulting difference matrices against the null of zero difference by sampling with replacement and generating 1,000 additional of such matrices. Finally, we compute the confidence interval ($\alpha=0.05$) for each element of the difference matrix to assess its statistical significance, taking into account multiple hypothesis testing issues (Bonferroni correction). 
Figure \ref{boot4} shows the results of the bootstrap validation procedure (statistically significant pairs in black). In general, the number of significant element is not large, especially for consecutive decades, indicating a fairly stable structure of the physics knowledge space. More importantly, the analysis discussed in Section \ref{timestep}, where past knowledge space is used in the regression, shows that changes in  knowledge relatedness do not affect the main conclusions on the drivers of research portfolio diversification.

\subsection{Alternative estimation strategies}
\subsubsection{Multidisciplinarity}\label{constrained}

\begin{figure}[b]
\centering 
{\includegraphics[scale=0.5]{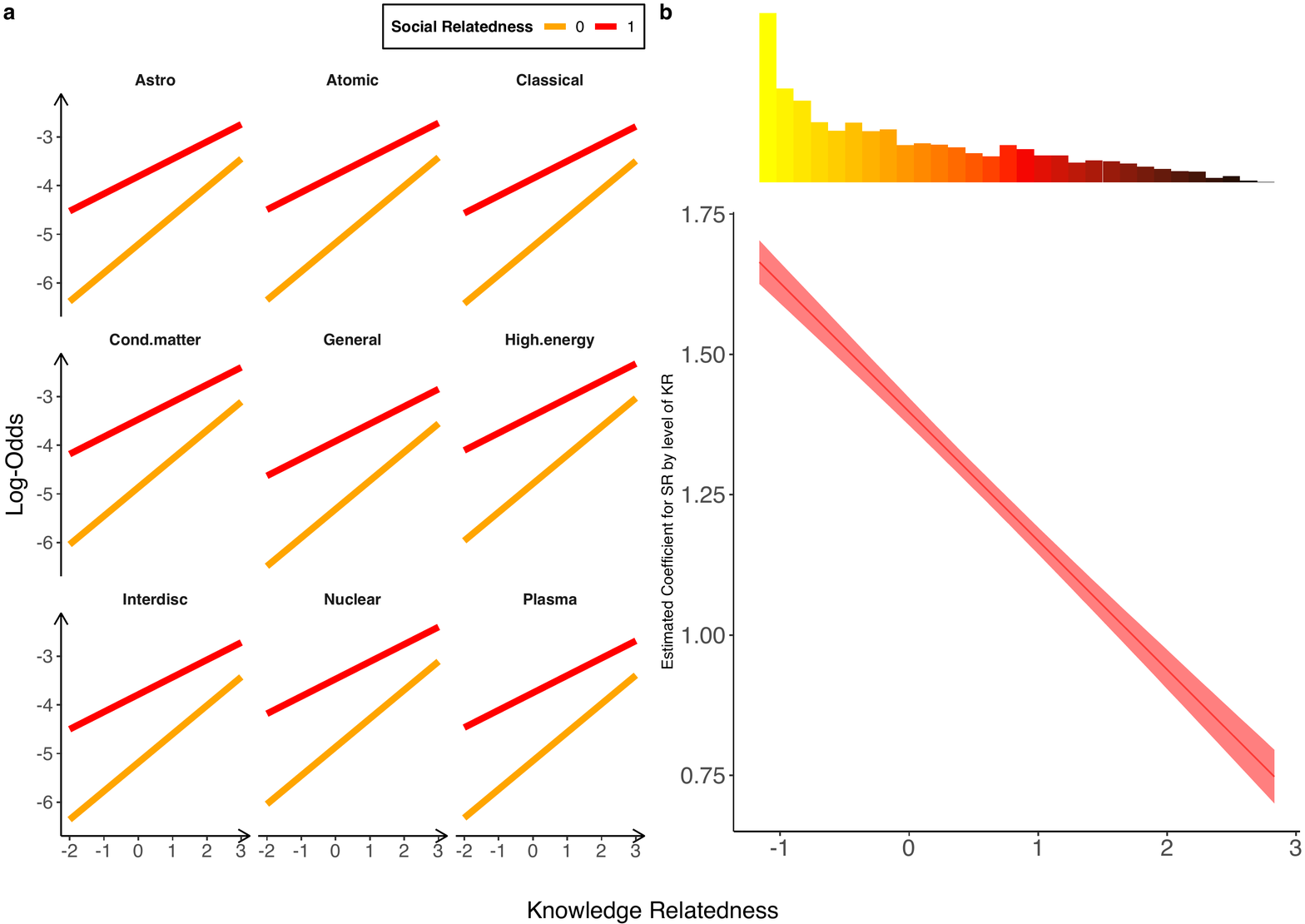}}
\caption{\textbf{Scientists' research portfolio diversification - constrained diversification.} (a) Log-odds as a function of (binary) social relatedness and (standardized) knowledge relatedness, accounting for multiple control variables. (i). (b) Estimated coefficient for social relatedness conditional on knowledge relatedness, and distribution of knowledge relatedness. The analysis is performed considering only truly unexplored sub-fields (see text). \label{odds}}
\end{figure}

Keeping track of diversification patterns for truly multidisciplinary scientists is a non-trivial task. Indeed, some scientists might have several core specializations leading to a positive bias in the previous estimates. To take into account this issue, we present an additional robustness check to validate further our empirical strategy: we assign each scientist to a single specialization - the one corresponding to the maximum value of $RSA$ -  but we constrain the choices of each scientists by eliminating from the regression the possibility to diversify in any of the PACS for which $RSA>0$. 
In other words, we take into account only truly unexplored sub-fields. Figure \ref{odds} confirms that scientists research portfolio diversification depends on social and knowledge relatedness, and the two measures interact with each other. Table \ref{strict} summarizes the results.

\begin{table}[b] \centering 
  \caption{Constrained diversification.} 
  \label{strict} 
                       \fontsize{11}{12}\selectfont

\begin{tabular}{@{\extracolsep{5pt}}lccc} 
\\[-1.8ex]\hline 
\hline \\[-1.8ex] 
 & \multicolumn{3}{c}{\textit{Dependent variable: P(diversification)}} \\ 
\cline{2-4} 
\\[-1.8ex] & \multicolumn{3}{c}{} \\ 
 & Baseline & Interactions & Robust SE \\ 
\\[-1.8ex] & (1) & (2) & (3)\\ 
\hline \\[-1.8ex] 
 Knowledge Relatedness& 0.507$^{***}$ & 0.586$^{***}$ & 0.586$^{***}$ \\ 
  & (0.005) & (0.006) & (0.007) \\ 
  Social Relatedness& 1.268$^{***}$ & 1.398$^{***}$ & 1.398$^{***}$ \\ 
  & (0.012) & (0.013) & (0.014) \\ 
  field core-Atomic & 0.036 & 0.029 & 0.029 \\ 
  & (0.025) & (0.025) & (0.023) \\ 
  field core-Classical & $-$0.034 & $-$0.043$^{*}$ & $-$0.043$^{*}$ \\ 
  & (0.026) & (0.026) & (0.024) \\ 
  field core-Cond.matter & 0.341$^{***}$ & 0.342$^{***}$ & 0.342$^{***}$ \\ 
  & (0.027) & (0.027) & (0.028) \\ 
  field core-General & $-$0.092$^{***}$ & $-$0.105$^{***}$ & $-$0.105$^{***}$ \\ 
  & (0.027) & (0.027) & (0.026) \\ 
  field core-High.energy & 0.426$^{***}$ & 0.418$^{***}$ & 0.418$^{***}$ \\ 
  & (0.030) & (0.030) & (0.027) \\ 
  field core-Interdisc & 0.040 & 0.023 & 0.023 \\ 
  & (0.026) & (0.026) & (0.024) \\ 
  Nuclear & 0.326$^{***}$ & 0.341$^{***}$ & 0.341$^{***}$ \\ 
  & (0.027) & (0.027) & (0.024) \\ 
  field core-Plasma & 0.063$^{*}$ & 0.058 & 0.058$^{*}$ \\ 
  & (0.036) & (0.036) & (0.032) \\ 
  {\#}\ of\ PACS & 0.761$^{***}$ & 0.753$^{***}$ & 0.753$^{***}$ \\ 
  & (0.007) & (0.007) & (0.006) \\ 
  {\#}\ of\ papers & 0.374$^{***}$ & 0.389$^{***}$ & 0.389$^{***}$ \\ 
  & (0.007) & (0.007) & (0.006) \\ 
  PACS target popularity & 1.478$^{***}$ & 1.473$^{***}$ & 1.473$^{***}$ \\ 
  & (0.006) & (0.006) & (0.006) \\ 
  $\Delta$ crowd & 0.193$^{***}$ & 0.192$^{***}$ & 0.192$^{***}$ \\ 
  & (0.006) & (0.006) & (0.006) \\ 
  {\#}\ of\ co-authors & $-$0.099$^{***}$ & $-$0.116$^{***}$ & $-$0.116$^{***}$ \\ 
  & (0.006) & (0.006) & (0.006) \\ 
  $\Delta$ PACS citations & $-$0.380$^{***}$ & $-$0.381$^{***}$ & $-$0.381$^{***}$ \\ 
  & (0.007) & (0.007) & (0.006) \\ 
  $\Delta$ field citations& 0.309$^{***}$ & 0.320$^{***}$ & 0.320$^{***}$ \\ 
  & (0.008) & (0.008) & (0.009) \\ 
  KR:SR &  & $-$0.230$^{***}$ & $-$0.230$^{***}$ \\ 
  &  & (0.009) & (0.010) \\ 
  Constant & $-$5.179$^{***}$ & $-$5.209$^{***}$ & $-$5.209$^{***}$ \\ 
  & (0.024) & (0.024) & (0.024) \\ 
 \hline \\[-1.8ex] 
Observations & 1,503,010 & 1,503,010 & 1,503,010 \\ 
Log Likelihood & $-$165,560.600 & $-$165,263.900 & $-$165,263.900 \\ 
Akaike Inf. Crit. & 331,157.100 & 330,565.900 & 330,565.900 \\ 
\hline 
\hline \\[-1.8ex] 
\textit{Note:}  & \multicolumn{3}{r}{$^{*}$p$<$0.1; $^{**}$p$<$0.05; $^{***}$p$<$0.01} \\ 
\end{tabular} 
\end{table} 

\subsubsection{Time dimension}\label{timestep}
\begin{figure}[b]
\centering 
\subcaptionbox{ {\tiny $logit(p_{t-1}) = \alpha + \beta KR_{t-2} + \gamma SR_{t-2}\ + \zeta (KR_{t-2} \times SR_{t-2}) + \delta field\ core $}} 
{\includegraphics[scale=0.3]{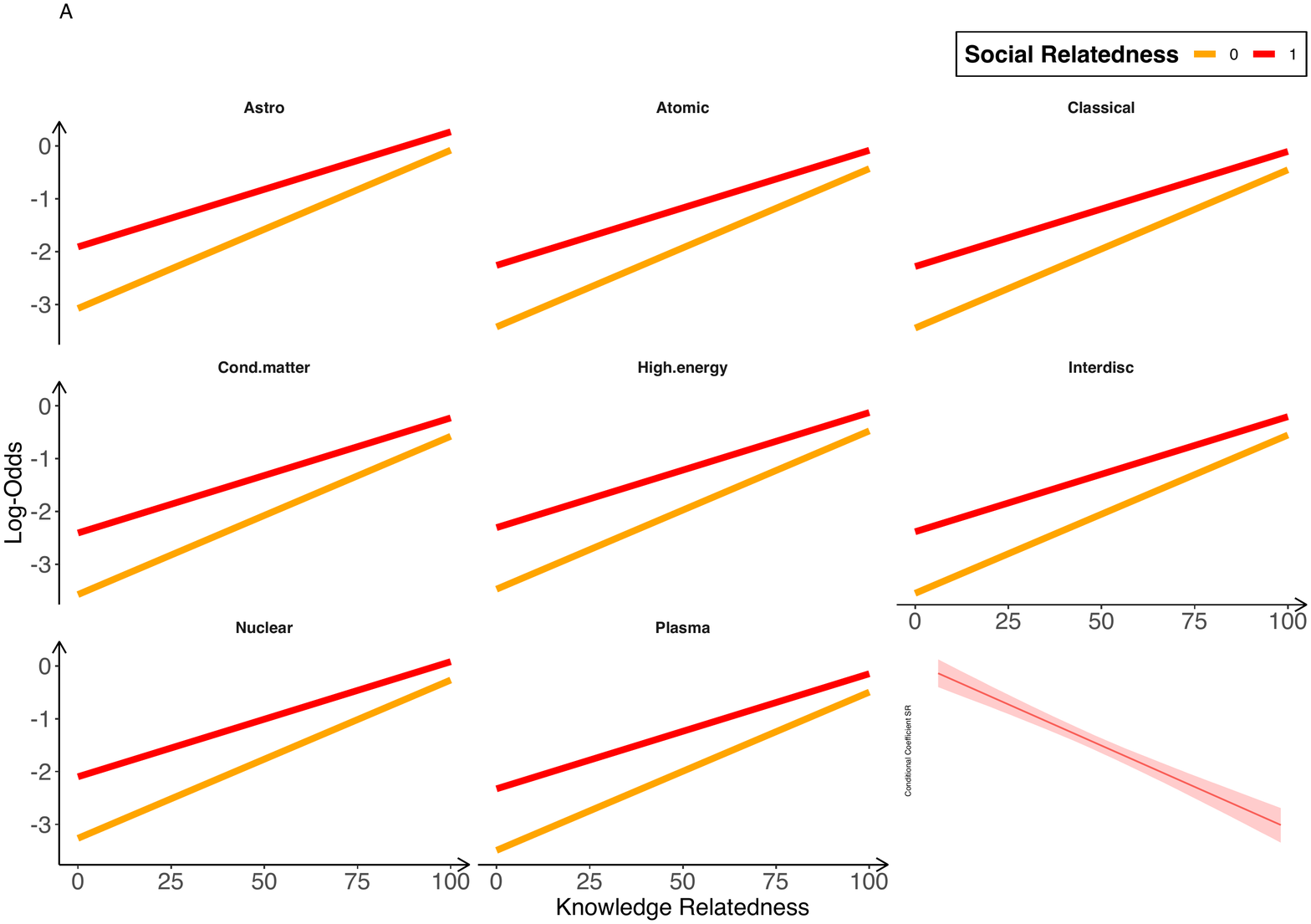}}
\subcaptionbox{ {\tiny $logit(p_{t}) = \alpha + \beta KR_{t-1} + \gamma SR_{t-1}\ + \zeta (KR_{t-1} \times SR_{t-1}) + \delta field\ core $}}
{\includegraphics[scale=0.3]{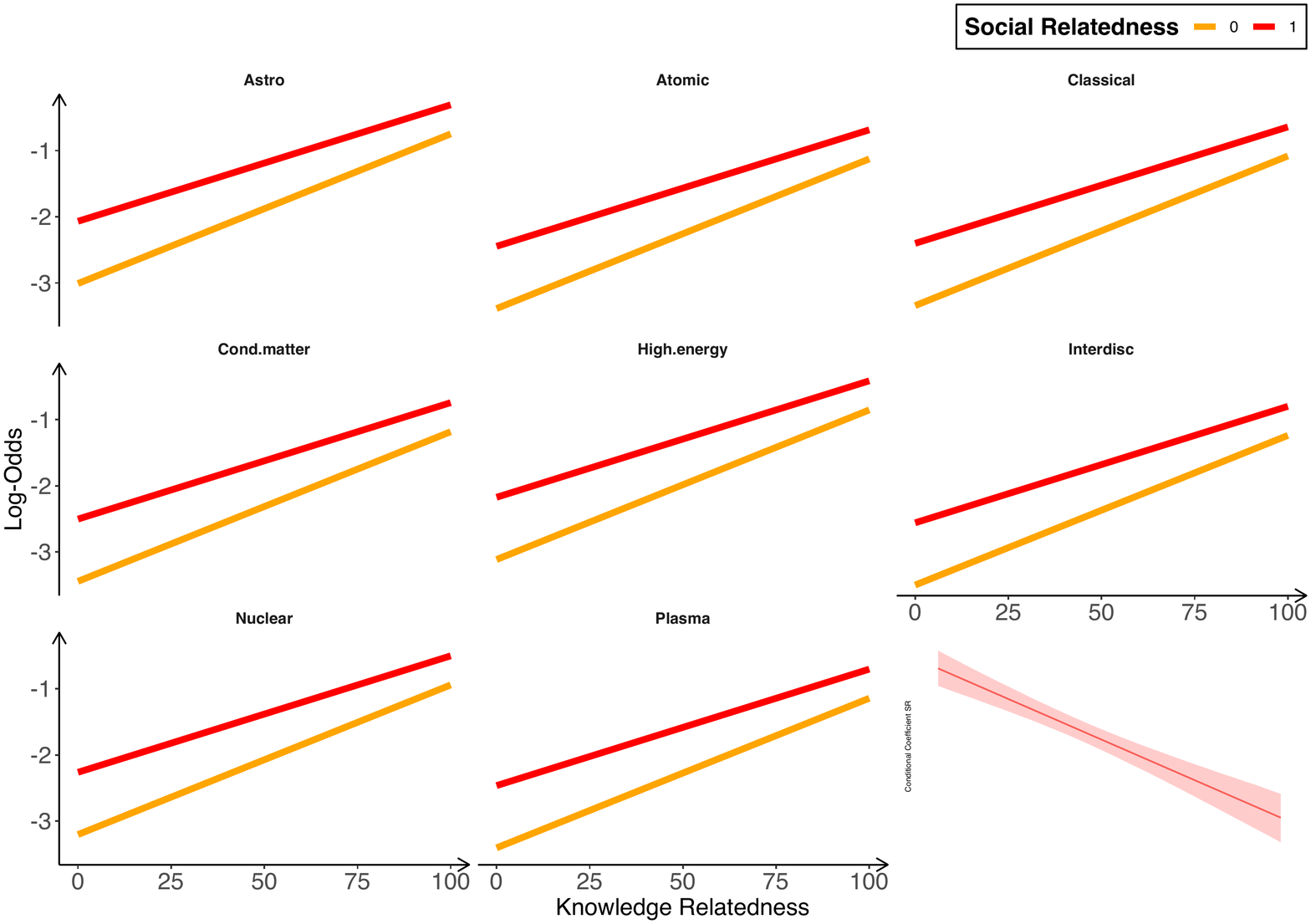}}
\subcaptionbox{  {\tiny $logit(p_{t}) = \alpha + \beta KR_{t-2} + \gamma SR_{t-2}\ +  \zeta (KR_{t-2} \times SR_{t-2}) + \delta field\ core $}}
{\includegraphics[scale=0.3]{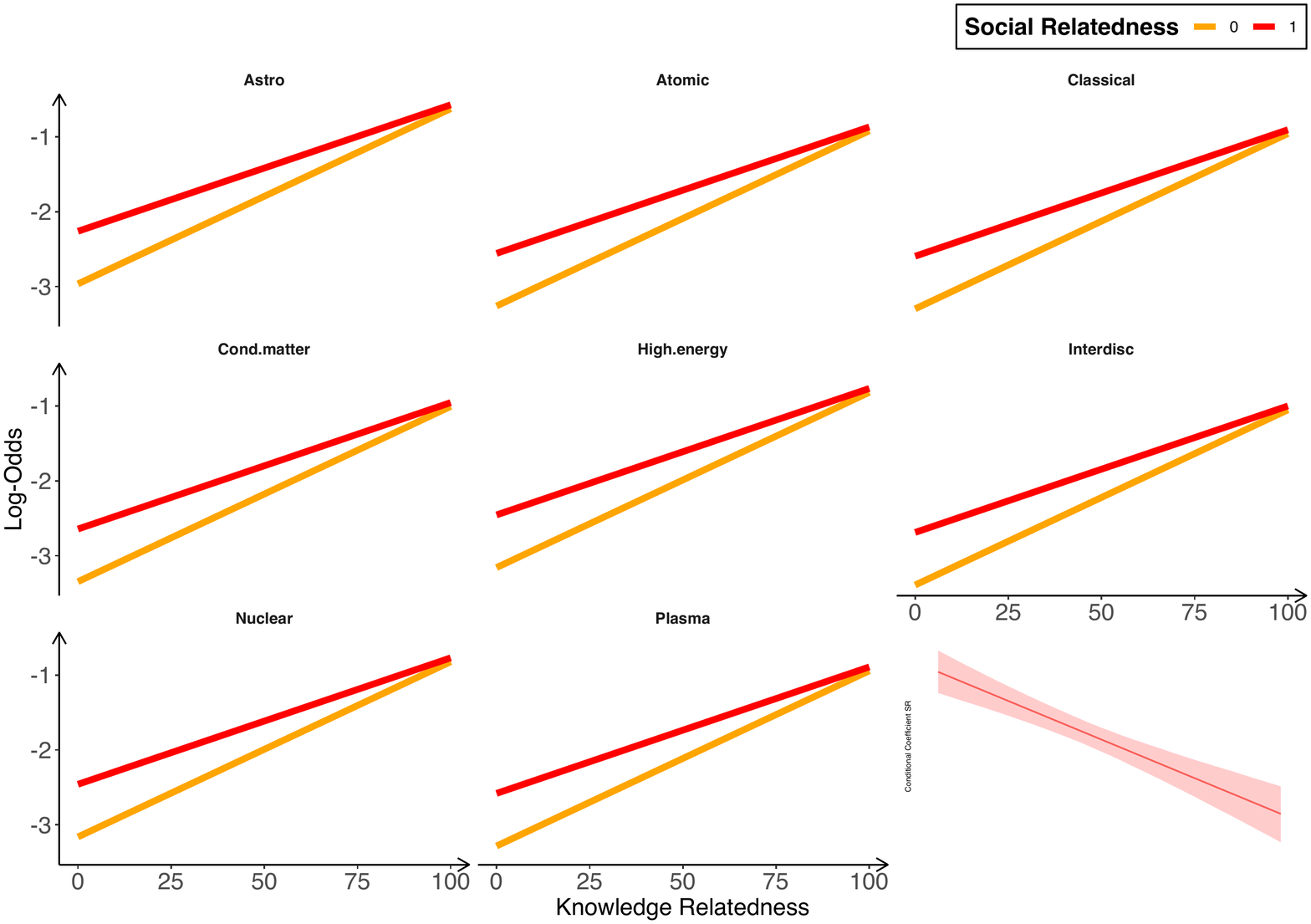}}
\caption{\textbf{Models with lagged variables.} Log-odds as function of social and (standardized) knowledge relatedness and (bottom right panel),
estimated coefficient for social relatedness conditional on (standardized) knowledge relatedness. \label{time}}
\end{figure}

The temporal dimension is of paramount importance when evaluating scientific activities, especially to disentangle the direction of causality. Over time, our measures of knowledge and social relatedness might be affected by scientists' research diversification themselves. We tackle this issue by running an additional robustness check to isolate the effect of our measures on scientists' diversification strategies. First, we split our dataset into three time periods (i.e., three decades: 1980-1989, 1990-1990, 2000-2009) and we identify 15,466 scientists active in all periods. Then, we compute our measures of knowledge and social relatedness for each period to predict authors' diversification in a given decade using relatedness measures of a past decade. As before, we use a logistic regression where our dependent variable is a binary one (being active in a sub-field different from specialization), but this time we use knowledge and social relatedness computed at time ${t-1}$ and ${t-2}$. 
Formally, we use three econometric specifications: 
\begin{equation}
ln(\frac{p_{t-1}}{1-p_{t-1}}) = \alpha + \beta KR_{t-2} + \gamma SR_{t-2}\ + \zeta (KR_{t-2} \times SR_{t-2}) + \delta field\ core 
\end{equation}
\begin{equation}
ln(\frac{p_{t}}{1-p_{t}}) = \alpha + \beta KR_{t-1} + \gamma SR_{t-1}\ + \zeta (KR_{t-1} \times SR_{t-1}) + \delta field\ core 
\end{equation}
\begin{equation}
ln(\frac{p_{t}}{1-p_{t}}) = \alpha + \beta KR_{t-2} + \gamma SR_{t-2}\ +  \zeta (KR_{t-2} \times SR_{t-2}) + \delta field\ core 
\end{equation}
where $t$ indicates the last decade (2000-2009). Such additional tests provide indication of the direction of causality since we take in account social and cognitive proximity prior to the scientists' choice to diversify. Moreover, we only consider sub-fields never explored before by each author so to approximate a quasi-experimental setting.
Results confirm the role played by knowledge and social relatedness as well as the negative interaction between our two measures (see Figure \ref{time} and Table \ref{lag}).

\begin{table}[b] \centering 
  \caption{Diversification (lag)} 
  \label{lag} 
                       \fontsize{11}{12}\selectfont

\begin{tabular}{@{\extracolsep{5pt}}lccc} 
\\[-1.8ex]\hline 
\hline \\[-1.8ex] 
 & \multicolumn{3}{c}{\textit{Dependent variable: P(diversification)}} \\ 
\cline{2-4} 
\\[-1.8ex] & $logit(p_{t-1})$ & \multicolumn{2}{c}{$logit(p_{t})$} \\ 
 & lag1 & lag1 & lag2 \\ 
\\[-1.8ex] & (1) & (2) & (3)\\ 
\hline \\[-1.8ex] 
 $KR_{t-2}$ & 0.030$^{***}$ &  & 0.023$^{***}$ \\ 
  & (0.0003) &  & (0.0003) \\ 
  $SR_{t-2}$ & 1.165$^{***}$ &  & 0.703$^{***}$ \\ 
  & (0.037) &  & (0.047) \\ 
  $KR_{t-1}$ &  & 0.023$^{***}$ &  \\ 
  &  & (0.0003) &  \\ 
  $SR_{t-1}$ &  & 0.941$^{***}$ &  \\ 
  &  & (0.035) &  \\ 
  field core-Atomic & $-$0.350$^{***}$ & $-$0.378$^{***}$ & $-$0.296$^{***}$ \\ 
  & (0.043) & (0.053) & (0.054) \\ 
  field core-Classical & $-$0.371$^{***}$ & $-$0.334$^{***}$ & $-$0.333$^{***}$ \\ 
  & (0.047) & (0.058) & (0.059) \\ 
  field core-Cond.matter & $-$0.495$^{***}$ & $-$0.434$^{***}$ & $-$0.383$^{***}$ \\ 
  & (0.039) & (0.049) & (0.049) \\ 
  field core-High.energy & $-$0.393$^{***}$ & $-$0.104$^{*}$ & $-$0.194$^{***}$ \\ 
  & (0.046) & (0.057) & (0.057) \\ 
  field core-Interdisc & $-$0.471$^{***}$ & $-$0.490$^{***}$ & $-$0.428$^{***}$ \\ 
  & (0.046) & (0.060) & (0.061) \\ 
  field core-Nuclear & $-$0.188$^{***}$ & $-$0.194$^{***}$ & $-$0.198$^{***}$ \\ 
  & (0.042) & (0.053) & (0.054) \\ 
  field core-Plasma & $-$0.416$^{***}$ & $-$0.395$^{***}$ & $-$0.319$^{***}$ \\ 
  & (0.050) & (0.062) & (0.063) \\ 
  $KR_{t-2}:SR_{t-2}$ & $-$0.008$^{***}$ &  & $-$0.007$^{***}$ \\ 
  & (0.001) &  & (0.001) \\ 
  $KR_{t-1}:SR_{t-1}$ &  & $-$0.005$^{***}$ &  \\ 
  &  & (0.001) &  \\ 
  Constant & $-$3.075$^{***}$ & $-$3.010$^{***}$ & $-$2.964$^{***}$ \\ 
  & (0.038) & (0.047) & (0.048) \\ 
 \hline \\[-1.8ex] 
Observations & 766,519 & 618,352 & 618,352 \\ 
Log Likelihood & $-$180,229.100 & $-$142,158.800 & $-$143,114.400 \\ 
Akaike Inf. Crit. & 360,480.300 & 284,339.500 & 286,250.900 \\ 
\hline 
\hline \\[-1.8ex] 
\textit{Note:}  & \multicolumn{3}{r}{$^{*}$p$<$0.1; $^{**}$p$<$0.05; $^{***}$p$<$0.01} \\ 
\end{tabular} 
\end{table}

\end{document}